\documentclass[atmp]{ipart_v1}

\Vol{19}
\Issue{3}
\Year{2015}
\firstpage{701}

\usepackage{t1enc}
\usepackage[latin1]{inputenc}
\usepackage[english]{babel}

\usepackage{amsthm}
\usepackage{yfonts}

\usepackage{bbm}
\usepackage{bm}
\usepackage{mathrsfs}
\usepackage{mathtools}
\usepackage{enumitem}
\usepackage{epstopdf}
\usepackage{subcaption}
\usepackage[all, knot]{xy}
\usepackage{tikz}
\usepackage{array}
\usepackage{hyperref}
\hypersetup{hidelinks}
\usepackage{booktabs}
\usepackage{amsfonts}
\usepackage{amsmath}
\usepackage{amssymb}
\usepackage{graphicx}

\usepackage{comment}

\def \be {\begin{equation}}
\def \ee {\end{equation}}
\def \bea {\begin{eqnarray}}
\def \eea {\end{eqnarray}}
\def \nn {\nonumber}

\def \rr {\raise.35ex\hbox{\small $\prime$}\kern-.17em{\mbox{\large $\imath$}}}

\def \dels {\partial\kern-.6em /\kern.1em}
\def \As {{A\kern-.5em / \kern.5em}}
\def \Ds {D\kern-.7em / \kern.5em}

\def \ks {k\kern-.5em /}
\def \ls {l\kern-.5em /}

\def \r {\rho}

\def \Z {{\mathcal{Z}}} 

\newcommand{\ci}[1]{}

\newcommand{\ba}{\begin{eqnarray}}
\newcommand{\ea}{\end{eqnarray}}
\newcommand{\bal}{\begin{align}}
\newcommand{\eal}{\end{align}}
\newcommand{\bay}[1]{\left(\begin{array}{#1}}
\newcommand{\eay}{\end{array}\right)}

\newcommand{\C}{{\mathcal{C}}} 
\newcommand{\sph}{\mathrm{sph}}
\newcommand{\Pcov}{P_{\mathrm{cov}}}
\newcommand{\PCL}{P_{\mathrm{CL}}}
\newcommand{\GM}{\mathrm{GM}}
\newcommand{\Ima}{\operatorname{Im}}
\newcommand{\Rea}{\operatorname{Re}}
\newcommand{\Lnc}{\ln_{\mathrm{cont}}}
\newcommand{\hyp}{\mathrm{hyp}}
\newcommand{\Fp}[1]{{}_2F_1\!\left(#1\right)}

\newcommand{\hide}[1]{}

\numberwithin{equation}{section}

\theoremstyle{plain}
\newtheorem{theorem}{Theorem}[section]

\newtheorem{proposition}[theorem]{Proposition}

  \usetikzlibrary{decorations.markings}

\begin{document}

\title{Covering Surfaces,\\ 
Conical-Defect Continuation, and Multi-Entropy at Arbitrary Genus}

\author[Chen-Te Ma]{Chen-Te Ma}

\begin{abstract}
\noindent
We study entanglement among three regions in a two-dimensional conformal field theory.
At each of the three points, copies of the theory join in equal groups.
For surfaces with multiple handles, the covering exponent equals the real part of a continued cone action.
Each handle beyond the first contributes $\mp\pi c/6$ to its imaginary part, where $c$ is the central charge.
The partition function supplies the opposite phase under this change of reference metric.
On a specified timelike path, interval phases cancel from the subtracted entropy.
Geometry reproduces the proposed gravity term; additional contributions can survive at large central charge.
\end{abstract}

\maketitle

\section{Introduction}
\label{sec:1}
\noindent
Suppose three observers share a pure quantum state. Each observer can measure everything in their own region. Even complete lists of the three sets of probabilities need not tell them whether they all share one label or each pair shares a separate label. What measurement can distinguish these possibilities, and how much of its answer is fixed by geometry?
We study three regions that together fill space in the vacuum of a two-dimensional conformal field theory (CFT). This is a quantum theory in one space and one time dimension with local symmetries that preserve angles. Using imaginary time, we can draw the vacuum calculation on a surface.
The key operation is to prepare identical copies, called \emph{replicas}, cut them along the regions, and join the exposed edges in a specified order. A path going around a cut endpoint can return to its original position on a different copy, or sheet. A \emph{twist} at the endpoint specifies which sheet it reaches.
This construction underlies the covering-space method~\cite{Lunin:2000yv} and the use of replica surfaces in entanglement calculations~\cite{Holzhey:1994we,Caraglio:2008pk,Calabrese:2009ez}.
\\

\noindent
The joined surface can have handles. Their number is its \emph{genus}: a sphere has none and a torus has one.
Once the gluing is specified, the calculation has two parts. The first compares two ways to measure length on the joined surface: the distances inherited from the copies and those measured by a smooth reference metric. The central charge $c$ fixes the response to this change of scale. The second part is the partition function of the original theory, which sums the weights of its field configurations on that surface. Geometry supplies the constant predicted by gravity for three regions. Whether that constant gives the whole entropy depends on the field theory.
\\

\noindent
For one region $A$, the reduced density operator $\rho_A$ predicts the probabilities of measurements in $A$ alone. Imagine preparing two copies and exchanging their $A$ degrees of freedom while leaving the rest in place.
The expectation of this exchange is $\mathrm{Tr}\,\rho_A^2$. In a basis where $\rho_A$ is diagonal, the trace sums over pairs with matching labels. An outcome of probability $p$ contributes $p^2$, the probability of seeing it twice in independent preparations.
Cycling $n$ copies gives $\mathrm{Tr}\,\rho_A^n$. Taking its natural logarithm and dividing by $1-n$ gives the R\'enyi entropy.
For a vacuum interval, conformal symmetry fixes the dependence on its length and on the smallest distance the calculation resolves~\cite{Holzhey:1994we}.
Twist fields implement the same exchange by local insertions at the interval endpoints~\cite{Caraglio:2008pk}.
For example, if $A$ itself is in a pure state, $\mathrm{Tr}\,\rho_A^2=1$, whereas an equal mixture of two orthogonal states gives $\mathrm{Tr}\,\rho_A^2=(1/2)^2+(1/2)^2=1/2$. The two-copy construction can distinguish these cases without first finding every eigenvalue of $\rho_A$.
For two disjoint intervals and integer $n>1$, the joined surface has handles. The answer can now depend on quantum data beyond the central charge~\cite{Calabrese:2009ez}.
Bringing two twist fields close together probes which fields can appear in their product. The operator-product expansion expresses this limit in terms of local fields of the theory~\cite{Ferrara:1973yt}.
With three twist insertions, symmetry fixes how the answer varies with their positions, leaving one number, the three-point coefficient. To compare that number between calculations, we first fix each twist's normalization by pairing it with its inverse at a known separation.
\\

\noindent
Three parties can ask a question that no single observer's entropy answers: does one item of information reach all three observers, or does each pair share its own? In Sec.~\ref{sec:7} we compare two pure states for which each observer has the same measurement statistics. Joining copies in a pattern that involves all three observers distinguishes the states.
A familiar example for three qubits is the tangle: its monogamy relation constrains how the correlations of one qubit can be shared with the other two~\cite{Coffman:1999jd}.
Multi-entropy uses a particular pattern to pair quantum-state amplitudes with their complex conjugates and sum the shared indices~\cite{Gadde:2022cqi}. This pairing and summation is called a contraction.
Other ways to pair the indices give polynomials that can be used to construct entanglement monotones~\cite{Gadde:2025csh} or to test whether a state separates into independent parts~\cite{Ma:2026wpv}.
Changing the basis in one observer's laboratory is a local unitary. An invariant gives the same answer in every such basis. A monotone must pass a further test: on average it cannot increase when the parties act locally and exchange classical messages. The gluing rule specifies an invariant; it does not by itself prove monotonicity. We therefore give both the rule for joining the copies and the normalization of the resulting number.
\\

\noindent
Replica calculations for systems at a critical point, where there is no intrinsic correlation length, can also be tested against lattice models~\cite{Harper:2024ker,Harper:2025uui}. The three single-region entropies provide a useful comparison: how much of the three-party result remains after accounting for them? For a specified subtraction, this remainder is called genuine multi-entropy~\cite{Iizuka:2025ioc,Iizuka:2025caq}.
Moving an endpoint changes the lengths of the regions that meet there. In the vacuum, these lengths drop out if we subtract half the sum of the three individual entropies, using the same short-distance prescription at every endpoint. The remainder depends on one number from a normalized three-point correlation. Two theories with the same central charge need not give the same number.
\\

\noindent
Holography gives us a number against which to test the calculation. The anti-de Sitter/conformal field theory (AdS/CFT) correspondence relates certain quantum theories to gravity in one higher dimension~\cite{Maldacena:1997re}. Sources on the boundary specify data for the bulk gravitational action~\cite{Gubser:1998bc,Witten:1998qj}.
In a static setting, the gravity prescription obtains the entropy of a region from the area of a minimal surface anchored on its boundary~\cite{Ryu:2006bv}. A replica derivation explains this relation when the required continuation of geometries exists~\cite{Lewkowycz:2013nqa}. In three-dimensional Einstein gravity, the boundary central charge fixes the ratio of the anti-de Sitter radius to Newton's constant~\cite{Brown:1986nw}.
For three regions, the proposed surface is a network of curves. One check is whether its replica symmetries agree with the permutations of the quantum copies~\cite{Gadde:2024taa}. Another, pursued here, is how much of the network's prediction can be calculated from the CFT without choosing a bulk geometry.
\\

\noindent
Three marked points on a sphere have a useful property: a conformal coordinate change can place them at $0$, $1$, and $\infty$. There is no remaining position to tune. The three cone angles determine a triangle. Reflections across its sides generate the identifications; the orientation-preserving transformations produce the sphere with its cones~\cite{Tsanov}.
In the negative-curvature case, the resulting metric can be written explicitly using hypergeometric functions on a disk. The same disk describes the smooth replica surface once we also specify the sheet reached by each loop. Different identifications give the sphere downstairs and the replica surface upstairs. With four marked points, one relative position remains even after fixing the other three. The differential equation then needs an additional coefficient, called an accessory parameter, to fit the global metric.
\\

\noindent
What the observers can measure also matters. Their allowed operators form an algebra. An operator in that algebra that commutes with all its members belongs to its \emph{center}; its possible values can label sectors. How those sectors are assigned to the regions can affect an entropy~\cite{Ma:2015xes,Huang:2016bkp,Ma:2016deg,Ma:2016pah}. Sec.~\ref{sec:7} gives an example in which all three parties carry the same label. Changing the number of possible labels changes their correlations while leaving the local central charge fixed. We specify both the available sectors and the state prepared across them.
\\

\noindent
Some states let us perform the replica sums exactly. A stabilizer state is fixed by a complete commuting set of operators made from generalized Pauli shifts and clocks. A shift advances a basis label cyclically; a clock multiplies each label by its assigned phase.
One way to organize the sum is by replica groups and their cosets, obtained by multiplying the elements of a subgroup by a fixed group element~\cite{Czech:2026dgm}. Another is a Gauss sum, which adds the phases of a quadratic form over a finite field~\cite{Czech:2026jil}. Both use information about how all the copies are joined. Given an underlying probability distribution, separate bounds and extrapolation formulas relate its integer R\'enyi entropies to its Shannon entropy~\cite{Zyczkowski:2003qyq}.
A second example uses the partial transpose, which interchanges the row and column labels of just one of two adjacent intervals~\cite{Calabrese:2012nk}. Its odd and even moments join the sheets differently. The odd moments reach every positive genus and can be measured by cycling the two intervals in opposite directions, as shown in Sec.~\ref{sec:4}. No continuation is needed for such an integer measurement. In contrast, knowing the measurements at integer indices does not choose a slope at index one~\cite{Czech:2026jil}.
\\

\noindent
The \emph{central charge} $c$ measures, among other things, how the theory responds when a metric is rescaled. This response is the Weyl anomaly. It gives one part of the three-point coefficient.
A handle shows what else can enter. Cut the handle across a circle. To glue it back, one sums over the quantum states that can propagate across that circle. This procedure, called \emph{sewing}, brings in the spectrum and the couplings of the theory~\cite{Headrick:2015gba,Belin:2017nze}. The result also depends on the complex structure (the shape of the surface) and, for theories with fermions, on their transport signs around loops. The central charge alone does not specify these data.
The other route to the geometric part starts in three-dimensional gravity. A conical defect is characterized by a missing angle around a worldline, just as a cone made by removing a wedge of paper has a missing angle at its tip.
Ref.~\cite{Chang:2016ftb} obtained a normalized cone exponent expressed through an integral of a gamma-function ratio. To reach the strengths of a surface with two or more handles, that function must be continued past a singular point. Going above or below it gives two answers. Related choices occur for complex solutions of the classical Liouville equation~\cite{Harlow:2011ny}. We compare both answers with the direct gluing calculation and ask whether the choice of path changes a physical observable.
For three adjacent regions in a pure vacuum state, the replica contraction defining multi-entropy~\cite{Gadde:2022cqi} places a twist at each of the three boundaries.
Ref.~\cite{Harper:2024ker} identifies the geometry, evaluates examples at small replica index, and gives a prediction from a network of shortest bulk curves for the constant left after subtracting the interval terms:
\begin{equation}
\kappa_{\mathrm{hol}}=\frac c2\ln\frac{2}{\sqrt3}.
\label{eq:holintro}
\end{equation}
Does the CFT fix that entire number from geometry? To compare the calculations, we normalize the twist fields, the partition function on the joined surface, and the exponent for the three defects in the same convention.
Only then can one decide whether there is a contribution from the particular CFT.
More precisely, a CFT respects local coordinate changes that preserve angles; a quantum anomaly can still make its partition function sensitive to the local length scale. We write $\C$ for the \emph{seed}, the theory on one copy. A local twist $\mathcal T_\sigma$ tells us that, after going around its insertion once, sheet $a$ becomes sheet $\sigma(a)$ for $a=1,\ldots,N$. On a connected cover, the same rule describes one copy of the seed on a larger surface.
\\

\noindent
A local field called a \emph{primary} transforms by simple powers of the coordinate derivatives under a conformal change of coordinates. Two numbers, $h$ and $\bar h$, give these powers. A dilation probes their sum; a rotation probes their difference.
The sum
\bea
\Delta=h+\bar h
\eea
is the scaling dimension, and their difference is the spin.
A scalar twist has
\bea
h=\bar h.
\eea
The stress tensor describes the response to a deformation of the metric. Its left- and right-moving parts carry central charges $c_L$ and $c_R$.
For example, bringing two left-moving stress tensors close together gives the singular terms
\begin{equation}
T_\C(z)T_\C(0)\sim\frac{c_L}{2z^4}+\frac{2T_\C(0)}{z^2}
+\frac{\partial T_\C(0)}{z}.
\end{equation}
The symbol $\sim$ here retains the singular terms as $z\to0$.
The antiholomorphic expression has $c_R$. We take
\bea
c_L=c_R=c,
\eea
so the replicated central charge in either sector is $Nc$.
Later we use $T(z)$ without the theory subscript for a \emph{projective connection}: an ordinary function in each coordinate chart that encodes the geometry and obeys the transformation law in Eq.~\eqref{eq:projectiveconnectionlaw}. It is distinct from the quantum stress-tensor operator $T_\C$.
\\

\noindent
The Euclidean partition function is the path integral
\begin{equation}
Z_\C[\Sigma,h]=\int\mathcal D\psi\,e^{-I_\C[\psi;h]},
\end{equation}
where $\psi$ runs over seed field configurations obeying the specified boundary or spin conditions, and the action $I_\C$ assigns their weights. Contributions from arbitrarily short distances can diverge. To define a \emph{renormalized} partition function, we subtract these divergences by a fixed prescription. Local terms added to the action for this purpose are called counterterms; one allowed finite term depends only on the topology and changes the overall normalization.
Even after the prescription is fixed, changing the metric's local length scale changes the quantum partition function. This response is the Weyl or trace anomaly~\cite{Polyakov:1981rd}. Knowing $c$ predicts the change between metrics, but still leaves the value at one starting metric to be calculated.
\\

\noindent
Unit two-point normalization means that a twist and its inverse obey
\begin{equation}
\langle\mathcal T_\sigma(z_1)\mathcal T_{\sigma^{-1}}(z_2)\rangle
=|z_1-z_2|^{-2\Delta_\sigma}
\end{equation}
with coefficient one. This provides a common calibration for the twist insertions. Dividing their normalized three-point function by its known position dependence leaves a number, $C_{123}$. Where its logarithm exists, it is the sum of the geometric exponent $P_{\mathrm{cov}}$ and the logarithm of the normalized cover partition function, on consistent branches.
The strength of a cone has an equally direct meaning. Draw a small circle at geodesic distance $\rho$ from its tip. If its circumference is $2\pi\beta\rho$ to leading order, the opening angle is $2\pi\beta$, and the missing angle is $2\pi(1-\beta)$.
Holography here refers to the proposed description of suitable large-$c$ CFT observables by gravity in three-dimensional anti-de Sitter space, denoted $\mathrm{AdS}_3$.
The gravity region is called the bulk, while the CFT lives on its conformal boundary.
In a semiclassical calculation, one evaluates a classical geometry and treats small fluctuations around it as corrections. A large central charge makes such an approximation possible in suitable theories. To justify using one specific geometry, one must also show that the other contributions are smaller.
\\

\noindent
The class of surfaces we need can now be specified exactly. We study a connected holomorphic map
\begin{equation}
\pi:\Sigma_g\longrightarrow\mathbb{CP}^{1}
\end{equation}
of degree $N$. This means that an ordinary point of the sphere has $N$ distinct points above it, one per sheet. At three points, the \emph{branch values}, some of these points merge. Over $z_j$ the sheets form $m_j$ groups of $k_j$ sheets each, with $m_jk_j=N$ for $j=1,2,3$. The groups have the same size at a given branch value; the size can differ from one value to the next. A circuit around $z_j$ cycles the sheets in each group. Together the three permutations connect all sheets, and their ordered product is the identity. We fix the individual permutations. A sum over covers or over permutations related by relabeling would require their multiplicities.
\\

\noindent
Here \emph{arbitrary genus} means that for every number of handles there is a cover in the class under study. It does not mean that every surface of that genus occurs. For instance, a long, thin rectangular torus and a square torus both have one handle but generally have different complex shapes. The set of inequivalent shapes at a given genus is its \emph{moduli space}. The permutations specify the gluing; even their cycle lengths need not determine the surface or its partition function.
\\

\noindent
We take a bosonic, nonchiral seed CFT with real central charge
\bea
c_L=c_R=c
\eea
and well-defined renormalized partition functions on the surfaces considered.
For a theory with fermions one must also say how a fermion is transported around each noncontractible loop. These consistent sign choices form a \emph{spin structure} and must be lifted to the cover. Insertions that change them locally must be included as well.
The central charge on the replicated base is $Nc$, whereas the theory on the connected cover has central charge $c$.
\\

\noindent
For $g>1$, let $h_g$ have Gaussian curvature $-1$, and let $h_{\mathrm{sph}}$ be the unit round-sphere metric.
We write
\begin{equation}
s_\C=Z_\C[S^2,h_{\mathrm{sph}}],\qquad
\Z_\C(\Sigma_g)=Z_\C[\Sigma_g,h_g]\,s_\C^{\,g-1}.
\label{eq:Zhatintro}
\end{equation}
The notation $\Z_\C(\Sigma_g)$ retains the surface, not just its genus.
Here $\C$ labels the theory, whereas $\mathbb C$ denotes the complex plane.
Why multiply by $g-1$ powers of the sphere partition function? The unnormalized theory admits a convention $Z[\Sigma]\mapsto\exp[b\chi(\Sigma)]Z[\Sigma]$, where $b$ is a constant and $\chi$ counts topology.
The sphere has $\chi=2$, while the cover has $\chi=2-2g$. The extra factor in $\Z_\C$ therefore contributes $\exp[2b(g-1)]$, exactly the inverse of the change on the cover.
We assume $s_\C\ne0$ and fix the local operator normalizations.
No reality assumption on $\Z_\C$ is required for the geometric comparison.
The same letter is occasionally conventional in different roles, which are distinguished by arguments and subscripts: $h_g$ is a metric but $h_j$ is a conformal weight; $R[h]$ is scalar curvature but $R$ alone is an outer coordinate cutoff; $K[h]$ is Gaussian curvature but $K(z_1,z_2,z_3)$ is a position factor.
The cone-strength sum $S$ differs from the entropy $S_n^{(q)}$.

\subsection{Summary of Results}
\noindent
Three questions organize the results. What does the shape of the cover contribute? Why does the cone calculation assign two exponents to one set of real angles? Which part of the entanglement does either exponent determine? Tab.~\ref{tab:sheetcounts} distinguishes two counts that recur in the answers: sheets are copies used in the calculation; handles belong to the surface formed after joining them.
\begin{table}[htbp]
\centering
{\small
\begin{tabular}{@{}lccc@{}}
\toprule
Gluing rule & Sheets $N$ & Group size $k_j$ & Handles $g$\\
\midrule
One five-sheet cycle at each point & $5$ & $5$ & $2$\\
Eight sheets, unequal orders & $8$ & $2,8,8$ & $2$\\
Three regions, grid with $n=2$     & $4$ & $2$ & $0$\\
Three regions, grid with $n=3$     & $9$ & $3$ & $1$\\
Three regions, grid with $n=4$     & $16$ & $4$ & $3$\\
\bottomrule
\end{tabular}\par}
\caption{Sheets count copies; handles count the topology of the joined surface. The first two covers both have two handles, although their sheets join differently. The group size at an endpoint is the length of each permutation cycle there; the ordered triple in the second row lists the sizes at the three endpoints. Secs.~\ref{sec:2} and \ref{sec:4} construct these covers. Sec.~\ref{sec:6} constructs the grids in the last three rows.}
\label{tab:sheetcounts}
\end{table}
\begin{itemize}
\item We define
\begin{equation}
\beta_j=\frac1{k_j},\qquad \eta_j=\frac{1-\beta_j}{2},\qquad
d=\frac{g-1}{N},\qquad
F(x)=\int_{1/2}^{x}\ln\frac{\Gamma(t)}{\Gamma(1-t)}\,\mathrm d t.
\label{eq:introdefs}
\end{equation}
Here $\Gamma$ is the Euler gamma function. Each $\beta_j$ is the fraction of a full turn left at cone $j$; the handle and sheet counts fix $d$. Together with the number of sheets, these are the geometric inputs to the following number.
For $g>1$, all nonzero arguments of $F$ in the expression lie between zero and one. The integral defining $F$ has a finite limit $F(0)$ as $x$ approaches zero from above. The covering exponent is
\begin{equation}
\boxed{\displaystyle
\Pcov=\frac{Nc}{6}\left\{
\sum_{j=1}^3\big[F(\beta_j)-F(\beta_j+d)+\beta_j\ln\beta_j\big]
+F(0)-F(d)+2d\right\}.}
\label{eq:mainintro}
\end{equation}
The unit-normalized fixed-permutation coefficient is
\begin{equation}
C_{123}=e^{\Pcov}\Z_\C(\Sigma_g).
\label{eq:factorintro}
\end{equation}
Eq.~\eqref{eq:factorintro} separates two calculations. The exponential is fixed by the branching and the central charge. The remaining factor is the seed theory evaluated on this particular surface. The separation is exact at finite $c$ in our normalization.
\item
We continue the normalized conical-defect expression from the spherical domain across
\bea
S=\sum_{j=1}^{3}\eta_j=1,
\eea
approaching $S=1+d\pm i0$ without other singularity crossings.
We obtain
\begin{equation}
\boxed{\displaystyle
\PCL^{(\pm)}=\Pcov\mp\frac{i\pi c}{6}(g-1).}
\label{eq:phaseintro}
\end{equation}
The difference between the real exponents is zero for all $g>1$, including $g=2$. The two paths have opposite imaginary parts. These statements concern the geometric exponent; the full correlation function also contains the fields' partition function.
Under the formal complex change $h_g\mapsto\exp(\mp i\pi)h_g$, the integrated trace anomaly supplies the displayed phase, with the sphere and local normalizations held fixed. When $h_g$ is only a reference metric, the partition function acquires the inverse phase. Sec.~\ref{sec:6} gives a separate check using timelike endpoint separations: individual interval entropies gain imaginary terms, but their contributions to the subtracted vacuum observable cancel on the specified path. Neither calculation identifies the cone phase with a count of ground states.
\item For tripartite multi-entropy, the replica cover is the Fermat curve, with $N=n^2$ and $g(n)=(n-1)(n-2)/2$.
Continuing the cone angles on the real spherical branch gives the constant in Eq.~\eqref{eq:holintro}. Here the limiting procedure is a derivative with respect to $n$ near one. This differs from taking a path around $S=1$ to find a phase.
The full endpoint-independent entropy is instead
\begin{equation}
\GM^{(3)}=\frac c2\ln\frac2{\sqrt3}-H_\C'(1),\qquad
H_\C(n)=\ln C_\C(n)-\PCL^{\sph}(n).
\label{eq:Hintro}
\end{equation}
The prime denotes a derivative at replica index one. Geometry gives the first term; the slope of the theory-dependent remainder supplies the second. We give a sufficient condition for this slope to grow more slowly than $c$ and a model where it grows faster. Thus, large central charge alone does not settle which term dominates.
\item At genus one the two branches meet, and the imaginary part vanishes. At genus zero one uses the real spherical formula directly. Substituting $g=0$ into the hyperbolic phase formula would use it outside its domain.
With four or more branch values, one can still normalize the twists. However, the cone action also depends on relative positions that cannot be removed by a coordinate change.
\item For a seed theory $\C_0$ with central charge $c_0>0$, fix a differentiable, tensor-product-compatible replica prescription and write $a(c_0)=H_{\C_0}'(1)$.
Tensoring $M$ copies with a direct sum of $D_M$ trivial topological sectors gives
\begin{equation}
\C_M=\C_0^{\otimes M}\otimes\mathcal T_{D_M},\qquad
c_M=Mc_0,\qquad H_{\C_M}'(1)=Ma(c_0)-\frac12\ln D_M.
\label{eq:largecintro}
\end{equation}
At fixed $c_0$, choosing $D_M=2^{M^2}$ gives $H_{\C_M}'(1)/c_M^2\to-\ln 2/(2c_0^2)$. The correction can grow faster than the geometric term, which is proportional to $c_M$.
The construction is reducible and uses an explicitly specified copying state; it is not a counterexample within the class of unique-vacuum CFTs.
Its purpose is to show which additional assumptions a large central-charge claim needs.
\end{itemize}

\noindent
A reader interested in what an observer can measure can start with the five-sheet example in Sec.~\ref{sec:4}, the phase cancellation in Sec.~\ref{sec:5}, and the replica measurements in Sec.~\ref{sec:6}. The derivation begins in Sec.~\ref{sec:2} with the covering map and its normalization. Secs.~\ref{sec:3} and \ref{sec:4} follow the two continuation paths and compare their exponents. Sec.~\ref{sec:5} accounts for their phases using the Weyl anomaly. Secs.~\ref{sec:6} and \ref{sec:7} derive the multi-entropy, compare its geometric part with gravity, and give an explicit test of the large-$c$ claim. Sec.~\ref{sec:8} discusses the remaining questions.

\section{Covering Maps}
\label{sec:2}
\noindent
A point on the joined surface carries two kinds of information: its position on the original sphere and the sheet it lies on. The covering map $\pi$ keeps the position and forgets the sheet. Following a circuit around a branch value tells us how the sheets are joined; it does not tell us the lengths of paths on them. For $g>1$, the developing map $u$ supplies lengths by mapping a small patch into a disk with a known metric. We use these two maps in turn, normalize the twists to obtain Eq.~\eqref{eq:mainintro}, and explain what extra information is needed when there are four branch values.

\subsection{Metrics, Curvature, and Uniformization}
\noindent
We use $\mathbb R$ for the real numbers, $\mathbb Z$ for the integers, and
\bea
\mathbb C=\{x+iy:x,y\in\mathbb R\}
\eea
for the complex plane, where
\bea
i^2=-1.
\eea
For $z=x+iy$, its complex conjugate is $\bar z=x-iy$ and its modulus is $|z|=\sqrt{x^2+y^2}$.
The Riemann sphere is
\bea
\mathbb{CP}^1=\mathbb C\cup\{\infty\}.
\eea
Equivalently, its points are homogeneous coordinates $[Z_0:Z_1]$, with $(Z_0,Z_1)\ne(0,0)$ and $[Z_0:Z_1]=[\lambda Z_0:\lambda Z_1]$ for every $\lambda\in\mathbb C\setminus\{0\}$.
The finite chart is $z=Z_1/Z_0$, and $[0:1]$ is infinity.
The notation $S^2$ denotes the underlying smooth sphere; specifying a metric supplies its lengths.
\\

\begin{figure}[htbp]
\centering
\includegraphics[width=.85\linewidth]{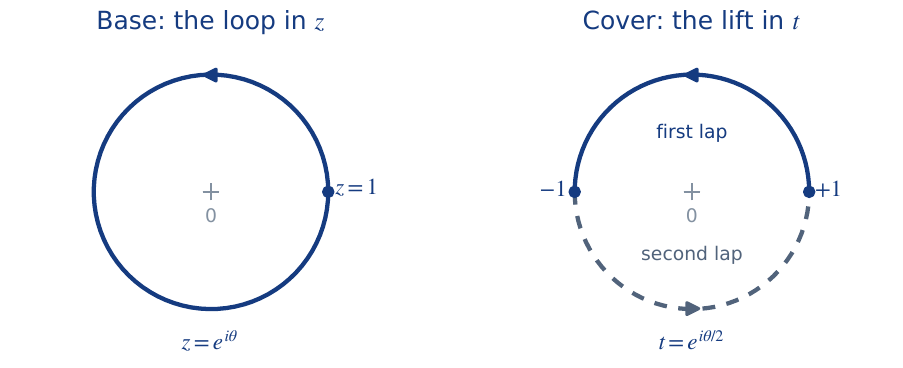}
\caption{Following a loop through the two sheets of $z=t^2$. Left: the point on the base goes around $0$ and returns to $z=1$. Right: starting at $t=+1$, the first lap follows the solid semicircle and ends at $t=-1$, the other point above $z=1$. Repeating the base loop follows the dashed semicircle back to $+1$. The branch point is the origin.}
\label{fig:sheetlift}
\end{figure}

\noindent
On a Riemann surface, a small neighborhood can be drawn in the complex plane. Where two such drawings overlap, the coordinate change is holomorphic, meaning complex differentiable. These drawings agree about angles, but an angle alone does not tell us the distance between two points. A compatible Riemannian metric assigns positive squared lengths and, in a local coordinate $z=x+iy$, takes the form
\begin{equation}
h=e^\varphi|\mathrm d z|^2=e^\varphi(\mathrm d x^2+\mathrm d y^2),\qquad
\mathrm d A_h=e^\varphi\mathrm d x\mathrm d y.
\label{eq:metricdefinition}
\end{equation}
The function $\varphi$ is the conformal exponent, and $\exp(\varphi)$ is the conformal factor.
Metrics related by $h'=\exp(\Phi) h$ have the same local angles; this is a Weyl transformation.
To measure a short path with the \emph{pullback} metric $\pi^*h$, first map the path by $\pi$, then measure its image with $h$. The factor that changes a squared length by $\exp(\Phi)$ changes the length itself by $\exp(\Phi/2)$.
As a local example, consider $z=t^2$. At $z=1$, the two points above it are $t=+1$ and $t=-1$. A single counterclockwise circuit $z=\exp(i\theta)$, $0\leq\theta\leq2\pi$, lifts from $+1$ along $t=\exp(i\theta/2)$ to $-1$. A second circuit brings the lift back to $+1$. The cover itself has no cut at which the field must jump: following the lift is continuous (Fig.~\ref{fig:sheetlift}).
\\

\noindent
We write
\bea
\mathrm d^2z=\mathrm d x\mathrm d y,\qquad \partial_z=\frac{\partial_x-i\partial_y}{2},\qquad  \partial_{\bar z}=\frac{\partial_x+i\partial_y}{2}.
\eea
Two differential operators will enter the calculation. The Laplacian measures the net second derivative of a scalar function $f$; on a curved surface we must measure those derivatives with the metric. The squared gradient measures how rapidly $f$ varies from point to point. With our conventions,
\begin{align}
\Delta_0f&=\partial_x^2f+\partial_y^2f=4\partial_z\partial_{\bar z}f,\\
\Delta_h f&=\frac1{\sqrt{\det h}}
\sum_{a=1}^{2}\sum_{b=1}^{2}\partial_a
\left(\sqrt{\det h}\,h^{ab}\partial_bf\right)=e^{-\varphi}\Delta_0f,\\
|\nabla_h f|^2&=\sum_{a=1}^{2}\sum_{b=1}^{2}h^{ab}\partial_af\,\partial_bf.
\label{eq:differentialoperators}
\end{align}
Here $(x^1,x^2)=(x,y)$, $(h^{ab})$ is the inverse matrix of $(h_{ab})$, and $\sqrt h$ elsewhere abbreviates $\sqrt{\det h}$.
In complex continuations, the quadratic gradient expression is continued as written, without inserting a complex conjugate.
\\

\noindent
Gaussian curvature $K[h]$ measures how angles in a small geodesic triangle differ from their values on a flat sheet. A geodesic follows the straightest path allowed by the metric: its tangent does not turn under parallel transport, and sufficiently short segments minimize distance. On a sphere, follow two meridians from the equator to the north pole, a quarter turn apart, and close the triangle along the equator. All three angles are right angles. Their sum exceeds $\pi$ because the sphere has positive curvature. For negative curvature, the sum in a small triangle falls below $\pi$. Both tests use measurements on the surface itself.
The scalar curvature $R[h]$, obtained by contracting the curvature tensor, satisfies $R[h]=2K[h]$ in two dimensions.
In the convention~\eqref{eq:metricdefinition},
\begin{equation}
K[h]=-\frac12e^{-\varphi}\Delta_0\varphi,\qquad
R[h]=-e^{-\varphi}\Delta_0\varphi.
\label{eq:curvaturedefinition}
\end{equation}
One derivation uses the orthonormal one-forms $\theta^1=\exp(\varphi/2)\mathrm d x$ and $\theta^2=\exp(\varphi/2)\mathrm d y$, so $h=(\theta^1)^2+(\theta^2)^2$.
The connection one-form, which describes how this orthonormal frame rotates under parallel transport, is $\varpi=(\varphi_y\mathrm d x-\varphi_x\mathrm d y)/2$.
Its exterior derivative is
\bea
\mathrm d\varpi=-\frac12\Delta_0\varphi\,\mathrm d x\wedge\mathrm d y=K\theta^1\wedge\theta^2,
\eea
proving Eq.~\eqref{eq:curvaturedefinition}.
The wedge product $\mathrm d x\wedge\mathrm d y$ is the oriented area two-form.
\\

\noindent
A round sphere means the standard sphere with the rotation-invariant metric inherited from Euclidean space.
For radius $\ell>0$, its spherical-coordinate and stereographic-coordinate forms are
\begin{equation}
h_{\sph,\ell}=\ell^2(\mathrm d\vartheta^2+\sin^2\vartheta\,\mathrm d\theta^2)
=\frac{4\ell^2|\mathrm d z|^2}{(1+|z|^2)^2},\qquad
z=\tan(\vartheta/2)e^{i\theta}.
\label{eq:roundsphere}
\end{equation}
The point $z=\infty$ completes this coordinate plane to $\mathbb{CP}^1$, the complex projective line or Riemann sphere.
The coordinate singularity at infinity is removed by $v=1/z$.
With $r=|z|$, the exponent obeys
\begin{equation}
\Delta_0\big[\ln(4\ell^2)-2\ln(1+r^2)\big]
=-\frac8{(1+r^2)^2},
\end{equation}
so $K[h_{\sph,\ell}]=+\ell^{-2}$ and $R=+2\ell^{-2}$.
For the Poincar\'e disk
\bea
\mathbb D=\{u\in\mathbb C:|u|<1\},
\eea
we have
\begin{equation}
h_{\mathbb D,\ell}=\frac{4\ell^2|\mathrm d u|^2}{(1-|u|^2)^2},\qquad
\Delta_0\big[\ln(4\ell^2)-2\ln(1-r^2)\big]=\frac8{(1-r^2)^2}.
\label{eq:diskmetric}
\end{equation}
Its Gaussian curvature is $-\ell^{-2}$.
The disk is open: its ideal boundary
\bea
\partial\mathbb D=\{u\in\mathbb C:|u|=1\}
\eea
is excluded.
To see what ``ideal'' means, the radial geodesic distance from $0$ to $u=r\exp(i\theta)$ is
\begin{equation}
\rho(r)=\int_0^r\frac{2\ell\,\mathrm d t}{1-t^2}
=\ell\ln\frac{1+r}{1-r}\longrightarrow+\infty\quad(r\to1^-).
\label{eq:diskboundarydistance}
\end{equation}
Although the unit circle is a finite distance away on the page, the radial integral diverges: no path of finite hyperbolic length from the center reaches it. The Poincar\'e disk therefore describes an entire hyperbolic plane, of curvature radius $\ell$. When we identify points of this disk to form a closed replica surface, the unit circle does not become a physical edge.
\\

\noindent
The upper half-plane can represent the same hyperbolic plane
\begin{equation}
\mathbb H=\{w\in\mathbb C:\Ima w>0\},\qquad
h_{\mathbb H,\ell}=\frac{\ell^2|\mathrm d w|^2}{(\Ima w)^2}.
\end{equation}
The Cayley transformation $w=i(1+u)/(1-u)$ is a holomorphic bijection from $\mathbb D$ to $\mathbb H$, with inverse $u=(w-i)/(w+i)$.
We directly obtain
\begin{equation}
\frac{\mathrm d w}{\mathrm d u}=\frac{2i}{(1-u)^2},\qquad
\Ima w=\frac{1-|u|^2}{|1-u|^2},\qquad
\frac{\ell^2|\mathrm d w|^2}{(\Ima w)^2}
=\frac{4\ell^2|\mathrm d u|^2}{(1-|u|^2)^2}.
\label{eq:cayleyisometry}
\end{equation}
Consequently, a half-plane uniformization and a disk uniformization related this way yield the same metric.
The Euclidean plane $h=|\mathrm d z|^2$ has $K=0$.
Thus, $+1$ and $-1$ set the curvature radius to one.
Under $h\mapsto\ell^2h$, lengths scale by $\ell$, areas by $\ell^2$, and curvature by $\ell^{-2}$; a real positive rescaling cannot change its sign.
\\

\noindent
An intrinsic way to compare the three geometries is to use geodesic polar coordinates. In a constant-curvature surface,
\begin{equation}
h=\mathrm d\rho^2+\mathscr S_K(\rho)^2\mathrm d\theta^2,\qquad
\mathscr S_K(\rho)=
\begin{cases}
\ell\sin(\rho/\ell),&K=+\ell^{-2},\\
\rho,&K=0,\\
\ell\sinh(\rho/\ell),&K=-\ell^{-2}.
\end{cases}
\label{eq:polarcurvature}
\end{equation}
These functions solve
\bea
\mathscr S_K''+K\mathscr S_K=0
\eea
with $\mathscr S_K(0)=0$ and $\mathscr S_K'(0)=1$, equivalently $K=-\mathscr S_K''/\mathscr S_K$.
A circle of geodesic radius $\rho$ has circumference $2\pi\mathscr S_K(\rho)$.
The spherical formula is used before the antipode, $0<\rho<\pi\ell$.
\\

\noindent
To picture \emph{uniformization}, start with a flat torus. One can draw it as a parallelogram with opposite sides identified. A path that leaves through one edge reappears at the opposite edge. Laying out copies of that parallelogram instead lets the path run on the plane without identifications. Uniformization gives the corresponding simply connected model for each Riemann surface, with a metric of constant curvature.
For compact connected oriented surfaces, the genus $g$ counts handles: a sphere has $g=0$, a torus $g=1$.
The Euler characteristic is
\bea
\chi(\Sigma_g)=2-2g.
\eea
Uniformization gives the following cases:
\begin{center}
\begin{tabular}{clcl}
\toprule
Genus & Universal Cover & $K$ & Compact Surface\\
\midrule
$0$ & Riemann sphere $\mathbb{CP}^1$ & $+1$ & Sphere\\
$1$ & Complex plane $\mathbb C$ & $0$ & $\mathbb C/(\mathbb Z+\tau\mathbb Z)$\\
$g>1$ & Disk $\mathbb D$ & $-1$ & $\mathbb D/\Gamma$\\
\bottomrule
\end{tabular}
\end{center}
A simply connected space has the property that every closed loop can be continuously contracted to a point.
The universal cover is a simply connected covering space: away from orbifold points, the projection is locally one-to-one, while globally it unwinds the noncontractible loops.
A quotient identifies points related by a specified group action.
In the torus row,
\begin{equation}
\Lambda_\tau=\mathbb Z+\tau\mathbb Z
=\{m+n\tau:m,n\in\mathbb Z\},\qquad \Ima\tau>0,
\end{equation}
is the period lattice, a discrete set of translations in $\mathbb C$.
The torus identifies $z$ with $z+\lambda$ for $\lambda\in\Lambda_\tau$; $\tau$ specifies its shape, while the metric still has an overall area.
In the last row, the transformations in $\Gamma$ preserve distances and orientation. No transformation except the identity fixes a point of the disk. Hence, identifying points by this group creates no cone tips.
We specify this group for the replica cover and define the quotient notation later.
At $g>1$, once the complex shape is fixed, there is exactly one compatible metric with $K=-1$. Changing the unit of length changes the magnitude of $K$ but not its sign. Varying the complex shape instead moves through the moduli of the surface. Uniformization guarantees a metric for each such shape; a formula for the metric and a value for the CFT partition function are further calculations.

\subsection{Gauss--Bonnet Theorem}
\noindent
For a smooth compact oriented surface without boundary, the Gauss--Bonnet theorem states
\begin{equation}
\int_{\Sigma_g}K[h]\,\mathrm d A_h=2\pi\chi(\Sigma_g)=4\pi(1-g),\qquad
\int_{\Sigma_g}R[h]\,\mathrm d  A_h=4\pi\chi(\Sigma_g).
\label{eq:GBsmooth}
\end{equation}
Curvature can move from one part of the surface to another as the metric changes. The theorem says that its total, including the area measure, cannot change as long as the topology stays the same.
A geometric explanation follows by triangulating into small geodesic triangles.
If triangle $t$ has interior angles $\alpha_{t,1},\alpha_{t,2},\alpha_{t,3}$, its integrated curvature is its angle excess $\sum_{r=1}^{3}\alpha_{t,r}-\pi$.
Summing over the $F_{\mathrm{tri}}$ faces gives
\begin{equation}
\int_{\Sigma_g}K\,\mathrm d  A_h
=\sum_{t=1}^{F_{\mathrm{tri}}}\left(\sum_{r=1}^{3}\alpha_{t,r}-\pi\right)
=2\pi V_{\mathrm{tri}}-\pi F_{\mathrm{tri}}
=2\pi(V_{\mathrm{tri}}-E_{\mathrm{tri}}+F_{\mathrm{tri}}).
\end{equation}
At a smooth vertex, the angles of all incident triangles sum to $2\pi$. Each face has three edges, and each edge borders two faces; counting edges both ways gives $3F_{\mathrm{tri}}=2E_{\mathrm{tri}}$. This leaves the Euler characteristic: vertices minus edges plus faces. For instance, splitting one triangle by adding an interior vertex adds one vertex, three edges and two faces, a net change of zero. When the curvature is constant and the total area is $A>0$, the theorem consequently gives
\bea
KA=4\pi(1-g).
\eea
This forces the sign of a constant $K$ to be positive, zero, or negative at $g=0,1,$ or $g>1$, respectively.
An arbitrary metric need not have constant or pointwise fixed-sign curvature; the Gauss--Bonnet theorem constrains only its integral.
\\

\noindent
A cone point of angle $2\pi\beta$ has local metric
\begin{equation}
e^A r^{2\beta-2}(\mathrm d  r^2+r^2\mathrm d \theta^2)
=\mathrm d \rho^2+\beta^2\rho^2\mathrm d \theta^2,\qquad
\rho=\frac{e^{A/2}}{\beta}r^\beta,\quad 0\leq\theta<2\pi.
\label{eq:conedefinition}
\end{equation}
Its small-circle circumference is $2\pi\beta\rho$.
The missing angle $2\pi(1-\beta)$ is the deficit; $\eta=(1-\beta)/2$ therefore describes a deficit $4\pi\eta$.
For a cone metric on the sphere with $\nu$ cone points and constant curvature $K_0$ away from them, the Gauss--Bonnet theorem includes the deficits:
\begin{equation}
K_0A_{\mathrm{orb}}+2\pi\sum_{j=1}^{\nu}(1-\beta_j)=4\pi,
\qquad K_0A_{\mathrm{orb}}=4\pi\left(1-\sum_{j=1}^{\nu}\eta_j\right).
\label{eq:GBcones}
\end{equation}
For example, removing small cone disks yields inner-boundary contributions of $-2\pi\beta_j$.
Applying the boundary version of the Gauss--Bonnet theorem to the punctured sphere, whose Euler characteristic is $2-\nu$, gives Eq.~\eqref{eq:GBcones}.
The boundary version adds the integral of geodesic curvature, the turning of a boundary tangent relative to parallel transport, and exterior corner angles if corners are present.
Thus, the cone formula is not obtained by treating the punctures as physical boundaries and discarding their contribution.
\\

\noindent
The Liouville equation is simply the constant-curvature equation in a conformal coordinate:
\begin{equation}
\Delta_0\varphi=-2K_0e^\varphi
\quad\hbox{away from cones}.
\label{eq:liouvillegeneral}
\end{equation}
The pointwise equation misses the curvature concentrated at a cone tip. To include the tips in an integral, we add a point source at each one. In a coordinate chart containing all $\nu$ cone points, this gives the distributional equation
\begin{equation}
\Delta_0\varphi=-2K_0e^\varphi
-4\pi\sum_{j=1}^{\nu}(1-\beta_j)\delta^{(2)}(z-z_j).
\label{eq:distributionalLiouville}
\end{equation}
The two-dimensional Dirac delta is defined by $\int_{\mathbb C}\delta^{(2)}(z-z_j)f(z)\mathrm d ^2z=f(z_j)$; the singular term follows from $\Delta_0\ln|z-z_j|=2\pi\delta^{(2)}(z-z_j)$. This delta is distinct from the quantity $\delta_j=\eta_j(1-\eta_j)$ used below.
\\

\noindent
A holomorphic branched cover has local form $z-z_j=b_{j,s}t^{k_j}$, $b_{j,s}\ne0$, at its $s$th preimage, where $s=1,2,\ldots,m_j$.
Its local degree, or ramification index, is $k_j$.
A generic target point has $N$ preimages, the degree of the cover.
Uniform ramification means that all preimages above a given branch value have the same index; different branch values can have different indices.
It implies
\bea
m_jk_j=N.
\eea
Start on one sheet and make a circuit around $z_j$. The new sheet label is $\sigma_j$ applied to the old one. This action of a closed path is its monodromy; repeat it until the starting label returns to give one cycle. If circuits about the different branch values can take us from any sheet to any other, the action is transitive, and the cover is connected.
\\

\noindent
To count handles, triangulate the base sphere so every branch value is a vertex. Each face and each open edge has $N$ copies upstairs. An ordinary vertex also has $N$ copies, but the vertex at $z_j$ has only $m_j$ because the sheets meet there. Subtracting the $N-m_j$ missing vertices from $N$ times the base Euler characteristic gives the Riemann--Hurwitz formula:
\begin{equation}
\chi(\Sigma_g)=N\chi(S^2)-\sum_{j=1}^{\nu}(N-m_j)
=2N-\sum_{j=1}^{\nu}m_j(k_j-1).
\label{eq:RHderivation}
\end{equation}
For angles
\bea
\beta_j=1/k_j,
\eea
the base is an orbifold: near each cone, it is a smooth disk modulo a rotation of finite order $k_j$.
Its orbifold Euler characteristic is
\begin{equation}
\chi_{\mathrm{orb}}=2-\sum_{j=1}^{\nu}(1-\beta_j),\qquad
\chi(\Sigma_g)=N\chi_{\mathrm{orb}}.
\end{equation}
At each preimage of cone $j$, $k_j$ copies of the angle $2\pi/k_j$ make one full turn. Away from these points, every area element on the sphere has $N$ lifts of equal area, and the points themselves have zero area. Hence, at $K_0=-1$ the cone sphere has area $A_{\mathrm{orb}}=4\pi(g-1)/N$.
As a check on the handle count, divide a sphere with three marked points into two triangular faces with three common edges. If five sheets meet in a single cycle at each marked point, the cover has ten faces, fifteen edges and just three vertices. Its Euler characteristic is $3-15+10=-2$, so the surface has two handles. No distances entered this count. Uniformization supplies those distances; analytic continuation of an area formula is a separate operation, as discussed in Sec.~\ref{sec:3}.
\par
\smallskip\noindent
The group size must be uniform \emph{at} a branch value, but it can change \emph{between} branch values. To see that at every positive genus, take a positive integer $p$ and consider
\[
w^{4p}=z^{2p}(z-1),\qquad \pi(z,w)=z .
\]
We take the smooth compact surface of this equation: branches meeting at a singular point of the plane drawing are separated, and the points at infinity are included. At an ordinary $z$ there are $4p$ distinct roots $w$. Label them by $s=0,\ldots,4p-1$ in their cyclic order. A circuit about zero advances $s$ by $2p$, a circuit about one advances it by one, and a circuit about infinity advances it by $-(2p+1)$, all modulo $4p$. The step by one reaches every sheet, so the cover is connected.
\par\smallskip\noindent
The first move pairs the sheets: it has $2p$ cycles of length two. Each of the other moves is a single cycle of length $4p$, since $2p+1$ and $4p$ have no common divisor larger than one. In Riemann--Hurwitz, the three contributions to ramification are $2p$, $4p-1$, and $4p-1$. Thus, $2g-2=-8p+2p+2(4p-1)=2p-2$ and $g=p$. At $p=2$ this gives the eight-sheet example in Tab.~\ref{tab:sheetcounts}: four pairs above zero and one eight-cycle above each of the other two values.

\subsection{Special functions and Analytic Conventions}
\noindent
Two functions with similar names do different jobs. The function \({}_2F_1\) solves the differential equation for the disk coordinate. The shorter name $F$ denotes an integral whose value enters the action. Both involve Euler's gamma function. We give their definitions here and specify how to follow their values along a path.
The Euler gamma function is
\begin{equation}
\Gamma(x)=\int_0^\infty t^{x-1}e^{-t}\mathrm d t,\qquad \Rea x>0,
\qquad \Gamma(x+1)=x\Gamma(x).
\end{equation}
At positive integers the gamma function extends the factorial: $\Gamma(m)=(m-1)!$ for $m=1,2,\ldots$. Its recurrence also defines it beyond the half-plane of convergence of the integral. The continued function has no zeros and is holomorphic except for isolated poles at $0,-1,-2,\ldots$; a function with only such isolated poles is called meromorphic. The action uses the ratio
\begin{equation}
\gamma(x)=\frac{\Gamma(x)}{\Gamma(1-x)},\qquad
\gamma(1-x)=\gamma(x)^{-1},\qquad
\gamma(1+x)=-x^2\gamma(x).
\label{eq:gammadefs}
\end{equation}
These definitions fix the special-function normalizations and branches used throughout the calculation.
\\

\noindent
The function $F$ in the conical action is a particular integrated logarithm, not the hypergeometric function:
\begin{equation}
F(x)=\int_{\frac{1}{2}}^{x}\ln\gamma(t)\,\mathrm d t,\qquad
F\bigg(\frac{1}{2}\bigg)=0,\quad F'(x)=\ln\gamma(x),\quad 0<x<1.
\label{eq:Fdefinition}
\end{equation}
For $0<t<1$, the ratio $\gamma(t)$ is positive, so the starting logarithm is real. At zero, $\gamma(t)\sim1/t$: although $F'(t)$ diverges, the area under $-\ln t$ from $0$ to $x$ is $x(1-\ln x)$ and tends to zero. This explains why $F(0)$ is finite.
Reflection gives $F(1-x)=F(x)$: the derivative of their difference vanishes by Eq.~\eqref{eq:gammadefs}, and they agree at $x=1/2$. Thus, $F(1)=F(0)$ as one-sided limits; neither value requires evaluating $F'$ at an endpoint.
\\

\noindent
The Gauss hypergeometric function is defined initially by its convergent power series
\begin{equation}
{}_2F_1(a,b;q;z)=\sum_{m=0}^{\infty}
\frac{(a)_m(b)_m}{(q)_m\,m!}z^m,\qquad |z|<1,
\label{eq:hypergeometricdefinition}
\end{equation}
where $q\notin\{0,-1,-2,\ldots\}$, and the Pochhammer symbol is the rising factorial
\begin{equation}
(a)_0=1,\qquad (a)_m=\prod_{r=0}^{m-1}(a+r)=\frac{\Gamma(a+m)}{\Gamma(a)},\quad m\geq1.
\end{equation}
The series gives a starting value. To find the function outside $|z|<1$, continue that value along a chosen path; different paths around a singular point can lead to different branches. The conventional principal branch takes $[1,\infty)$ as its cut in the $z$ plane. The series parameter $q$ is unrelated to the central charge $c$.
The Riemann zeta function and Euler's constant, used in the replica expansion, are
\begin{equation}
\zeta(s)=\sum_{m=1}^{\infty}m^{-s},\quad\Rea s>1,
\qquad \gamma_E=\lim_{M\to\infty}\left(\sum_{m=1}^{M}\frac1m-\ln M\right).
\label{eq:zetadefinition}
\end{equation}
Only $\zeta(m)$ at integers $m\geq2$ is needed, so this convergent definition suffices; no zeta regularization is used.
\\

\noindent
Analytic continuation extends a holomorphic function along a path by matching its values in overlapping neighborhoods. A familiar example is the angle of a complex number. After one turn around zero, the number returns to its starting value. However, its continuously followed angle has increased by $2\pi$ on a counterclockwise path. Zero is a branch point of the logarithm. A branch cut removes a line so one can choose a single-valued angle on what remains. For a nonzero complex $w$,
\begin{equation}
\ln_{\mathrm{cont}}w=\ln|w|+i\operatorname{Arg}_{\mathrm{cont}}w.
\end{equation}
The principal logarithm chooses $-\pi<\operatorname{Arg}w<\pi$ away from its negative-real cut.
The notation $x\pm i0$ means the limit $x\pm i\varepsilon$ as $\varepsilon\to0^+$; it specifies the side of the cut.
Accordingly, we obtain complex values of $F$ by continuing its integrand and integration path together.
The symbols $\Rea$ and $\Ima$ denote the real and imaginary parts.
Finally, ${\mathcal O}(t^p)$ denotes a remainder bounded by a constant times $|t|^p$ in the stated limit.

\subsection{Ramification}
\noindent
For three branch values, the count in Eq.~\eqref{eq:RHderivation} simplifies. Each ordinary cell lifts to $N$ copies, but at $z_j$ there are just $m_j$ vertices. Expressing this count in terms of the fractions $\beta_j=m_j/N$ gives
\begin{align}
2g-2&=-2N+\sum_{j=1}^{3}m_j(k_j-1)
=N\left(1-\sum_{j=1}^{3}\beta_j\right),\label{eq:RH}\\
\sum_{j=1}^{3}\beta_j&=1-2d,\qquad S\equiv\sum_{j=1}^{3}\eta_j=1+d.
\end{align}
We next need the strength of the twist field at an endpoint. The stress tensor measures the response to a change of coordinates; its leading singularity near the branch value gives the twist weight. Adding the contributions of its cycles gives
\begin{equation}
h_j=\bar h_j=\frac{m_jc}{24}\left(k_j-\frac1{k_j}\right)
=\frac{Nc}{24}(1-\beta_j^2)
=\frac{Nc}{6}\delta_j,\qquad
\delta_j=\eta_j(1-\eta_j).
\label{eq:weights}
\end{equation}
For completeness, a local inverse branch is $t=(z-z_j)^{1/k_j}$.
The stress-tensor transformation on one sheet contains the Schwarzian term
\begin{equation}
\frac{c}{12}\{t,z\}=\frac{c(1-k_j^{-2})}{24(z-z_j)^2},\qquad
\{t,z\}=\frac{t'''}{t'}-\frac32\left(\frac{t''}{t'}\right)^2.
\end{equation}
The double pole has the same coefficient on each sheet of a cycle. Counting the $k_j$ sheets in each of the $m_j=N/k_j$ cycles therefore gives $h_j$ in Eq.~\eqref{eq:weights}. Regular terms in the lifted stress tensor cannot alter this leading pole. The connection to a measurement of the twist comes from the conformal Ward identity: inserting the stress tensor gives the response of a correlation function to a small coordinate change.
For
\begin{equation}
G_3=\left\langle\prod_{j=1}^{3}\mathcal T_{\sigma_j}(z_j)\right\rangle,
\qquad T_{\mathrm{rep}}=\sum_{a=1}^{N}T_\C^{(a)},
\end{equation}
it reads, away from a zero of $G_3$,
\begin{equation}
\frac{\left\langle T_{\mathrm{rep}}(z)\prod_{j=1}^{3}\mathcal T_{\sigma_j}(z_j)\right\rangle}{G_3}
=\sum_{j=1}^{3}\left[\frac{h_j}{(z-z_j)^2}
+\frac{\partial_{z_j}\ln G_3}{z-z_j}\right].
\end{equation}
The two poles have different meanings. The coefficient of the double pole gives the conformal weight in Eq.~\eqref{eq:weights}; the coefficient of a simple pole tells us how the correlation changes when an insertion moves. For a scalar twist, the scaling dimension is $\Delta_j=2h_j$. These dimensions fix the position factor in the three-point correlation:
\begin{equation}
K(z_1,z_2,z_3)=\prod_{1\leq j<k\leq3}|z_j-z_k|^{-2(h_j+h_k-h_l)},
\label{eq:K}
\end{equation}
where $l=6-j-k$ is the remaining index in $\{1,2,3\}$.
One can recover the powers in this expression by inversion, $z\mapsto1/z$. Write the exponent of $|z_j-z_k|$ as $-p_{jk}$ for $1\leq j<k\leq3$.
Inversion divides this distance by $|z_j z_k|$, so the two factors touching point 1 supply the power $p_{12}+p_{13}$ of $|z_1|$.
A primary of dimension $\Delta_1$ requires that power to be $2\Delta_1$.
Writing the corresponding conditions at points 2 and 3 and eliminating $p_{13}$ and $p_{23}$ gives $p_{12}=\Delta_1+\Delta_2-\Delta_3$, just as in Eq.~\eqref{eq:K}.
This fixes how the three distances enter but leaves their common coefficient undetermined.
\\

\noindent
For $g>1$, the orbifold sphere with angles $2\pi\beta_j$ has a hyperbolic cone metric
\begin{equation}
h_{\mathrm{orb}}=e^{\varphi}|\mathrm d z|^2,\qquad
\Delta_0\varphi=2e^{\varphi},\qquad K_{\mathrm{orb}}=-1.
\label{eq:liouville}
\end{equation}
Here $\Delta_0=4\partial_z\partial_{\bar z}$. Its pullback
\bea
h_g=\pi^*h_{\mathrm{orb}}
\eea
is smooth.
Indeed, in a local coordinate $t$ with
\bea
z-z_j=b_jt^{k_j},
\eea
we obtain
\begin{equation}
e^{\varphi}|\mathrm d z|^2\sim
e^{A_j}k_j^2|b_j|^{2\beta_j}
|t|^{2k_j-2-4k_j\eta_j}|\mathrm d t|^2,
\qquad 2k_j-2-4k_j\eta_j=0.
\label{eq:smoothlift}
\end{equation}
The cancellation of powers means that the length scale at $t=0$ is finite and nonzero. Geometrically, a full turn in $t$ covers the cone $k_j$ times, so its opening angle upstairs is $k_j(2\pi/k_j)=2\pi$. The local disk coordinate also makes the higher-order terms smooth.
The Gauss--Bonnet theorem, including the cone deficits $4\pi\eta_j$, yields
\begin{equation}
-A_{\mathrm{orb}}+4\pi S=4\pi,
\qquad A_{\mathrm{orb}}=4\pi d,
\qquad A_{\Sigma_g}=4\pi(g-1).
\label{eq:areas}
\end{equation}
The area is positive if and only if $d>0$.
\\

\noindent
The restriction $g>1$ is essential in Eq.~\eqref{eq:areas}.
With a curvature radius $\ell>0$, the general relation for a smooth cover is
\begin{equation}
K A_{\Sigma_g}=4\pi(1-g),\qquad
A_{\Sigma_g}=\begin{cases}4\pi\ell^2,&g=0,\ K=+\ell^{-2},\\
A>0\ \text{free},&g=1,\ K=0,\\
4\pi(g-1)\ell^2,&g>1,\ K=-\ell^{-2}.
\end{cases}
\end{equation}
Thus, inserting $g=0$ into $4\pi(g-1)$ yields a negative value outside that formula's domain; it is not the area of the spherical cover.
For a sphere with cones, the same statement is $K_0A_{\mathrm{orb}}=4\pi(1-S)$.
As total cone strength grows, more of the sphere's required curvature concentrates at its tips. When $S>1$, their deficits alone exceed $4\pi$, so the smooth parts must carry negative curvature.
At $S=1$, the Gauss--Bonnet theorem fixes zero integrated curvature but leaves the flat scale free.
Analytically continuing a signed expression such as $4\pi(1-S)$ is a separate operation from measuring area with a real positive metric.

\subsection{Hypergeometric Developing Map}
\noindent
To compute distances, we can choose coordinates with the three marked points at $0,1,\infty$. The unknown $\varphi$ depends on both $z$ and $\bar z$. The constant-curvature equation lets us combine its derivatives into a quantity $T$ that depends only on $z$ away from the marked points:
\begin{equation}
T(z)=\frac12\partial_z^2\varphi-\frac14(\partial_z\varphi)^2.
\end{equation}
The term projective connection refers to its transformation law.
If $w=w(z)$ is a holomorphic coordinate change, we get
\begin{equation}
T_z(z)=(w'(z))^2T_w(w(z))+\frac12\{w,z\},\qquad
\{w,z\}=\frac{w'''}{w'}-\frac32\left(\frac{w''}{w'}\right)^2.
\label{eq:projectiveconnectionlaw}
\end{equation}
The last expression is the Schwarzian derivative.
It vanishes for a M\"obius map $w=(az+b)/(cz+d)$ with $ad-bc\ne0$, and measures the failure of a general map to have this fractional-linear form.
Its chain rule is $\{f\circ w,z\}=(w')^2\{f,w\}+\{w,z\}$.
Instead of solving the nonlinear equation for $\varphi$ directly, we solve a linear equation for two independent functions. Their ratio gives a disk coordinate. Choosing a different basis for the same two solutions only changes that coordinate by a fractional linear transformation.
The cone strengths determine the double poles of $T$. Coefficients at simple poles that remain free after the local and infinity conditions are called accessory parameters. Where such parameters occur, we choose them so that following a path around each cone preserves the metric.
The field equation gives $\partial_{\bar z}T=0$ away from the cones. The pole data, together with regularity at infinity, can therefore determine $T$.
The double-pole coefficient at $z_j$ is $\delta_j$.
The condition $T(z)\sim\delta_3/z^2$ at infinity fixes both simple-pole coefficients, so
\begin{equation}
T(z)=\frac{\delta_1}{z^2}+\frac{\delta_2}{(z-1)^2}
+\frac{\delta_1+\delta_2-\delta_3}{z(1-z)}.
\label{eq:projective}
\end{equation}
There is no free accessory parameter in this three-point problem.
In detail, the possible simple poles are $A/z+B/(z-1)$.
Expanding at large $z$, the absence of a $z^{-1}$ term requires $A+B=0$, and the coefficient of $z^{-2}$ requires $\delta_1+\delta_2+B=\delta_3$.
Substituting $B=-A$ gives Eq.~\eqref{eq:projective}.
These two conditions are why the three-point problem closes without additional moduli.
\\

\noindent
Let $y''+Ty=0$, and set
\begin{equation}
a=\eta_1+\eta_2-\eta_3,\qquad b=S-1=d,\qquad q=2\eta_1.
\end{equation}
Factoring $y=z^{\eta_1}(1-z)^{\eta_2}Y$ cancels the double poles and gives
\begin{equation}
z(1-z)Y''+[q-(a+b+1)z]Y'-abY=0.
\end{equation}
One can see the cancellation directly by setting $v(z)=z^{\eta_1}(1-z)^{\eta_2}$.
The coefficient of $Y'$ in $y''+Ty=0$ is $2v'/v=2\eta_1/z-2\eta_2/(1-z)$.
The coefficient of $Y$ is $v''/v+T=-ab/[z(1-z)]$, because $\eta_j(\eta_j-1)+\delta_j=0$ for $j=1,2$.
Multiplying by $z(1-z)$ gives the displayed equation and identifies its hypergeometric parameters.
We can obtain the power-series solution without guessing the special function.
We write $Y(z)=\sum_{m=0}^{\infty}v_mz^m$ with $v_0=1$.
Equating the coefficient of $z^m$ in the differential equation gives
\begin{equation}
(m+1)(m+q)v_{m+1}=(m+a)(m+b)v_m,
\qquad m=0,1,2,\ldots.
\label{eq:hypergeomrecurrence}
\end{equation}
Thus, we obtain $v_m=(a)_m(b)_m/[(q)_m m!]$.
The second local exponent of $Y$ at zero is $1-q$; substituting $Y=z^{1-q}\widetilde Y$ shifts the parameters to $(a-q+1,b-q+1;2-q)$.
A convenient basis is therefore
\begin{align}
y_1&=z^{\eta_1}(1-z)^{\eta_2}\Fp{a,b;q;z},\nn\\
y_2&=z^{1-\eta_1}(1-z)^{\eta_2}
\Fp{a-q+1,b-q+1;2-q;z}.
\label{eq:hypergeom}
\end{align}
The Wronskian, the determinant measuring linear independence of the two solutions and their first derivatives, is $y_1y_2'-y_1'y_2=\beta_1$.
The ratio of two independent solutions has Schwarzian derivative $2T$, so it contains the local metric information encoded by $T$. To use that ratio as a disk coordinate, distances must also agree after a trip around any cone. The returned coordinate can differ from the starting coordinate, provided the transformation between them preserves disk distances. Such a transformation is called an isometry. This requirement and the curvature scale fix the constants $A,B>0$ below:
\begin{equation}
u(z)=\sqrt{\frac{B}{A}}\,\frac{y_2(z)}{y_1(z)},\qquad
e^{\varphi}=\frac{4|u'(z)|^2}{(1-|u(z)|^2)^2}.
\label{eq:developing}
\end{equation}
Near zero, $u(z)\sim\sqrt{B/A}\,z^{1/k_1}$. On the base, one circuit about zero rotates $u$ by $2\pi/k_1$, leaving measured lengths unchanged.
On a chosen sheet, use $z=b_1t^{k_1}$ instead. Then $u$ is proportional to $t$ to leading order, with a nonzero coefficient: the apparent singularity at the tip disappears in this coordinate. Requiring lengths to agree after circuits about the other two cones fixes the remaining normalization.
The Wronskian is constant because the differential equation has no first-derivative term; evaluating it at zero gives $(1-\eta_1)-\eta_1=\beta_1$.
For $w=y_2/y_1$, we obtain $w'=\beta_1/y_1^2$ and $w''/w'=-2y_1'/y_1$.
Hence, we obtain $\{w,z\}=-2y_1''/y_1=2T$, which verifies the Schwarzian statement without assuming the metric.
Multiplying $w$ by $\sqrt{B/A}$ does not change its Schwarzian. The ratio $B/A$ chooses the disk coordinate in which continuation around the cones acts by isometries. The factor $4$ in the disk metric fixes the curvature radius to one. When that same metric is written as $4/f^2$ below, consistency requires $AB\beta_1^2=1$.
\\

\noindent
A developing map takes a patch of the surface to the constant-curvature model so that paths have the same length before and after mapping. Here the model is the disk: on each local branch, Eq.~\eqref{eq:developing} says $u^*h_{\mathbb D,1}=h_{\mathrm{orb}}$. Returning around a cone can change the disk coordinates by an isometry, a transformation that preserves distances. Thus the coordinates can change while the length assigned to every path remains single-valued.
{
To recover the whole surface from local disk coordinates, start with a hyperbolic triangle of angles $\pi/k_1,\pi/k_2,\pi/k_3$ in $\mathbb D$. Reflect it across a side and repeat: the images tile the disk. One reflection reverses orientation, while a pair of reflections preserves it. The orientation-preserving transformations form the triangle group $\Delta_\triangle$. Two adjacent triangles make a fundamental region, a region containing one representative of a generic orbit. Identifying its edges doubles the vertex angles and gives the sphere with three cone points. In terms of the vertex rotations, the group is
\[
\Delta_\triangle=\langle x_1,x_2,x_3\mid x_1^{k_1}=x_2^{k_2}=x_3^{k_3}=x_1x_2x_3=1\rangle .
\]
Each $x_j$ represents a circuit about $z_j$. Assigning the rotation $x_j$ its sheet permutation $\sigma_j$ gives an action $\rho:\Delta_\triangle\to S_N$, where $S_N$ is the group of permutations of $N$ objects. Now mark sheet $1$. A circuit lifts to a closed path from that sheet precisely when its permutation brings the mark back to itself. Such elements form the subgroup $\Gamma=\{\gamma\in\Delta_\triangle:\rho(\gamma)(1)=1\}$.
\\

\noindent
Two group elements lead to the same sheet if and only if they lie in the same left coset of $\Gamma$, a set of the form $\gamma\Gamma$. All $N$ sheets can be reached, so there are $N$ cosets: $[\Delta_\triangle:\Gamma]=N$. A rotation of this tiling that fixes an interior disk point is, up to a change of triangle, a power of a vertex rotation. No nontrivial such power fixes a sheet, since every cycle at that vertex has length $k_j$. The subgroup $\Gamma$ therefore identifies no disk point with itself by a nonidentity element. Its quotient is the smooth surface on the left:
\[
\pi:\ \Sigma_g=\mathbb D/\Gamma\ \longrightarrow\ \mathbb D/\Delta_\triangle,
\qquad [u]_\Gamma\longmapsto [u]_{\Delta_\triangle} .
\]
The brackets tell us which copies of a point to identify. Using $\Gamma$ on the left keeps track of the sheet; using the larger group $\Delta_\triangle$ on the right forgets it and gives the sphere $\mathbb{CP}^1$ with three cones. Thus two distinct points upstairs can have the same image under $\pi$. The coordinate $u(z)$, built from two hypergeometric solutions, goes in the other direction locally: it chooses a lift into the disk. Its branches supply coordinates for the metric; no single branch labels every point of the replica surface.
\\

\noindent
For a genus-two example, take five sheets and three branch orders $k_1=k_2=k_3=5$. Each triangle has three angles of $\pi/5$. At curvature $-1$, its area is its angle deficit, $\pi-3\pi/5=2\pi/5$, by Gauss--Bonnet. Two triangles form the base orbifold, with area $4\pi/5$; the five-sheet cover has area $4\pi$. The same result follows on the smooth surface from $-\mathrm{Area}=2\pi(2-2g)$ at $g=2$. The two different area counts directly check which space the disk coordinate describes.
}

\subsection{Connection Coefficients and Finite Cone Constants}
\noindent
The powers of the local solutions determine the cone angles, but an angle does not specify a unit of length. We also need the finite scale multiplying the metric near each tip. To find it, require that an observer measure the same length after carrying the local solutions around any cone. In the basis at zero, write
\begin{equation}
f=A|y_1|^2-B|y_2|^2,\qquad e^{\varphi}=\frac4{f^2},\qquad A,B>0.
\label{eq:fmetric}
\end{equation}
After a circuit about zero, $y_1$ and $y_2$ acquire different phases. Each squared magnitude in $f$ returns to its original value, whereas a mixed term $y_1\bar y_2$ would retain its relative phase. This fixes the form of $f$; the curvature equation fixes the product of its two coefficients. Indeed, differentiating gives
\begin{equation}
f f_{z\bar z}-f_zf_{\bar z}=-AB\beta_1^2,
\qquad \partial_z\partial_{\bar z}\varphi=\frac{2AB\beta_1^2}{f^2}.
\end{equation}
Comparison with Eq.~\eqref{eq:liouville} fixes $AB\beta_1^2=1$.
\\

\noindent
Let $p\sim(1-z)^{\eta_2}$ and $q_1\sim(1-z)^{1-\eta_2}$ be normalized solutions near one.
Explicitly, using $q-a-b=\beta_2$ shows
\begin{align}
p(z)&=z^{\eta_1}(1-z)^{\eta_2}
\Fp{a,b;a+b-q+1;1-z},\\
q_1(z)&=z^{\eta_1}(1-z)^{1-\eta_2}
\Fp{q-a,q-b;q-a-b+1;1-z}.
\label{eq:onebasis}
\end{align}
The prefactor $z^{\eta_1}$ approaches one here, so each new solution has leading coefficient one. We now have two descriptions of the same two solutions, one adapted to zero and the other to one. The four \emph{connection coefficients} tell us how to express each member of the first pair in terms of the second. The integral below gives the change of basis:
\begin{equation}
y_1=Up+Vq_1,\qquad y_2=\widetilde U p+\widetilde Vq_1,
\end{equation}
with coefficients, real on the chosen interval $0<z<1$,
\begin{align}
U&=\frac{\Gamma(q)\Gamma(q-a-b)}{\Gamma(q-a)\Gamma(q-b)},&
V&=\frac{\Gamma(q)\Gamma(a+b-q)}{\Gamma(a)\Gamma(b)},\\
\widetilde U&=\frac{\Gamma(2-q)\Gamma(q-a-b)}{\Gamma(1-a)\Gamma(1-b)},&
\widetilde V&=\frac{\Gamma(2-q)\Gamma(a+b-q)}{\Gamma(a-q+1)\Gamma(b-q+1)}.
\label{eq:connection}
\end{align}
These coefficients follow from Euler's integral.
To specify every function entering this derivation, we define the beta function by
\begin{equation}
\mathrm B(x,y)=\int_0^1 t^{x-1}(1-t)^{y-1}\mathrm d t
=\frac{\Gamma(x)\Gamma(y)}{\Gamma(x+y)},\qquad \Rea x,\Rea y>0.
\label{eq:betaintegral}
\end{equation}
The gamma-function identity follows by changing variables to the sum and ratio in the product of the two defining gamma integrals. Expanding $(1-zt)^{-a}$ inside the integral and integrating each power of $t$ gives
\begin{equation}
\Fp{a,b;q;z}=\frac{\Gamma(q)}{\Gamma(b)\Gamma(q-b)}
\int_0^1 t^{b-1}(1-t)^{q-b-1}(1-zt)^{-a}\,\mathrm d t,
\label{eq:Eulerhypergeom}
\end{equation}
initially for $|z|<1$ and $\Rea q>\Rea b>0$.
When $\Rea(q-a-b)>0$, its value at $z=1$ is another beta integral and equals $U$.
In fact, $q_1/p$ tends to zero at this endpoint, so the limit of $y_1/p$ isolates the coefficient $U$ of the first solution.
The Euler transformation
\begin{equation}
\Fp{a,b;q;z}=(1-z)^{q-a-b}\Fp{q-a,q-b;q;z}
\label{eq:Eulertransformation}
\end{equation}
can be checked by substitution into the differential equation and matching the analytic solution of value one at zero.
Applying the same beta-integral evaluation to the transformed function when $\Rea(a+b-q)>0$ fixes $V$.
These two evaluations use different ranges of the parameters. Their gamma-function expressions can then be continued to the common physical values, where the exponent difference $q-a-b=\beta_2$ is not an integer.
Applying this argument to the second zero-centered solution gives $\widetilde U,\widetilde V$.
Thus, the local bases fix the connection constants.
\\

\noindent
Around $z=1$ the two local solutions acquire phases $\exp(2\pi i\eta_2)$ and $\exp\big(2\pi i(1-\eta_2)\big)$. The squared magnitude of either solution stays the same, but their mixed product $p\bar q_1$ gains the relative phase $\exp(-2\pi i\beta_2)$.
For a nontrivial cone, this phase is not one. If the mixed term survived in $f$, a return to the same point could give a different length. Its coefficient must therefore vanish. In terms of the connection coefficients, this requirement reads
\begin{equation}
AUV=B\widetilde U\widetilde V,\qquad
A^2=\frac{\widetilde U\widetilde V}{\beta_1^2UV}.
\end{equation}
We introduce
\bea
\gamma(x)=\frac{\Gamma(x)}{\Gamma(1-x)}.
\eea
Using
$\Gamma(2-q)=(1-q)\Gamma(1-q)$ and $\gamma(1-x)=\gamma(x)^{-1}$ reduces this expression to
\begin{equation}
A^2=\frac{\gamma(S-1)\gamma(S-2\eta_2)\gamma(S-2\eta_3)}
{\gamma(2\eta_1)^2\gamma(S-2\eta_1)}.
\label{eq:A2}
\end{equation}
Due to $f\sim A|z|^{2\eta_1}$, the position-independent finite part at the first cone is
\bea
A_1^{(0)}=\ln4-\ln A^2.
\eea
Cyclically,
\begin{equation}
A_j^{(0)}=\ln4-
\ln\left[\frac{\gamma(d)\gamma(\beta_k+d)\gamma(\beta_l+d)
\gamma(\beta_j)^2}{\gamma(\beta_j+d)}\right].
\label{eq:coneconstants}
\end{equation}
The constants $A_j^{(0)}$ are what remains at the cones after subtracting the singular logarithm. Two metrics with the same cone angle can assign different lengths to a circle at a fixed small coordinate radius.
\\

\noindent
To return to finite $z_j$, we use $x=(z-z_j)(z_k-z_l)/[(z-z_l)(z_k-z_j)]$, which sends $(z_j,z_k,z_l)$ to $(0,1,\infty)$.
Its expansion at $z_j$ is
\begin{equation}
x=\alpha_j(z-z_j)+{\mathcal O}\big((z-z_j)^2\big),\qquad
\alpha_j=\frac{z_k-z_l}{(z_j-z_l)(z_k-z_j)}.
\end{equation}
In the following local expansions, the displayed ${\mathcal O}(1)$ terms tend to zero.
Substituting this expansion into
\bea
\varphi_x=-4\eta_j\ln|x|+A_j^{(0)}+{\mathcal O}(1),
\eea
and transforming the metric rather than only its scalar coefficient, gives
\begin{align}
\varphi_z&=\varphi_x+\ln|x'|^2\nonumber\\
&=-4\eta_j\ln|z-z_j|+A_j^{(0)}
+(2-4\eta_j)\ln|\alpha_j|+{\mathcal O}(1).
\end{align}
Due to $2-4\eta_j=2\beta_j$, the finite part is
\begin{equation}
A_j=A_j^{(0)}-\beta_j\ln\frac{|z_j-z_k|^2|z_j-z_l|^2}{|z_k-z_l|^2}.
\label{eq:finitepositions}
\end{equation}
We will henceforth place all cones at finite positions, with infinity unbranched.

\subsection{Finite Weyl Anomaly and Regulator Caps}
\noindent
We now compare the partition function computed with two different rulers on the same surface. For a uniform change of scale, only curvature enters. If the ruler changes by different factors at neighboring points, its variation also enters. The Weyl anomaly gives both terms for a smooth metric~\cite{Polyakov:1981rd}; we then explain how to treat the branch points. For a smooth reference metric $h$ on a closed surface $\Sigma$ and $h'=\exp(\Phi)h$, our convention is~\cite{Polyakov:1981rd}
\begin{equation}
\ln Z_\C[e^\Phi h]-\ln Z_\C[h]
=\frac c{96\pi}\int_\Sigma\mathrm d^2x\sqrt h\,
\big[(\nabla\Phi)^2+2R[h]\Phi\big].
\label{eq:Weyl}
\end{equation}
The first term records changes of scale from one point to the next. The second couples the change of scale to curvature. To obtain the finite change starting from the infinitesimal trace anomaly~\cite{Lunin:2000yv,Polyakov:1981rd}, turn on the new scale gradually by setting $h_t=\exp(t\Phi)h$ for $0\leq t\leq1$. Then
\begin{equation}
\frac{\mathrm d}{\mathrm d t}\ln Z[h_t]=\frac c{48\pi}\int_\Sigma\mathrm d^2x\sqrt{h_t}R[h_t]\Phi,
\qquad \sqrt{h_t}R[h_t]=\sqrt h(R[h]-t\Delta_h\Phi).
\end{equation}
Integrating over $t$ from zero to one and integrating by parts on the closed surface yields Eq.~\eqref{eq:Weyl}. We now apply it to a particular pair of metrics on the punctured cover. The physical metric $g_{\mathrm{pull}}=\pi^*|\mathrm d z|^2$ measures lengths inherited from the sheets. The reference metric $h_g=\pi^*(\exp(\varphi)|\mathrm d z|^2)$ uses the same covering map but measures the image in the cone metric. Their ratio is
\begin{equation}
g_{\mathrm{pull}}=e^{-\varphi\circ\pi}h_g.
\label{eq:relative}
\end{equation}
In a sheet coordinate $t$, both metrics contain the same factor $|\mathrm dz/\mathrm dt|^2$: both were pulled back by the same map. Dividing them cancels that factor before any integral is taken. Keeping an additional $\ln|\pi'|^2$ would count the coordinate change twice.
\\

\noindent
The integral is singular at the branch values, so we temporarily omit small disks $|z-z_j|\leq\epsilon_j$ and the exterior of $|z|=R$. This is the regulator; $D_{\epsilon,R}$ is the region left for integration. At fixed branch orders, the metric exponent near the removed points has the form
\begin{align}
\varphi&=-4\eta_j\ln|z-z_j|+A_j+{\mathcal O}(1),\quad z\to z_j,\nn\\
\varphi&=-4\ln|z|+A_\infty+{\mathcal O}(1),\quad z\to\infty.
\label{eq:asymptotics}
\end{align}
The removed regions are filled by the same flat caps that define the bare twists.
These caps are a temporary way to define the path integral. Their seams are not boundaries of the quantum system, so one must track what survives when the disks shrink.
Thus, we get
\bea
D_{\epsilon,R}=\{z\in\mathbb C:|z|<R,\ |z-z_j|>\epsilon_j\text{ for }j=1,2,3\}.
\eea
The inner disks are disjoint.
The limits are taken at fixed positions and fixed monodromy data.
If several preimages lie above $z_j$, the local coordinates are indexed by $s=1,2,\ldots,m_j$; the estimates apply to each of these finitely many disks.
\\

\noindent
A small cap could, in principle, leave a finite term if its integrand grew fast enough as it shrank. Here its area shrinks faster than the only divergent factor, a logarithm. We can see this in a coordinate on the smooth sheet.
Above $z_j$, we choose $z-z_j=b t^k$ exactly and let $\rho=(\epsilon_j/|b|)^{1/k}$.
We then replace the degenerate pulled-back metric inside $|t|<\rho$ by
\begin{equation}
g_{\mathrm{cap}}=k^2|b|^2\rho^{2k-2}|\mathrm d t|^2.
\end{equation}
It matches the exterior continuously.
If $h_g=\exp\bigl[\sigma(t,\bar t)\bigr]|\mathrm d t|^2$ is smooth, we then get
\bea
\Phi_{\mathrm{cap}}=L_\rho-\sigma
\eea
with
\bea
L_\rho=\ln(k^2|b|^2)+(2k-2)\ln\rho.
\eea
As the disk shrinks, the diverging part of $L_\rho$ is constant across it, so it contributes no gradient. The remaining gradient stays bounded and is integrated over an area proportional to $\rho^2$. This gives ${\mathcal O}(\rho^2)$ for the kinetic term; the curvature term is ${\mathcal O}\big(\rho^2(1+|\ln\rho|)\big)$ because it contains the undifferentiated $L_\rho$.
At infinity the same argument in $v=1/z$, with cap $R^4|\mathrm d v|^2$, gives ${\mathcal O}\big(R^{-2}(1+\ln R)\big)$ per sheet.
There are $m_j$ such disks above $z_j$ and $N$ disks at infinity. Adding their contributions gives
\begin{equation}
W_{\mathrm{caps}}={\mathcal O}\!\left(\sum_{j=1}^{3}m_j\epsilon_j^{2/k_j}(1+|\ln\epsilon_j|)
+NR^{-2}(1+\ln R)\right)\longrightarrow0.
\label{eq:capbound}
\end{equation}
The estimate is for a fixed cover.
It does not give uniform control when $k_j$ or $N$ also tends to infinity.
Because Eq.~\eqref{eq:Weyl} uses the reference curvature, it adds no extra delta function from the physical cap seam.
For example, smoothness bounds $|\nabla_0\sigma|$ and $|R[h_g]|\exp(\sigma)$ by finite constants on a small coordinate disk. Consequently, we obtain
\begin{equation}
\int_{|t|<\rho}|\nabla_0\Phi_{\mathrm{cap}}|^2\mathrm d^2t\leq C_1\rho^2,
\qquad
\left|\int_{|t|<\rho}\sqrt{h_g}R[h_g]\Phi_{\mathrm{cap}}\mathrm d^2t\right|
\leq C_2\rho^2(1+|\ln\rho|).
\end{equation}
Substituting $\rho\propto\epsilon_j^{1/k_j}$ and adding all preimages gives Eq.~\eqref{eq:capbound}.
Continuity of the Weyl factor at a seam permits a smooth approximation with the same limit; a jump of its normal derivative does not create a new term in the quadratic gradient integral.
Pushing the anomaly down over the $N$ local sheets and using $R[h_g]=-2$ gives
\begin{equation}
W=\frac{Nc}{96\pi}\int_{D_{\epsilon,R}}\bigl[|\nabla_0\varphi|^2+4\varphi e^\varphi\bigr]\mathrm d^2z
+{\mathcal O}(1).
\end{equation}
A small patch away from the cones has $N$ copies, each making the same contribution to the geometric integral. The fields, however, can pass from one sheet to another along the cuts. Their partition function belongs to the one connected surface, not to $N$ independent copies.
Here the remainder is the vanishing cap contribution.
Equivalently, we have
\begin{equation}
\lim_{\epsilon_j\to0,\ R\to\infty}
\left[W-\frac{Nc}{96\pi}\int_{D_{\epsilon,R}}
(|\nabla_0\varphi|^2+4\varphi e^\varphi)\mathrm d^2z\right]=0.
\end{equation}
The subsequent ${\mathcal O}(1)$ cutoff remainders vanish.
The local developing map makes the errors shrink as positive powers of the cutoff radius. Differentiating a power with respect to a cone strength can introduce a logarithm, but the product still tends to zero.
All nonzero finite contributions, including the area, are displayed explicitly.

\subsection{Regularized Action and Variation Identity}
\noindent
Instead of evaluating the whole anomaly integral at once, we ask how it changes when just one cone angle varies. The field equation cancels the interior variation, leaving data at the boundary circles. We keep both the integral over the retained region, $E=\int_D|\nabla_0\varphi|^2\mathrm d^2z$, and the boundary integral, $J=\oint_{\partial D}\varphi\partial_n\varphi\,\mathrm d s$. Integration by parts with $\Delta_0\varphi=2\exp(\varphi)$ gives
\begin{equation}
W=\frac{Nc}{96\pi}(2J-E)+{\mathcal O}(1).
\end{equation}
Here $D=D_{\epsilon,R}$ and $\partial_n$ differentiates in the direction pointing \emph{out of the retained region}. That direction is $-\partial_r$ on a hole and $+\partial_r$ on the outer circle; this sign fixes the boundary terms.
Evaluating the inner and outer circles with these orientations gives
\begin{equation}
\frac{J}{8\pi}=4\ln R-A_\infty
+\sum_{j=1}^{3}\eta_jA_j-4\sum_{j=1}^{3}\eta_j^2\ln\epsilon_j+{\mathcal O}(1).
\label{eq:J}
\end{equation}
The leading integrand on the $j$th circle is $(-4\eta_j\ln\epsilon_j+A_j)(4\eta_j/\epsilon_j)\epsilon_j\mathrm d\theta$.
Its integral divided by $8\pi$ is $-4\eta_j^2\ln\epsilon_j+\eta_jA_j$.
The local expansion described above makes the omitted integral vanish even after multiplication by $\ln\epsilon_j$.
The outer circle similarly gives $4\ln R-A_\infty$.
\\

\noindent
An on-shell action is an action evaluated on a solution of its field equation, here the Liouville equation, rather than on an arbitrary trial field.
We define the dimensionless on-shell action
\begin{equation}
I_{\hyp}=\lim_{\epsilon_j\to0,\,R\to\infty}
\left[\frac{E}{16\pi}+\frac1{4\pi}\int_D e^\varphi\mathrm d^2z
+A_\infty-2\ln R-\sum_{j=1}^{3}\eta_jA_j
+2\sum_{j=1}^{3}\eta_j^2\ln\epsilon_j\right].
\label{eq:Ireg}
\end{equation}
Near a cone, $E/(16\pi)$ has divergence $-2\eta_j^2\ln\epsilon_j$; the outer divergence is $2\ln R$.
The displayed subtractions cancel both.
The omitted area tends to zero as $ {\mathcal O}(\epsilon_j^{2\beta_j})$ at cone $j$ and as $ {\mathcal O}(R^{-2})$ at infinity. By Eq.~\eqref{eq:areas}, the remaining area divided by $4\pi$ tends to $d$. Solving Eq.~\eqref{eq:Ireg} for $E$ and using Eq.~\eqref{eq:J} now gives
\begin{equation}
W=-\frac{Nc}{6}I_{\hyp}+\frac{Nc}{6}d
+\frac{Nc}{3}\ln R-\frac{Nc}{3}\sum_{j=1}^{3}\eta_j^2\ln\epsilon_j+{\mathcal O}(1).
\label{eq:Wreg}
\end{equation}
The finite term
\bea
\frac{Ncd}{6}=\frac{c(g-1)}{6}
\eea
must be retained.
It is absent for a flat torus.
\\

\noindent
There is a second way to evaluate the same action once the metric is known. The kinetic integral taken on its own diverges at the holes. Integrating by parts transfers its divergent pieces to the boundary and leaves the following convergent integral over the interior:
\begin{equation}
\mathcal V_\varphi=\int_{\mathbb C}\varphi(z,\bar z)e^{\varphi(z,\bar z)}\mathrm d^2z.
\end{equation}
All cone points are finite, and infinity is unbranched.
Near a cone, the absolute radial integrand is bounded by a constant times $r^{2\beta_j-1}(1+|\ln r|)$, which is integrable since $\beta_j>0$.
At infinity, it is ${\mathcal O}(r^{-3}\ln r)$ after including the radial measure.
Integration by parts gives $E=J-2\mathcal V_\varphi+{\mathcal O}(1)$ in the hyperbolic case.
Inserting Eq.~\eqref{eq:J} into Eq.~\eqref{eq:Ireg} cancels every boundary divergence and yields
\begin{equation}
I_{\hyp}=d+\frac{A_\infty}{2}
-\frac12\sum_{j=1}^{3}\eta_jA_j
-\frac{\mathcal V_\varphi}{8\pi}.
\label{eq:finitebulkaction}
\end{equation}
This formula evaluates the action directly from the developing metric.
For curvature $K_0=+1$, the spherical field equation gives $E=J+2\mathcal V_\varphi+{\mathcal O}(1)$.
The spherical action has a negative area term, and its counterpart is
\begin{equation}
I_{K_0}=d+\frac{A_\infty}{2}
-\frac12\sum_{j=1}^{3}\eta_jA_j
+\frac{K_0}{8\pi}\mathcal V_\varphi,
\qquad K_0\in\{-1,+1\},\qquad d=-\frac{K_0A_{\mathrm{orb}}}{4\pi}.
\label{eq:finitebulkboth}
\end{equation}
Here $I_{-1}=I_{\hyp}$ and $I_{+1}=I_+$.
\\

\noindent
The variational identity can be checked before removing the regulators.
Let
\bea
I_{\mathrm{bulk}}=\frac{E}{16\pi}+\frac{1}{4\pi}\int_D e^\varphi\mathrm d^2z.
\eea
At fixed positions and cutoffs, its field variation is
\begin{align}
\delta I_{\mathrm{bulk}}
&=\frac1{8\pi}\oint_{\partial D}\delta\varphi\,\partial_n\varphi\,\mathrm d s
+\frac1{8\pi}\int_D\delta\varphi\,(-\Delta_0\varphi+2e^\varphi)\mathrm d^2z\nonumber\\
&=-\delta A_\infty+
\sum_{j=1}^{3}\left(\eta_j\delta A_j-4\eta_j\ln\epsilon_j\,\delta\eta_j\right)+{\mathcal O}(1).
\label{eq:bulkvariation}
\end{align}
The bulk integral vanishes by the equation of motion.
On an inner circle, the estimates above give
\bea
\delta\varphi&=&-4\ln\epsilon_j\,\delta\eta_j+\delta A_j+{\mathcal O}(1),
\nn\\
\partial_n\varphi&=&4\eta_j/\epsilon_j+\epsilon_j^{-1}{\mathcal O}(1).
\eea
Both displayed ${\mathcal O}(1)$ remainders tend to zero, and they vanish even after multiplication by the cutoff logarithms in the boundary integral.
The explicit factor $\epsilon_j^{-1}$ is canceled by $\mathrm d s=\epsilon_j\mathrm d\theta$.
Thus, Eq. \eqref{eq:bulkvariation} has a vanishing remainder.
The outer circle contributes $-\delta A_\infty$.
The explicit boundary terms in Eq.~\eqref{eq:Ireg} vary as
\begin{equation}
\delta A_\infty-\sum_{j=1}^{3}(A_j\delta\eta_j+\eta_j\delta A_j)
+4\sum_{j=1}^{3}\eta_j\ln\epsilon_j\,\delta\eta_j.
\end{equation}
Adding the interior and boundary variations cancels the cutoff logarithms. The terms proportional to changes in $A_j$ and $A_\infty$ also cancel. What survives is the change of the action with each cone strength, with its coefficient fixed by the corresponding finite metric constant:
\begin{equation}
\mathrm d I_{\hyp}=-\sum_{j=1}^{3} A_j\,\mathrm d\eta_j.
\label{eq:variation}
\end{equation}
For example, changing $\eta_1$ by $\mathrm d\eta_1$ while holding the other two strengths fixed changes the finite action by $-A_1\,\mathrm d\eta_1$ to first order. The strengths in this derivative are continuous parameters of the cone equation. We set them to their permitted covering values only after finding the action; no intermediate surface with a fractional sheet count is needed.
The position dependence integrates to
\begin{equation}
I_{\hyp}=I_{\hyp}^{(0)}+I_{\mathrm{pos}},\qquad
I_{\mathrm{pos}}=\sum_{1\leq j<k\leq3}(\delta_j+\delta_k-\delta_l)\ln|z_j-z_k|^2.
\label{eq:Ipos}
\end{equation}
Here $l=6-j-k$.
For example, $\partial_{\eta_1}I_{\mathrm{pos}}=\beta_1\ln(|z_{12}|^2|z_{13}|^2/|z_{23}|^2)$, as required by Eq.~\eqref{eq:finitepositions}.
\\

\noindent
For $0<x<1$, $F'(x)=\ln\gamma(x)$ and $F(1-x)=F(x)$. Both $F(0)$ and $F(1)$ are finite. Using $\partial_{\eta_j}d=1$ and $\partial_{\eta_j}\beta_k=-2\delta_{jk}$, direct differentiation shows that
\begin{equation}
I_{\hyp}^{(0)}=-d\ln4+F(d)+\sum_{j=1}^{3}\big[F(\beta_j+d)-F(\beta_j)\big]+C_0
\label{eq:Iwithconstant}
\end{equation}
has the derivatives~\eqref{eq:variation} with Eq.~\eqref{eq:coneconstants}.
For a direct check, we put $f(x)=\ln\gamma(x)$.
The derivative of Eq.~\eqref{eq:Iwithconstant} at fixed $\eta_k$ for $k\ne j$ is
\begin{equation}
\partial_{\eta_j}I_{\hyp}^{(0)}=-\ln4+f(d)
+\sum_{k=1}^{3}f(\beta_k+d)-2f(\beta_j+d)+2f(\beta_j)
=-A_j^{(0)}.
\end{equation}
The sign of the final term can be checked at a single cone: differentiating $-F(\beta_j)$ gives $+2f(\beta_j)$ because $\beta_j=1-2\eta_j$. Keeping this term reproduces the finite scale at that cone.
The variation fixes how the action changes with angle, but not its additive constant $C_0$. We use a limit in which the metric becomes flat, and its action can be evaluated independently. As $d\to0^+$, the area $4\pi d$ shrinks to zero, so first divide the metric by $d$, setting $\widetilde\varphi=\varphi-\ln d$.
Its area is $4\pi$, while its equation is $\Delta_0\widetilde\varphi=2d \exp(\widetilde\varphi)$.
The boundary terms shift the action by $-d\ln d$, which vanishes.
The area term also vanishes in the limit.
This is a limit of the continuous cone-metric family used to fix the action constant.
It is not a continuous change of the genus of a fixed smooth compact cover.
In the limit, we get
\begin{equation}
\varphi_{\mathrm{flat}}=A_0-4\sum_{j=1}^{3}\bar\eta_j\ln|z-z_j|,
\qquad \sum_{j=1}^{3}\bar\eta_j=1.
\label{limitingfield}
\end{equation}
The harmonic field has $E=J$.
Substitution into Eq.~\eqref{eq:Ireg} gives
\begin{equation}
I_{\mathrm{flat}}=\frac12\left(A_0-\sum_{j=1}^{3}\bar\eta_jA_j^{\mathrm{flat}}\right)
=2\sum_{1\leq j<k\leq3}\bar\eta_j\bar\eta_k\ln|z_j-z_k|^2.
\end{equation}
At $S=1$, we have $\delta_j+\delta_k-\delta_l=2\bar\eta_j\bar\eta_k$, so this equals $I_{\mathrm{pos}}$.
Hence, we obtain $I_{\hyp}^{(0)}\to0$ and $C_0=-F(0)$.
We have derived
\begin{equation}
\boxed{\displaystyle I_{\hyp}^{(0)}=-d\ln4+F(d)-F(0)
+\sum_{j=1}^{3}\big[F(\beta_j+d)-F(\beta_j)\big].}
\label{eq:Ifinal}
\end{equation}

\subsection{Finite Sphere Normalization}
\noindent
There are two sphere metrics in the normalization. The circular cutoff gives one by joining two flat disks. The partition function $s_\C$ uses the other, the round sphere. They are topologically the same sphere, but a short arc generally has different lengths in the two metrics. Their finite conversion factor consequently survives the cancellation of cutoff powers. For the two unit disks the metric is
\begin{equation}
h_D=f(r)|\mathrm d z|^2,\qquad
f(r)=\begin{cases}1,&r\leq1,\\r^{-4},&r\geq1.\end{cases}
\end{equation}
The exterior is the second flat disk written in the coordinate $1/z$: its area element looks different in $z$ because inversion changes the coordinate scale. We now express the metric of the unit round sphere in the same $z$ coordinate:
\bea
h_{\sph}=4\frac{|\mathrm d z|^2}{(1+r^2)^2}=e^\Phi h_D
\eea
with
\begin{equation}
\Phi(r)=\begin{cases}
\ln4-2\ln(1+r^2),&r\leq1,\\
\ln4+4\ln r-2\ln(1+r^2),&r\geq1.
\end{cases}
\label{eq:diskspherePhi}
\end{equation}
The reference curvature is concentrated at the seam:
\bea
\sqrt{h_D}R[h_D]\mathrm d^2x=4\delta(r-1)\mathrm d r\mathrm d\theta.
\eea
All $8\pi$ of the integrated scalar curvature lies on the joining circle. The two metrics give the same length along that circle, so their relative exponent there is $\Phi(1)=0$. The curvature term in the anomaly therefore vanishes; the kinetic term, which measures variation away from the circle, survives.
To see how the curvature is localized, write $h_D=\exp(\sigma_D)|\mathrm d z|^2$. Then $\sqrt{h_D}R[h_D]=-\Delta_0\sigma_D$. Here $r\sigma_D'(r)=0$ for $r<1$ and $r\sigma_D'(r)=-4$ for $r>1$, so
\bea
\Delta_0\sigma_D=-4\frac{\delta(r-1)}{r}.
\eea
Integrating with the polar measure $r\mathrm d r\mathrm d\theta$ gives the stated seam curvature.
This also shows why dropping the curvature term is legitimate in this particular conversion: its coefficient $\Phi$ vanishes at the support of that curvature.
\\

\noindent
Inside and outside the seam, respectively, $\partial_r\Phi=-4r/(1+r^2)$ and $\partial_r\Phi=4/[r(1+r^2)]$. The angular integral gives $2\pi$ and the area element supplies $r\,\mathrm d r$, so
\begin{align}
\int_{\mathbb C}|\nabla_0\Phi|^2\mathrm d^2z
&=32\pi\left[\int_0^1\frac{r^3\mathrm d r}{(1+r^2)^2}
+\int_1^\infty\frac{\mathrm d r}{r(1+r^2)^2}\right]\nonumber\\
&=64\pi\left(\frac12\ln2-\frac14\right)
=16\pi(\ln4-1).
\label{eq:sphereintegral}
\end{align}
The equality of the radial integrals follows by the substitution $r=1/u$ in the second one. In the first, $v=1+r^2$ gives
\bea
\frac12\int_1^2(v^{-1}-v^{-2})\mathrm d v=\frac12\ln2-\frac14,
\eea
completing the explicit integration.
We let $q_\C=Z_\C[S^2,h_D]$.
We then get
\begin{equation}
s_\C=q_\C\exp\left[\frac c6(\ln4-1)\right],\qquad
Z_{\mathrm{base}}(R)=q_\C R^{c/3}.
\label{eq:spherenorm}
\end{equation}
To obtain the second formula, rescale all lengths of the two-disk sphere by $R$. Its metric becomes $R^2h_D$, so its constant Weyl exponent is $2\ln R$. The gradient term vanishes, and the scalar curvature integrates to $8\pi$; Eq.~\eqref{eq:Weyl} therefore gives $(c/3)\ln R$ in the logarithm of its partition function.
A comparison makes it clear why the first formula differs. The two unit disks have total area $2\pi$, whereas the round sphere has area $4\pi$. Uniformly doubling the area would change $\ln Z$ by $(c/6)\ln2$. The round metric scales different regions by different amounts and gives $(c/6)(\ln4-1)$ instead. Total area alone does not determine the anomaly.
Writing $q_\C=Q^c$ is possible when $c\ne0$, but it only parametrizes an arbitrary normalization. We retain $q_\C$, which also makes sense for the $c=0$ sector used below.

\subsection{Two-Point Normalization and Cancellation of all Cutoffs}
\noindent
A twist can be multiplied by an arbitrary constant unless we calibrate it. We use a twist and its inverse placed one unit apart: their normalized two-point function must be one. Applying that same calibration to each of the three insertions lets us remove the cutoff without an arbitrary finite factor.
For a single length-$k$ cycle at $z=0,1$, we use the spherical cover
\begin{equation}
z=\frac{t^k}{1+t^k},\qquad
h_{\sph}=\frac{4|\mathrm d t|^2}{(1+|t|^2)^2}.
\end{equation}
With $\beta=1/k$, the descended positive-curvature cone metric is
\begin{equation}
e^{\varphi_+}=\frac4{k^2}
\frac{|z|^{2\beta-2}|1-z|^{2\beta-2}}
{( |z|^{2\beta}+|1-z|^{2\beta})^2},\qquad
\Delta_0\varphi_+=-2e^{\varphi_+}.
\label{eq:football}
\end{equation}
Its finite constants at zero and one are $A_0=A_1=\ln4-2\ln k$, and its area is $4\pi/k$.
\\

\noindent
The spherical action has a minus sign in its area term.
With two equal cutoffs, its explicit definition is
\begin{equation}
\begin{split}
I_+=\lim_{\epsilon\to0,\,R\to\infty}\bigg[
\frac1{16\pi}\int_D|\nabla_0\varphi_+|^2\mathrm d^2z
&-\frac1{4\pi}\int_D e^{\varphi_+}\mathrm d^2z\\
&+A_\infty-2\ln R-\eta(A_0+A_1)+4\eta^2\ln\epsilon\bigg].
\end{split}
\end{equation}
At $\eta=0$, $\varphi_+=\ln4-2\ln(1+|z|^2)$.
After removing the inner disks, the kinetic term is $2\ln R-1+{\mathcal O}(R^{-2})$, the negative area term is $-1+{\mathcal O}(R^{-2})$, and $A_\infty=\ln4$.
Thus, the sum gives $I_+(0)=\ln4-2$.
For equal strengths $\eta=(1-1/k)/2$, the variational identity gives $\mathrm d I_+/\mathrm d\eta=-2\ln4+4\ln k$.
With $\beta=1/k=1-2\eta$, this becomes $\mathrm d I_+/\mathrm d\beta=\ln4+2\ln\beta$.
Integrating from $\beta=1$ yields
\begin{equation}
I_+=\frac1k(\ln4-2\ln k-2).
\end{equation}
Repeating the boundary calculation, with two equal cutoffs $\epsilon$, gives
\begin{align}
W_{2,k}&=-\frac{kc}{6}I_+-\frac c6+\frac{kc}{3}\ln R
-\frac{c(k-1)^2}{6k}\ln\epsilon+{\mathcal O}(1)\nonumber\\
&=\frac c3\ln k+\frac c6(1-\ln4)+\frac{kc}{3}\ln R
-\frac{c(k-1)^2}{6k}\ln\epsilon+{\mathcal O}(1).
\label{bcoeff}
\end{align}
The ${\mathcal O}(1)$ remainder in Eq.~\eqref{bcoeff} tends to zero.
Eqs.~\eqref{eq:Bk} and \eqref{eq:Nkm} give the leading regulated coefficients with omitted multiplicative factors tending to one; retaining those factors gives the same normalized limit.
Using $B_k=\exp(W_{2,k})s_\C/Z_{\mathrm{base}}^k$ and Eq.~\eqref{eq:spherenorm}, the bare coefficient at unit separation is
\begin{equation}
B_k=k^{c/3}\epsilon^{-c(k-1)^2/(6k)}q_\C^{\,1-k}.
\label{eq:Bk}
\end{equation}
For $m$ disjoint $k$-cycles, the two-point cover has $m$ connected components, so its coefficient is $B_k^m$.
The unit-normalization factor, with a consistent choice of square root, is
\begin{equation}
\mathcal N_{k,m}=B_k^{-m/2}
=k^{-mc/6}\epsilon^{mc(k-1)^2/(12k)}q_\C^{\,m(k-1)/2}.
\label{eq:Nkm}
\end{equation}
Sheets untouched by the permutation contribute equally with and without the two twists, so their factors cancel. For the remaining sheets, the unnormalized pair contributes $B_k^m$; assigning its inverse square root to each insertion gives a pair of coefficient one. This fixes the magnitudes in the Euclidean normalization. Opposite rephasings of a twist and its inverse would still leave their two-point function unchanged. We fix that freedom by using the same sheet-gluing convention for all three insertions and in both calculations.
\\

\noindent
The regulated three-point answer is
\begin{equation}
\left(\prod_{j=1}^{3}\mathcal N_{k_j,m_j}\right)
\frac{e^W Z_\C[\Sigma_g,h_g]}{Z_{\mathrm{base}}(R)^N}.
\label{eq:bareassembly}
\end{equation}
Each cutoff has a different origin. The outer radius $R$ regulates the unbranched region at infinity, while $\epsilon_j$ regulates only the $j$th insertion. Their coefficients must cancel separately; finite contributions from the area and the sphere conversion need not cancel.
Let $\mathcal W_{\mathrm{reg}}$ denote the logarithm of the prefactor multiplying $Z_\C[\Sigma_g,h_g]$ in Eq.~\eqref{eq:bareassembly}, with the same normalization branches as above.
Substituting all three contributions gives
\begin{align}
\mathcal W_{\mathrm{reg}}={}&-\frac{Nc}{6}I_{\hyp}+\frac{Nc}{6}d
-\frac c6\sum_{j=1}^{3}m_j\ln k_j\nonumber\\
&+\left(\frac{Nc}{3}-\frac{Nc}{3}\right)\ln R\nonumber\\
&+\sum_{j=1}^{3}\left[-\frac{Nc}{3}\eta_j^2
+\frac{m_jc(k_j-1)^2}{12k_j}\right]\ln\epsilon_j\nonumber\\
&+\left[-N+\frac12\sum_{j=1}^{3}m_j(k_j-1)\right]\ln q_\C+{\mathcal O}(1).
\label{eq:normalizationledger}
\end{align}
The second line compares the outer circle on the cover with the same circle on each of the $N$ independent base sheets. The third line compares a small circle around a branch value with the small circle used to calibrate that twist. The outer dependence cancels immediately. The inner cancellation can be checked one branch value at a time:
\begin{equation}
\frac{m_jc(k_j-1)^2}{12k_j}=\frac{Nc}{3}\eta_j^2,
\end{equation}
so every $\epsilon_j$ power cancels. Finally, we obtain
\begin{equation}
-N+\frac12\sum_{j=1}^{3}m_j(k_j-1)=g-1,
\end{equation}
which leaves $q_\C^{g-1}$.
After removing $K=\exp(-NcI_{\mathrm{pos}}/6)$, we obtain
\begin{equation}
C_{123}=q_\C^{g-1}Z_\C[\Sigma_g,h_g]
\exp\left\{\frac{Nc}{6}\left[-I_{\hyp}^{(0)}+d+\sum_{j=1}^{3}\beta_j\ln\beta_j\right]\right\}.
\label{eq:assembly}
\end{equation}
Replace $q_\C$ by $s_\C$ using Eq.~\eqref{eq:spherenorm}.
The geometric exponent is
\begin{equation}
\Pcov=\frac{Nc}{6}\left[-I_{\hyp}^{(0)}+\sum_{j=1}^{3}\beta_j\ln\beta_j+d(2-\ln4)\right].
\label{eq:PbeforeI}
\end{equation}
There is a useful check on the constant in the final exponent. The area term in Eq.~\eqref{eq:assembly} gives $d$. The change from the two-disk sphere to the round sphere adds $d(1-\ln4)$, and the action in Eq.~\eqref{eq:Ifinal} contributes $+d\ln4$ through $-I_{\hyp}^{(0)}$. The logarithms of four cancel, leaving $2d$. Inserting the remaining angle terms now gives Eqs.~\eqref{eq:mainintro} and \eqref{eq:factorintro} in the same normalization.

\subsection{Extension to more Branch Values}
\noindent
With four marked points, one can choose a conformal coordinate that puts three at $0$, $1$, and $\infty$. The fourth then has a definite coordinate $x$, the cross ratio in this ordering. Moving it from $x=1/3$ to $x=1/2$ changes the shape even though the sheets and their gluing permutations can stay the same. The cutoff cancellation near each point still works. The metric action must now depend on the relative position as well as the angles.
We consider a connected degree-$N$ cover with $\nu\geq3$ finite branch values $z_1,z_2,\ldots,z_\nu$, uniform ramification $k_j$ at each value, and $m_j=N/k_j$ preimages, for $j=1,2,\ldots,\nu$.
The fixed permutations obey $\sigma_1\sigma_2\cdots\sigma_\nu=1$ and generate a transitive action.
We set $\beta_j=1/k_j$, $\eta_j=(1-\beta_j)/2$, and assume $g>1$.
We then get
\begin{align}
2g-2&=N\left(\nu-2-\sum_{j=1}^{\nu}\beta_j\right),\\
d&=\frac{g-1}{N}=\sum_{j=1}^{\nu}\eta_j-1>0.
\label{eq:manyRH}
\end{align}
The hyperbolic cone sphere again has area $4\pi d$, and its pullback is smooth.
We define its action $I_{\hyp}^{(\nu)}(\mathbf z,\boldsymbol\eta)$ by Eq.~\eqref{eq:Ireg}, with every three-cone sum replaced by the explicit range $j=1,2,\ldots,\nu$:
\begin{equation}
\begin{split}
I_{\hyp}^{(\nu)}=\lim_{\boldsymbol\epsilon\to0,\,R\to\infty}\bigg[
&\frac1{16\pi}\int_{D_{\boldsymbol\epsilon,R}}|\nabla_0\varphi|^2\mathrm d^2z
+\frac1{4\pi}\int_{D_{\boldsymbol\epsilon,R}}e^\varphi\mathrm d^2z+A_\infty-2\ln R\\
&-\sum_{j=1}^{\nu}\eta_jA_j+2\sum_{j=1}^{\nu}\eta_j^2\ln\epsilon_j\bigg].
\end{split}
\label{eq:manyaction}
\end{equation}
Here $D_{\boldsymbol\epsilon,R}$ omits all $\nu$ inner disks.
The finite-cone and infinity asymptotics are those of Eq.~\eqref{eq:asymptotics} for $j=1,2,\ldots,\nu$.
The preceding boundary calculation and each local two-point normalization remain unchanged.
In particular, we have
\begin{equation}
-N+\frac12\sum_{j=1}^{\nu}m_j(k_j-1)=g-1,
\end{equation}
so the residual sphere factor is still $s_\C^{g-1}$.
The full fixed-permutation correlator therefore factorizes as
\begin{align}
\left\langle\prod_{j=1}^{\nu}\mathcal T_{\sigma_j}(z_j)\right\rangle
&=\exp\big[\mathcal P_{\mathrm{cov}}^{(\nu)}(\mathbf z)\big]\Z_\C(\Sigma_g),
\label{eq:manyfactor}\\
\mathcal P_{\mathrm{cov}}^{(\nu)}(\mathbf z)
&=\frac{Nc}{6}\left[-I_{\hyp}^{(\nu)}(\mathbf z,\boldsymbol\eta)
+\sum_{j=1}^{\nu}\beta_j\ln\beta_j+d(2-\ln4)\right].
\label{eq:manyP}
\end{align}
Unlike $\Pcov$, the symbol $\mathcal P_{\mathrm{cov}}^{(\nu)}(\mathbf z)$ includes the entire position dependence of the geometric factor.
At $\nu=3$, it equals $\Pcov+\ln K$.
Eq.~\eqref{eq:manyP} gives a prescription even when no closed formula for the metric is available. To get a number for four points, for instance, one must still solve the cone equation at their chosen relative positions and insert that solution into the action.
\\

\noindent
The new position data are cross ratios, combinations invariant under M\"obius coordinate changes.
For example, four distinct points define
\begin{equation}
x=\frac{(z_1-z_2)(z_3-z_4)}{(z_1-z_3)(z_2-z_4)}.
\end{equation}
Three points can be moved to $0,1,\infty$ by a M\"obius map; for $\nu$ points, the remaining $\nu-3$ independent complex positions are moduli.
One can hold the first three points fixed and move the fourth without changing how many sheets meet anywhere. This changes the cross ratio $x$, so the local branching orders no longer suffice to evaluate the action.
The extra calculation required at $\nu>3$ can be identified precisely.
With infinity unbranched, the projective connection is
\begin{equation}
T(z)=\sum_{j=1}^{\nu}\left[\frac{\delta_j}{(z-z_j)^2}
+\frac{a_j}{z-z_j}\right],\qquad
\delta_j=\eta_j(1-\eta_j).
\end{equation}
Its regularity in the coordinate $1/z$ requires $T(z)={\mathcal O}(z^{-4})$, giving three constraints:
\begin{equation}
\sum_{j=1}^{\nu}a_j=0,\qquad
\sum_{j=1}^{\nu}(\delta_j+z_ja_j)=0,\qquad
\sum_{j=1}^{\nu}(2z_j\delta_j+z_j^2a_j)=0.
\end{equation}
For three branch values, these three constraints leave no coefficient free. For four, they leave one. Its value is fixed by requiring that a length measured in the disk agree before and after following each independent loop. Solving this global condition determines the remaining accessory parameters for general $\nu$, and hence the action. The local cancellation of cutoffs and the conversion to the round sphere still apply before this last step is solved.

\section{Analytic Continuation of Conical-Defect Expression}
\label{sec:3}
\begin{figure}[!htb]
\centering
\includegraphics[width=.72\linewidth]{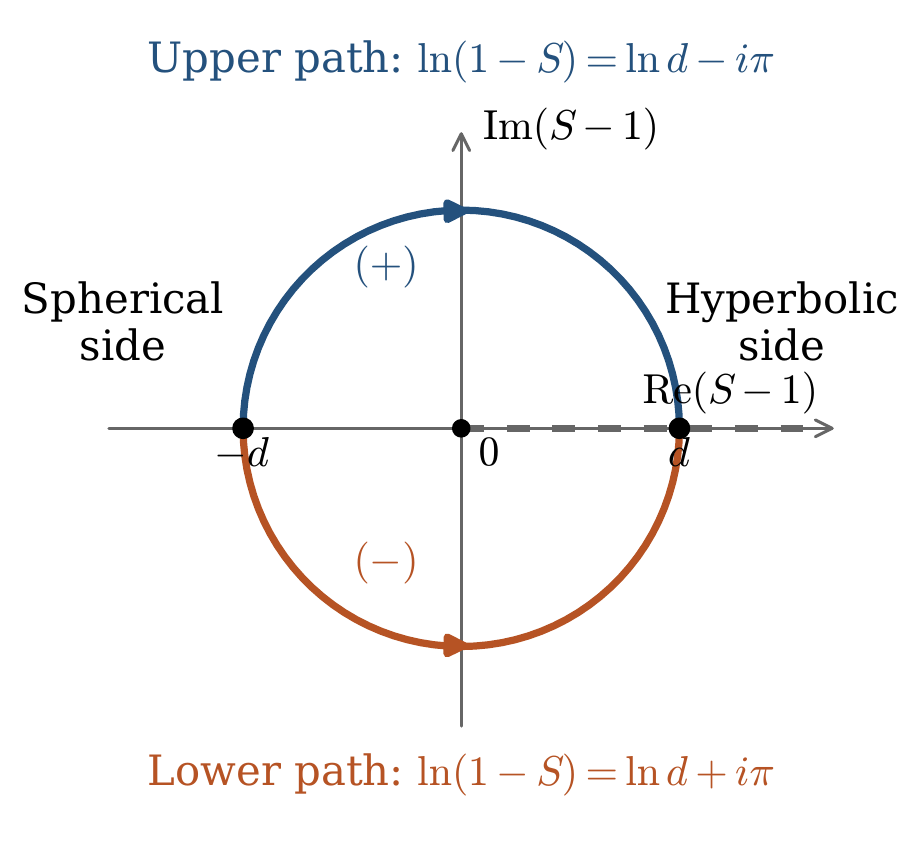}
\caption{Two ways to reach the same three real cone strengths, shown in the $S-1$ plane. Both start at $-d$ on the spherical side and end at $d$ on the hyperbolic side. The positive real ray marks the cut. On the upper path, $1-S$ reaches the negative real axis from below, so its continuously followed angle is $-\pi$. On the lower path, the angle is $+\pi$. The same path choice also fixes $F(S)$.}
\label{fig:continuation}
\end{figure}

\noindent
The direct covering calculation used a real metric with negative curvature. A second formula starts on a sphere with positive curvature and three cone tips. How can the two give the same real exponent? We fix the twist normalization and treat the cone strengths as continuous variables. At one value the spherical area vanishes; the continuation passes around that value in the complex plane.
The starting exponent is Eq.~(4.32) of Ref.~\cite{Chang:2016ftb}, with its two-point normalization already included. The defects live on the base, where all $N$ copies contribute to the stress tensor. This explains the factor $Nc$, the same central charge that gives the twist weights in Eq.~\eqref{eq:weights}. In our notation, the formula reads
\begin{align}
\PCL=\frac{Nc}{6}\bigg\{&\sum_{j=1}^{3}\Big[F(2\eta_j)-F(S-2\eta_j)
+(1-2\eta_j)\ln(1-2\eta_j)\Big]\nonumber\\
&+F(0)-F(S)-2(1-S)\ln(1-S)\bigg\}.
\label{eq:CL}
\end{align}
In the original calculation, a classical geometry supplied this expression. Here we will use the expression as an analytic function of the cone angles. Matching it to the Weyl anomaly tests that function; the match does not assert that one geometry dominates a gravity path integral.
We start where the geometry and the action are real: $0<\eta_j<1/2$ for $j=1,2,3$, $S<1$, and $S-2\eta_j>0$ at each tip. The last condition says that each strength is less than the sum of the other two, the triangle inequality. As $S$ increases to one, the spherical area $4\pi(1-S)$ shrinks to zero. Fig.~\ref{fig:continuation} shows the two detours that let us pass this point without passing through the singularity. We continue the parameters of the formula, not an individual surface, through a change in topology. This is a different continuation from crossing a triangle inequality in Ref.~\cite{Chang:2016ftb}.
\\

\noindent
At covering values, $2\eta_j=1-\beta_j$ and $S-2\eta_j=\beta_j+d$. Reflection of $F$ gives
\begin{equation}
\PCL=\frac{Nc}{6}\left\{
\sum_{j=1}^{3}\big[F(\beta_j)-F(\beta_j+d)+\beta_j\ln\beta_j\big]
+F(0)-F(1+d)+2d\ln(-d)\right\}.
\label{eq:CLreduced}
\end{equation}
The handle count tells us where the function must be evaluated. For three hyperbolic branch values, $2d=1-\sum_{j=1}^{3}\beta_j$ with every $\beta_j>0$, so $0<d<1/2$. Also, $\beta_j+d<1$ for each $j=1,2,3$, because $\beta_j\leq 1-2d$.
Thus, at the endpoint, the terms $F(\beta_j)$ and $F(\beta_j+d)$ are evaluated on their original real interval. On the paths considered here, only $F(1+d)$ and $\ln(-d)$ require a choice of side at the cut. Choosing signs for those two terms independently would not continue the original function.
\\

\noindent
One can exhibit such a path for every covering endpoint in this section. Denote its strengths by $\eta_j^\star$, set $S_\star=1+d$, and hold their proportions $\lambda_j=\eta_j^\star/S_\star$ fixed for $j=1,2,3$. Along the path put $\eta_j(S)=\lambda_j S$. The arguments whose behavior needs checking are then
\[
2\eta_j(S)=2\lambda_j S,\qquad
S-2\eta_j(S)=(1-2\lambda_j)S,\qquad
1-2\eta_j(S)=1-2\lambda_j S,\qquad j=1,2,3 .
\]
For real $0<S\le S_\star$, the first two grow from zero to $1-\beta_j$ and $\beta_j+d$, respectively. Both endpoint values lie strictly between zero and one. The third argument stays positive. Start at $S=1-d$ on this real line, follow it toward one, and take a sufficiently small semicircle above or below one before returning to the real line on the chosen side of the cut. By continuity, the detour encounters no singularity of these three arguments. Only $F(S)$ and $\ln(1-S)$ need a branch choice.
For example, on the five-sheet genus-two cover all three strengths end at $2/5$. At $S_\star=6/5$ the displayed arguments are $4/5$, $2/5$, and $1/5$, yet $1-S_\star=-1/5$. The intermediate strengths are arguments of a function. Only the endpoint is required to describe this five-sheet cover.
\\

\noindent
The sign can be read directly in Fig.~\ref{fig:continuation}. On the upper detour $\Ima S>0$, so $\Ima(1-S)<0$. At the endpoint $1-S=-d-i0$ the angle followed from positive $1-S$ is $-\pi$, giving $\ln(1-S)=\ln d-\mathrm{i}\pi$. There is also a contribution from $F$, since its derivative contains a logarithm. To find its sign, use the recurrence
\begin{equation}
\gamma(1+x)=-x^2\gamma(x),
\end{equation}
following its logarithm along the same upper path. For $0<x<d$ this gives
\begin{equation}
F'(1+x+i0)=F'(x)+2\ln x-i\pi.
\end{equation}
The recurrence gives the magnitude of $\gamma(1+x)$, but its minus sign does not tell us which detour ends at an argument of $-\pi$ and which at $+\pi$. To find out, follow $\gamma$ near its zero at $1$. For a small complex displacement $\zeta$,
\bea
\gamma(1+\zeta)=-\zeta+{\mathcal O}(\zeta^2).
\eea
Along the upper semicircle in Fig.~\ref{fig:continuation}, $\zeta$ goes from the negative to the positive real axis with its argument decreasing from $\pi$ to zero.
Thus, the continuous argument of $-\zeta$ decreases from zero to $-\pi$.
This determines the $-i\pi$ in $F'$; the lower path gives $+i\pi$.
The path fixes the sign without introducing a second integration constant. On a detour of radius $\rho$, the integration length shrinks proportionally to $\rho$, while $F'$ grows only logarithmically. The integral is ${\mathcal O}\big(\rho(1+|\ln\rho|)\big)$ and tends to zero as $\rho\to0$. We can therefore start from the common value $F(1)=F(0)$. Using $\int_0^d\ln x\,\mathrm d x=d\ln d-d$ gives
\begin{equation}
F(1+d+i0)=F(d)+2d\ln d-2d-i\pi d.
\label{eq:Fcontinue}
\end{equation}
The explicit logarithm and the integral $F$ enter the action with different coefficients. Combining their contributions on this same upper path gives
\begin{align}
-F(1+d+i0)+2d\ln(-d-i0)
&=-F(d)-2d\ln d+2d+i\pi d\nonumber\\
&\hspace{1.2em}+2d\ln d-2i\pi d\nonumber\\
&=-F(d)+2d-i\pi d.
\label{eq:cutcancellation}
\end{align}
On the upper path, the explicit logarithm supplies $-2i\pi d$. The term $-F(1+d+i0)$ supplies $+i\pi d$, leaving $-i\pi d$ in the braces. The lower path reverses both signs and leaves the real terms unchanged. Multiplying by the original prefactor gives
\begin{equation}
\PCL^{(\pm)}=\frac{Nc}{6}\left\{
\sum_{j=1}^{3}\big[F(\beta_j)-F(\beta_j+d)+\beta_j\ln\beta_j\big]
+F(0)-F(d)+2d\right\}\mp\frac{i\pi Nc d}{6}.
\label{eq:CLfinal}
\end{equation}
The notation $\Lnc$ denotes a logarithm followed continuously from a specified starting value. The superscript $(+)$ names the path in the upper half of the $S$ plane. For $c>0$ and $g>1$, it gives the \emph{negative} imaginary part: $1-S$ goes below the origin when $S$ goes above one. The lower path gives the complex conjugate of the upper result.

\section{Comparison}
\label{sec:4}
\noindent
The two calculations now use identical twist and sphere normalizations. One measured lengths with a real metric. The other followed a cone action along either side of a singularity. We compare the exponents for $g>1$ and then change the branching while holding the genus at two. The flat torus~\cite{Ma:2026wpv} and the positively curved sphere require separate checks.

\subsection{Equality of Geometric Factors}
\begin{proposition}
Let a connected cover of genus $g>1$ have three branch values. At each branch value, suppose all sheet cycles have the same length. Normalize the twists by their two-point functions and the partition function by the round-sphere factor in Eq.~\eqref{eq:Zhatintro}. Then the direct covering exponent and the two continued cone exponents obey
\begin{equation}
\Pcov=\Rea\PCL^{(+)}=\Rea\PCL^{(-)},\qquad
\Ima\PCL^{(\pm)}=\mp\frac{\pi c}{6}(g-1).
\label{eq:comparison}
\end{equation}
\end{proposition}
\noindent
The equality can be checked without knowing a single energy level of the CFT. Every real term in Eq.~\eqref{eq:CLfinal} occurs in Eq.~\eqref{eq:mainintro}, including the $2d$ from the common normalization. The remaining term is imaginary and contains $Nd=g-1$. The number of sheets has dropped out of that term: at fixed $c$, each handle beyond the first contributes the same angle on a given path. No sign or reality assumption about the seed partition function enters this comparison. Restoring the position dependence and the seed gives the full correlation:
\begin{equation}
\langle\mathcal T_{\sigma_1}(z_1)\mathcal T_{\sigma_2}(z_2)
\mathcal T_{\sigma_3}(z_3)\rangle
=K(z_1,z_2,z_3)e^{\Rea\PCL^{(\pm)}}\Z_\C(\Sigma_g).
\label{eq:complete}
\end{equation}
This equation separates three effects that are measured differently. Moving an endpoint changes the position factor $K$. Changing the reference metric changes both the geometric conversion and the partition function. Changing the field theory can change $\Z_\C$ even when the branch positions, sheet permutations and central charge agree. We have fixed the permutations themselves; averaging over relabelings would require a separate sum and normalization.
The order of operations matters. If $P=p+i\theta$, then $|\exp(P)|=\exp(p)$, whereas $\Rea\exp(P)=\exp(p)\cos\theta$. For example, at $g=2$ and $c=3$ the phase angle is $\mp\pi/2$: the real part of the exponentiated cone answer vanishes even though its magnitude $\exp(\Pcov)$ is positive. Eq.~\eqref{eq:comparison} equates the real parts \emph{before} exponentiation.
\\

\noindent
Consider $w^5=z(z-1)$ and include the point above infinity to make a smooth closed surface. The covering map keeps $z$ and forgets $w$. At $z=2$, for example, $w^5=2$ has five distinct roots: the five points above that one base point. Going around either $0$ or $1$ multiplies $w$ by $\exp(2\pi i/5)$; going around infinity acts by its inverse square. Each permutation is a single five-cycle. Thus $N=5$, $k_1=k_2=k_3=5$, and Riemann--Hurwitz gives $2g-2=-10+3(5-1)=2$. Substituting $d=1/5$ and $\beta_j=1/5$ in the two calculations gives
\[
\frac{\Pcov}{c}
=\frac56\left[F(0)+2F(1/5)-3F(2/5)
+\frac25-\frac35\ln 5\right]\quad(c\ne0),
\qquad \PCL^{(\pm)}=\Pcov\mp\frac{i\pi c}{6}.
\]
The difference $\Rea\PCL^{(\pm)}-\Pcov$ is exactly zero. The nonzero term for $c\ne0$ is the \emph{imaginary} part of the continued exponent. Numerically, direct integration of Eq.~\eqref{eq:Fdefinition} and continuation of $F(S)$ along the two semicircles give $\Pcov/c\simeq-1.31240$ and $\PCL^{(\pm)}/c\simeq-1.31240\mp0.52360\,i$. For an example at every positive genus, replace $5$ by any odd integer $m\geq3$. This leaves all three permutations full cycles, since $2$ and $m$ are coprime. Now $2g-2=-2m+3(m-1)=m-3$, giving $m=2g+1$. This construction gives an example at each positive genus; the genus does not fix all possible shapes.
\\

\noindent
Now change the branching, keeping the two handles. The eight-sheet cover of Sec.~\ref{sec:2} has $(\beta_1,\beta_2,\beta_3)=(1/2,1/8,1/8)$ and $d=1/8$. In Eq.~\eqref{eq:mainintro}, the first cone gives $F(1/2)-F(5/8)$, while each of the other two gives $F(1/8)-F(1/4)$ before the logarithms are included. With $F(1/2)=0$ and $F(5/8)=F(3/8)$, the result is
\[
\frac{\Pcov}{c}=\frac43\left[F(0)+F(1/8)-2F(1/4)-F(3/8)
+\frac14-\frac54\ln2\right]\simeq-1.77192,\qquad c\ne0 .
\]
The five-sheet value was $-1.31240$ in the same units. The real geometric exponent can therefore distinguish these covers. Their upper-path imaginary parts are identical, $-\pi c/6$, since $Nd=1$ in each case; the lower-path signs reverse. On \emph{each} cover separately the difference $\Rea\PCL^{(\pm)}-\Pcov$ remains zero.
\\

\noindent
The same sheet pattern occurs in a familiar entanglement calculation. For two adjacent intervals $A_1,A_2$, form the matrix $\rho_{A_1A_2}^{T_2}$ by exchanging the row and column indices belonging to $A_2$ while leaving the $A_1$ indices alone. This is the partial transpose of their joint reduced density matrix. Ref.~\cite{Calabrese:2012nk} represents its moment $\mathrm{Tr}[(\rho_{A_1A_2}^{T_2})^m]$ by three twists. If $\sigma$ advances one step around the $m$ sheets, the endpoint permutations in spatial order are $\sigma,\sigma^{-2},\sigma$. For odd $m$, stepping by two visits every sheet. After relabeling the three endpoints, these are precisely the permutations of the cyclic cover above. In particular, $m=5$ produces genus two.
\\

\noindent
This moment still includes the seed partition function. At finite cutoff, the partially transposed matrix is Hermitian, so every integer moment is real. Logarithmic negativity uses the sum of the absolute values of its eigenvalues, obtained in the replica method from the even moments. Knowing one odd moment does not give that sum.
\\

\noindent
One need not prepare the partially transposed matrix as a state; indeed, it can have negative eigenvalues. Instead prepare $m$ identical copies of $\rho=\rho_{A_1A_2}$, and let $V_X$ cycle the $m$ copies of subsystem $X$, for $X=A_1,A_2$. A cycle contracts the column index of one copy with the row index of the next, giving the trace of a matrix raised to the $m$th power. Thus, for integer $m\ge2$,
\[
\mathrm{Tr}[(\rho^{T_2})^m]
=\mathrm{Tr}[(\rho^{T_2})^{\otimes m} V_{A_1}V_{A_2}]
=\mathrm{Tr}[\rho^{\otimes m} V_{A_1}V_{A_2}^{-1}] .
\]
In the second equality, the transpose on $A_2$ is moved across the trace. The transpose of a permutation matrix is its inverse, so the operation cycles one interval forward and the other backward. Its Hermitian part, the average of the operation and its adjoint, measures the displayed real moment. At $m=2$, the two directions coincide, and the result is $\mathrm{Tr}\rho^2$. At every odd $m=2g+1$, this operation realizes the cyclic replica family above. It measures the complete moment, including its CFT dependence.
\\

\noindent
For a two-qubit example, take the pure state $(|00\rangle+|11\rangle)/\sqrt{2}$. Its partially transposed density matrix interchanges the two basis labels and multiplies by $1/2$. Its three symmetric eigenvectors have eigenvalue $1/2$; the antisymmetric one has eigenvalue $-1/2$. Thus at any odd order $m=2g+1$ with $g\ge1$ the moment is $3(1/2)^{2g+1}+(-1/2)^{2g+1}=4^{-g}$. The fifth moment is $1/16$. By contrast, the sum of the absolute eigenvalues is always $2$, giving logarithmic negativity $\ln2$. This is a finite quantum system illustrating the two measurements; the odd-order formula here is not the CFT partition function of a genus-$g$ surface.

\subsection{Torus and Sphere Limits}
\noindent
At the boundary $d=0$, there are two ways to check the calculation. One can vary the cone angles continuously and rescale the shrinking metric as before Eq.~\eqref{limitingfield}.
One can also start with branching orders that already give a torus and calculate with its flat metric. These procedures agree on the exponent. Neither procedure turns a fixed surface with two or more handles into a torus.
On a flat torus the scalar curvature is zero. Changing its overall size by a constant factor contributes neither a curvature term nor a gradient term in Eq.~\eqref{eq:Weyl}. The partition function can still depend on the shape of the torus, but changing its area in this way does not change it. At the flat point, the exponent reduces to~\cite{Ma:2026wpv}
\begin{equation}
\Pcov\big|_{g=1}=\frac{Nc}{6}\sum_{j=1}^{3}\beta_j\ln\beta_j
=-\frac c6\sum_{j=1}^{3}m_j\ln k_j,
\qquad \Ima\PCL=0.
\label{eq:torus}
\end{equation}
The three flat triangles can be found without solving for a metric. Order the integers $2\leq k_1\leq k_2\leq k_3$. Flatness requires $1/k_1+1/k_2+1/k_3=1$, so $k_1=2$ or $3$. If $k_1=3$ all three equal $3$; if $k_1=2$ the other two reciprocals must sum to $1/2$, giving $(k_2,k_3)=(4,4)$ or $(3,6)$.
Thus, up to permutation, the signatures are $(3,3,3)$, $(2,4,4)$ and $(2,3,6)$~\cite{Ma:2026wpv}. At the minimal degrees $N=3,4,6$, Eq.~\eqref{eq:torus} gives $-(c/2)\ln3$, $-c\ln2$, and $-(c/6)(4\ln2+3\ln3)$, respectively~\cite{Ma:2026wpv}. The orders specify the angles; one must still give the permutations to specify a cover.
\\

\noindent
For $g=0$, the corresponding calculation uses a positive-curvature cone metric.
It is useful to derive its action explicitly because this is the real branch entering the replica limit.
We work first in the continuous spherical domain $0<\eta_j<1/2$, $S<1$, and $S-2\eta_j>0$ for $j=1,2,3$ and set $t=1-S>0$.
The last inequalities ensure the spherical three-cone solution in this domain.
Its area is $4\pi t$.
The same linear differential equation gives the coordinates. What changes is how its two solutions measure distance: on a sphere the squared magnitudes add, whereas on the disk they enter with opposite signs [Eq.~\eqref{eq:fmetric}]. The spherical expression is
\begin{equation}
f_+=A_+|y_1|^2+B_+|y_2|^2,\qquad
e^{\varphi_+}=\frac4{f_+^2},\qquad
A_+B_+\beta_1^2=1.
\label{eq:sphericalform}
\end{equation}
Here $A_+,B_+>0$, and the basis is Eq.~\eqref{eq:hypergeom} evaluated at $b=-t$.
Direct differentiation gives
\bea
\partial_z\partial_{\bar z}\varphi_+=-\frac{2}{f_+^2},
\eea
hence
\bea
\Delta_0\varphi_+=-2e^{\varphi_+}.
\eea
Cancellation of the mixed term near one now requires $A_+UV+B_+\widetilde U\widetilde V=0$.
In comparison with Eq.~\eqref{eq:A2}, this changes the sign of the ratio defining $A_+^2$:
\begin{equation}
A_+^2=-\frac{\gamma(S-1)\gamma(S-2\eta_2)\gamma(S-2\eta_3)}
{\gamma(2\eta_1)^2\gamma(S-2\eta_1)}.
\end{equation}
The recurrence
\bea
-\gamma(-t)=\frac{\gamma(1-t)}{t^2}
\eea
makes positivity and the finite cone constants explicit.
With $\{j,k,l\}=\{1,2,3\}$, we obtain
\begin{equation}
A_{j,+}^{(0)}=\ln4-
\ln\left[\frac{\gamma(S)\gamma(S-2\eta_k)\gamma(S-2\eta_l)}
{t^2\gamma(2\eta_j)^2\gamma(S-2\eta_j)}\right].
\label{eq:sphericalconstants}
\end{equation}
All gamma-ratio arguments in this expression lie between zero and one, so the displayed logarithm is real.
Its position dependence is still Eq.~\eqref{eq:finitepositions}.
\\

\noindent
We define $I_+$ by replacing the positive area term in Eq.~\eqref{eq:Ireg} by a negative one.
The field equation again removes the bulk variation, leaving
\bea
\mathrm d I_+=-\sum_{j=1}^{3}A_{j,+}\,\mathrm d\eta_j.
\eea
Integration of Eq.~\eqref{eq:sphericalconstants} gives its position-independent part:
\begin{align}
I_+^{(0)}={}&F(S)+\sum_{j=1}^{3}\big[F(S-2\eta_j)-F(2\eta_j)\big]-F(0)
\nonumber\\
&+2t\ln t+t(\ln4-2).
\label{eq:sphericalaction}
\end{align}
To check the integration, we write $f(x)=\ln\gamma(x)$ as before and differentiate at fixed $\eta_k$ for $k\ne j$:
\begin{align}
\partial_{\eta_j}I_+^{(0)}={}&f(S)+\sum_{k=1}^{3}f(S-2\eta_k)
-2f(S-2\eta_j)-2f(2\eta_j)\nonumber\\
&-2\ln t-\ln4=-A_{j,+}^{(0)}.
\end{align}
The additive constant is fixed by setting cone strengths to zero.
Then $t\to1$, all $F$ terms in Eq.~\eqref{eq:sphericalaction} cancel and $I_+^{(0)}\to\ln4-2$.
On the two-cone boundary $(\eta,\eta,0)$, continuity gives $I_+^{(0)}=\beta(\ln4+2\ln\beta-2)$ with $\beta=1-2\eta$, in agreement with the independently normalized two-point cover.
\\

\noindent
For spherical covering data, $d=-t=-1/N$, and the smooth reference surface is the unit sphere.
The anomaly calculation now uses its positive scalar curvature and the negative area term of $I_+$; the local cap estimates and twist normalizations are unchanged.
Consequently, its finite geometric exponent is
\begin{align}
P_{\mathrm{cov},+}
&=\frac{Nc}{6}\left[-I_+^{(0)}+\sum_{j=1}^{3}\beta_j\ln\beta_j
-t(2-\ln4)\right]\nonumber\\
&=\frac{Nc}{6}\left\{\sum_{j=1}^{3}
\big[F(2\eta_j)-F(S-2\eta_j)+\beta_j\ln\beta_j\big]
+F(0)-F(S)-2t\ln t\right\}.
\label{eq:sphericalcomparison}
\end{align}
This is precisely the real spherical expression~\eqref{eq:CL}, derived from the positive-curvature covering calculation.
On a spherical cover, $Z_\C[\Sigma_0,h_{\sph}]=s_\C$, and the factor $s_\C^{g-1}=s_\C^{-1}$ cancels it. Hence $\Z_\C(\Sigma_0)=1$ and $C_{123}=\exp(\PCL^{\sph})$ for these fixed spherical covers.
The two-point limit or the two-cone boundary gives $P_{\mathrm{cov},+}=0$, as required.
The continuous cone action also defines a geometric function away from the discrete covering values; it does not define a seed partition function at noninteger genus or determine a continuation of the full replica amplitude.

\section{Imaginary Factor and Weyl Anomaly}
\label{sec:5}
\noindent
The cone calculation assigns different imaginary parts to two paths; the calculation with a real covering metric assigns neither. To see why the answers are consistent, keep the physical surface and all distances on it fixed. Change only the reference metric used to split the answer into a geometric factor and a partition function. We first work out how each factor changes under a real rescaling, then follow that change to the two complex scales selected by the cone calculation in Sec.~\ref{sec:3}.
Let $\mathcal A[\Phi;h]$ denote the right-hand side of Eq.~\eqref{eq:Weyl}. It is the change in $\ln Z$ when the squared lengths measured by $h$ are multiplied by $\exp(\Phi)$. For two smooth real rescalings of a closed surface,
\begin{equation}
\mathcal A[\Phi+\Psi;h]
=\mathcal A[\Phi;h]+\mathcal A[\Psi;e^\Phi h].
\label{eq:Weylcocycle}
\end{equation}
This is the Weyl cocycle relation. Read the right-hand side in order: the first rescaling starts from $h$, and the second starts from the metric produced by the first. The total must match a direct change between the same initial and final metrics.
To verify it, the two-dimensional conformal transformation of curvature gives \cite{Polyakov:1981rd}
\begin{equation}
\mathcal A[\Psi;e^\Phi h]=\frac c{96\pi}
\int_\Sigma\mathrm d^2x\sqrt h\,
\big[(\nabla\Psi)^2+2R[h]\Psi-2\Psi\Delta_h\Phi\big].
\end{equation}
Integration by parts replaces the last term with $2\nabla\Psi\cdot\nabla\Phi$, which completes the square in $\mathcal A[\Phi+\Psi;h]$ after adding $\mathcal A[\Phi;h]$.
There is no physical boundary term because the replica surface is closed; we include the regulator caps before taking the limit.
In particular, we write the fixed physical metric as $g_{\mathrm{phys}}=\exp(\Phi) h$ and change the reference metric to $h'=\exp(\Psi) h$.
Its new relative exponent is $\Phi-\Psi$. Eq.~\eqref{eq:Weylcocycle} implies
\begin{equation}
\mathcal A[\Phi-\Psi;h']-\mathcal A[\Phi;h]
=-\mathcal A[\Psi;h],\qquad
Z_\C[h']=e^{\mathcal A[\Psi;h]}Z_\C[h].
\label{eq:generalreferencecancel}
\end{equation}
For example, suppose the new reference multiplies the partition function by $\exp(a)$. Converting the new reference lengths back to the original physical lengths multiplies the geometric factor by $\exp(-a)$. The physical amplitude is unchanged. This cancellation extends to a complex rescaling wherever both factors can be continued along the same path; it does not fix the phase of the original partition function.
The quadratic gradient term is continued bilinearly, as $\sum_{a=1}^{2}\sum_{b=1}^{2}h^{ab}\partial_a\Phi\partial_b\Phi$; it is not replaced by a modulus squared involving complex conjugation.
\\

\noindent
First, change every squared length by the same factor $\exp(\omega)$. Because $\omega$ is constant, neighboring points see no difference in the rescaling and the kinetic term in Eq.~\eqref{eq:Weyl} is zero. What remains is $c\omega\int_{\Sigma_g}R\,\mathrm dA/(48\pi)$. Gauss--Bonnet sets this integral of scalar curvature to $4\pi\chi(\Sigma_g)$, giving
\begin{equation}
\ln Z_\C[e^\omega h]-\ln Z_\C[h]
=\frac{c\chi(\Sigma_g)}{12}\omega.
\label{eq:constantWeyl}
\end{equation}
For real $\omega$, this is the response to changing every squared length by the same positive factor. Because the right-hand side is linear in $\omega$, its continuation along a complex scale rotation is explicit. Interpreting it as a change of the full partition function assumes that the partition function can be followed along that path.
A real example fixes the sign before any continuation. Double all reference lengths on a genus-two surface. Then $\omega=2\ln2$ and $\chi=-2$, so $\ln Z_\C$ changes by $-(c/3)\ln2$. The factor that converts this reference back to the fixed physical metric must add precisely $(c/3)\ln2$ to its own exponent.
\\

\noindent
To identify the sign, we write a fixed physical amplitude as
\begin{equation}
C=e^{P[h]}Z_\C[\Sigma_g,h]\,\mathcal N_{\mathrm{fixed}},
\end{equation}
where $\mathcal N_{\mathrm{fixed}}$ contains the fixed sphere and local operator normalizations.
A change of reference metric requires
\begin{equation}
P[e^\omega h]-P[h]=-\frac{c\chi(\Sigma_g)}{12}\omega.
\label{eq:Ptransformation}
\end{equation}
This minus sign is essential: an increase in $\ln Z$ from changing the reference must be offset by a decrease in the geometric exponent. To identify the complex rescaling in the cone calculation, recall the spherical equation $\Delta_0\varphi_{\sph}=-2\exp(\varphi_{\sph})$. Its area $4\pi(1-S)$ becomes negative if the formula is continued to $S>1$. In contrast, the area measured by the real hyperbolic metric is positive. The following shift of the metric logarithm realizes that sign in the field equation:
\begin{equation}
\varphi_{\mathrm{CL}}^{(\pm)}=\varphi_{\hyp}\mp i\pi,
\qquad e^{\varphi_{\mathrm{CL}}^{(\pm)}}=-e^{\varphi_{\hyp}}.
\end{equation}
The left-hand side of the spherical equation contains derivatives, so a constant shift of $\varphi$ leaves it unchanged. On the right-hand side, the shift multiplies the exponential by $-1$, exactly the sign needed for the hyperbolic equation. This formal substitution identifies an anomaly response; it does not establish convergence of a path integral on a negative metric. The final tensor on the smooth cover is the same on both paths, $h_{\mathrm{CL}}^{(\pm)}=\exp(\mp i\pi)h_g=-h_g$. Its two continuously chosen scale logarithms differ, however. Using each one in Eq.~\eqref{eq:Ptransformation} gives
\begin{equation}
P[h_{\mathrm{CL}}^{(\pm)}]-P[h_g]
=-\frac{c(2-2g)}{12}(\mp i\pi)
=\mp\frac{i\pi c}{6}(g-1).
\label{eq:anomalyphase}
\end{equation}
For example, on the upper path $\omega=-i\pi$. At genus two Eq.~\eqref{eq:constantWeyl} gives $+i\pi c/6$ in $\ln Z_\C$, while Eq.~\eqref{eq:anomalyphase} gives $-i\pi c/6$ in the geometric exponent. The second sign is exactly the one found by continuing the gamma-function expression. For any $g>1$, the changes in the two factors are
\begin{equation}
\Delta\ln Z_\C=\pm\frac{i\pi c}{6}(g-1),\qquad
\Delta P=\mp\frac{i\pi c}{6}(g-1),\qquad
\Delta\ln C=0.
\label{eq:phasecancel}
\end{equation}
A phase angle remembers the path; its exponential only remembers that angle modulo $2\pi$. At $g=2$ and $c=6$ the angles are $-\pi$ and $+\pi$, but both amplitudes acquire the factor $-1$. Each reference partition function supplies a second $-1$, leaving the physical amplitude unchanged. More generally, the two geometric phase factors coincide exactly when $c(g-1)/6$ is an integer. The cancellation in the \emph{full} amplitude does not require this condition: it works on each path separately.
A positive Euclidean covering metric never takes either path. Continuing the geometric exponent while keeping the seed factor fixed would define a different quantity. We have used no assumption that the original $\Ima\ln Z_\C$ vanishes.
\\

\noindent
There is a separate situation in which we do know the original partition function is real. Suppose cutting the Euclidean surface along a curve separates it into a piece $M$ and its reflected copy, with the same reflected metric and field data. If the theory has reflection positivity, the path integral on $M$ prepares a state $|\Psi_M\rangle$ and the integral on the reflected piece prepares its adjoint. Sewing along the cut sums over a complete set of boundary states and gives $Z_\C=\langle\Psi_M|\Psi_M\rangle$. This is positive if the prepared state is nonzero, and its real logarithm then has zero imaginary part. This is a sufficient condition on the geometry and the preparation. The handle count alone gives no such argument, and the cancellation in Eq.~\eqref{eq:phasecancel} does not use it.
\\

\noindent
Why is the coefficient topological? Removing or adding a handle changes the integrated curvature, whereas bending the metric at fixed genus leaves that integral unchanged. This coefficient does not count ground states: on a torus it vanishes even when the partition function depends strongly on the spectrum. A chiral gravitational anomaly is a different response: it obstructs coordinate invariance when the left- and right-moving central charges differ. Here $c_L=c_R$; the Weyl anomaly remains because it concerns a change of length scale.

\section{R\'enyi  Multi-Entropy and Replica Limit}
\label{sec:6}
\noindent
There are two different questions to ask about the vacuum of three adjacent regions. First, keep the number of copies fixed and move the endpoints to timelike separations: which phases survive after subtracting the single-region entropies? Then keep the endpoints fixed and ask how the answer changes as the replica index approaches one: what constant survives in the entropy? Tab.~\ref{tab:operations} identifies the variable in each operation. We start with counts at integer index, obtain the first two corrections to the geometric entropy near one, and state what extra information would determine the full entropy at large $c$. The same counting argument also applies to more parties.

\begin{table}[htbp]
\centering
{\small
\begin{tabular}{@{}>{\raggedright\arraybackslash}p{.24\linewidth}>{\raggedright\arraybackslash}p{.34\linewidth}>{\raggedright\arraybackslash}p{.325\linewidth}@{}}
\toprule
Operation & What varies & What it determines\\
\midrule
Follow a sheet & The point $z$ moves around an endpoint on a fixed surface. & Which sheet a path reaches.\\[3pt]
Vary the cone angles & The total cone strength $S$ passes above or below $1$. & The phase of the geometric factor on that branch.\\[3pt]
Move the endpoints & The three squared separations rotate together at fixed $n$ and real cutoff. & The phases of the physical interval factors.\\[3pt]
Take the replica limit & The normalized replica amplitude is continued to $n=1$. & Minus the slope of its logarithm gives the entropy.\\
\bottomrule
\end{tabular}\par}
\caption{Four operations involving different variables. A circuit on a fixed surface reads off its gluing. A detour in cone strength compares two geometric exponents. Moving the endpoints can change the correlation itself, while varying the replica index asks for its derivative.}
\label{tab:operations}
\end{table}

\subsection{Replica Permutations}
\noindent
Multi-entropy specifies how to reconnect copies of $q$ parties, or subsystems~\cite{Gadde:2022cqi}. To obtain a number that depends on the reconnection, divide the joined path integral by the product of the same path integrals before gluing. The resulting normalized ratio is the replica invariant $\mathcal R_n^{(q)}$~\cite{Gadde:2022cqi}.
For two parties, this rule cycles $n$ copies and gives the ordinary R\'enyi construction~\cite{Gadde:2022cqi}. For $q$ parties it uses $N=n^{q-1}$ copies. Here $n$ labels the order of the measurement; it need not equal the number of copies, and one party's density matrix need not tell us the result.
We use the finite-index convention of Ref.~\cite{Harper:2024ker},
\begin{equation}
\mathcal R_n^{(q)}=\frac{Z_n^{(q)}}{Z_1^{\,n^{q-1}}},\qquad
S_n^{(q)}=\frac{\ln\mathcal R_n^{(q)}}{n^{q-2}(1-n)},\qquad
S^{(q)}=-\partial_n\ln\mathcal R_n^{(q)}\big|_{n=1}.
\label{eq:multidef}
\end{equation}
The finite-$n$ prefactor differs from the original convention in Ref.~\cite{Gadde:2022cqi}. Why do the two give the same entropy at one? At $n=1$ the gluing is trivial, so $\mathcal R_1^{(q)}=1$ and its logarithm vanishes. If the continuation is differentiable, the logarithm to first order is $(n-1)\,\partial_n\ln\mathcal R_n^{(q)}|_{n=1}$, while the extra power of $n$ tends to one. Dividing by $1-n$ leaves the minus sign in the last expression.
\\

\begin{figure}[htbp]
\centering
\includegraphics[width=.96\linewidth]{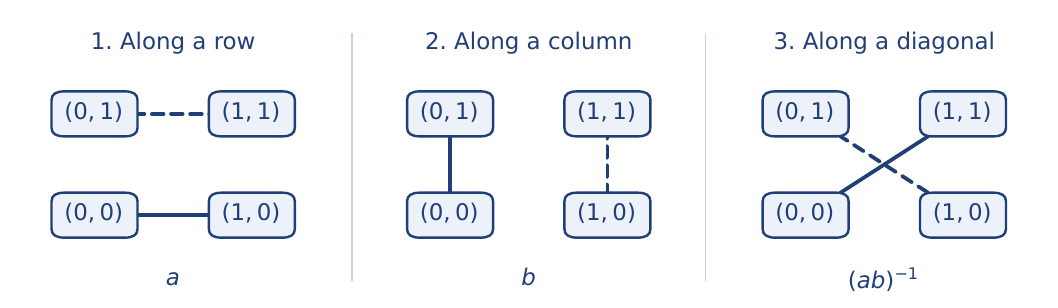}
\caption{The four sheets of the smallest three-party replica, $n=2$. Each panel shows the same sheets. A line joins the two sheet labels exchanged by a circuit about that endpoint; solid and dashed lines distinguish the two pairs. The crossing in the last panel is not a fifth sheet. Starting at $(0,0)$, loops about the first two endpoints already reach every label, so the assembled surface is connected.}
\label{fig:replicapairs}
\end{figure}

\noindent
For three parties, visualize the sheets as an $n$-by-$n$ grid. A sheet is labeled $(r,s)$ with $r,s=0,\ldots,n-1$. The labels form $\mathbb Z_n^2$: both coordinates wrap around after $n$ steps. Let $a(r,s)=(r+1,s)$ move one step along a row and $b(r,s)=(r,s+1)$ move one step along a column.
The triple
\begin{equation}
\sigma_1=a,\qquad\sigma_2=b,\qquad\sigma_3=(ab)^{-1}
\end{equation}
has product one: the last move undoes one row step and one column step. From any chosen sheet, row and column steps reach every other one, so the surface is connected. Each move returns to its starting label after $n$ steps and splits the $n^2$ sheets into $n$ cycles.
At $n=2$, Fig.~\ref{fig:replicapairs} lets us count the six exchanges: each of the three endpoints exchanges two pairs of sheets. These six units of ramification give a sphere for a cover of degree four. For larger $n$, the same counting rule applies.
For any integer $n\geq2$, the same count at all three endpoints gives
\begin{equation}
N=n^2,\qquad k_j=m_j=n,\qquad
g(n)=\frac{(n-1)(n-2)}2.
\label{eq:Fermatgenus}
\end{equation}
\par
\par\noindent
An algebraic realization is the compactified Fermat curve
\begin{equation}
x^n+y^n=1,\qquad z=x^n.
\label{eq:Fermat}
\end{equation}
At an ordinary $z$ there are $n$ choices of the root $x$ of $z$ and, independently, $n$ choices of the root $y$ of $1-z$. These are the $n^2$ sheets of the grid. At $0$, $1$ and $\infty$, the roots meet as prescribed.
Which genera can this three-party grid actually reach? Rearranging Eq.~\eqref{eq:Fermatgenus} gives $8g+1=(2n-3)^2$. Thus, a nonnegative integer $g$ occurs on the grid precisely when $8g+1$ is a perfect square, in which case $n=(3+\sqrt{8g+1})/2$ for $n\ge2$. The first four indices $n=2,3,4,5$ give $0,1,3,6$ handles. For $g=2$ the test gives $17$, which is not a square. The fifth partial-transpose moment in Sec.~\ref{sec:4} still reaches genus two because it joins the sheets differently. The first hyperbolic grid has sixteen sheets at $n=4$. These are counts at integer indices; interpolating the polynomial does not construct an intermediate surface.
The branch count can also be read directly from the curve's equation. Above $z=0$, $y^n=1$ has $n$ solutions, while $z=x^n$ has local degree $n$ at each one. At $z=1$ the roles of $x$ and $y$ are exchanged.
In projective coordinates $X^n+Y^n=W^n$, the points at infinity have $W=0$ and $(Y/X)^n=-1$, giving $n$ distinct points.
Near each, $W/X$ is a local coordinate, and $z=(X/W)^n$ has a pole of order $n$.
Thus, the total ramification is $3n(n-1)$, and we have
\begin{equation}
2g(n)-2=-2n^2+3n(n-1)=n(n-3),
\end{equation}
which reproduces Eq.~\eqref{eq:Fermatgenus}.

\subsection{Integer-Index Multi-Entropy}
\noindent
Let $x_{ij}=x_i-x_j$ denote an endpoint separation and let $\epsilon$ be the length at which we regulate each endpoint. Substituting $k_j=m_j=n$ in the twist weights gives $h_n=\bar h_n=c(n^2-1)/24$ and $\Delta_n=c(n^2-1)/12$. Conformal symmetry fixes all dependence on the three separations. In the unit two-point convention, the only undetermined number is $C_n$:
\begin{equation}
\mathcal R_n^{(3)}=C_n
\left(\frac{|x_{12}x_{23}x_{31}|}{\epsilon^3}\right)^{-\Delta_n}.
\label{eq:Rmulti}
\end{equation}
For $n\geq4$, $\beta_j=1/n$ and $d=(n-3)/(2n)$, so
\begin{align}
\Pcov(n)=\frac{cn^2}{6}\bigg[&3F\bigg(\frac{1}{n}\bigg)-3F\!\left(\frac{n-1}{2n}\right)
-\frac{3\ln n}{n}+F(0)\nonumber\\
&-F\!\left(\frac{n-3}{2n}\right)+\frac{n-3}{n}\bigg],\label{eq:Pmultin}\\
C_n&=e^{\Pcov(n)}\Z_\C(\Sigma_n),\qquad\Sigma_n:\ x^n+y^n=1.
\end{align}
This equality holds in either direction. Calculating the normalized seed partition function on the specified Fermat surface predicts $C_n$. Conversely, if the three-point measurement gives $C_n$, then $\Z_\C(\Sigma_n)=C_n\exp[-\Pcov(n)]$. Knowing the covering geometry alone supplies only the conversion factor between the two. Substituting the factorization into the entropy gives, at integer $n\ge4$,
\begin{equation}
S_n^{(3)}=\frac c{12}\left(1+\frac1n\right)
\ln\frac{|x_{12}x_{23}x_{31}|}{\epsilon^3}
-\frac{\Pcov(n)+\ln\Z_\C(\Sigma_n)}{n(n-1)}.
\label{eq:Snfull}
\end{equation}
\noindent The dependence on the cover can be seen even at fixed genus: the seven-sheet cyclic cover $w^7=z(z-1)$ and the sixteen-sheet grid $x^4+y^4=1$ both have genus three.
Eq.~\eqref{eq:mainintro} uses $(N,\beta_j,d)=(7,1/7,2/7)$ and $(16,1/4,1/8)$, respectively, for every $j=1,2,3$.
It gives $\Pcov/c\simeq-1.94376$ and $-3.89040$ for $c\ne0$. Yet Eq.~\eqref{eq:comparison} gives the same imaginary part, $\mp\pi c/3$, in both continued exponents. Counting handles fixes this angle; it does not fix the real geometric contribution.\par\smallskip\noindent
At $n=2$, $\eta=1/4$ and the spherical expression reduces to
\begin{equation}
\frac{\PCL^{\sph}(2)}{c}=\frac23\left[F(0)-4F\bigg(\frac{1}{4}\bigg)-\frac12\ln2\right].
\end{equation}
The duplication identity
\bea
\gamma(x)\gamma\bigg(x+\frac{1}{2}\bigg)=2^{1-4x}\gamma(2x),
\eea
gives $F(0)-4F(1/4)=-(\ln 2)/4$ after integrating over $0<x<1/4$ and using the symmetry of $F$ about $1/2$.
Thus, we get $C_2=2^{-c/2}$.
At $n=3$, the periods of the torus form a hexagonal lattice. Its shape, called the \emph{equianharmonic modulus}, can be written $\tau=\exp(i\pi/3)$. Other choices of period basis describe the same torus: $\tau$ and $(a\tau+b)/(r\tau+s)$ are equivalent for integers $a,b,r,s$ with $as-br=1$.
In this case, we have
\begin{equation}
C_3=3^{-3c/2}Z_\C[T^2,\tau=e^{i\pi/3}].
\label{eq:n3}
\end{equation}
The torus already shows where knowledge of $c$ stops: two theories can give different values of $Z_\C$ at the same torus shape.
The product formula for $C_n$ also makes sense when $C_n=0$, but its logarithm then does not exist. Away from zeros, we use one continuous branch throughout, so $\ln C_n=\Pcov(n)+\ln\Z_\C(\Sigma_n)$ on the hyperbolic covers.
Near one, the specified analytic continuation is nonzero in a small enough neighborhood because $C_1=1$.
A general multipartite contraction need not have the positivity guarantee that holds for $\mathrm{Tr}\rho_A^n$.
\\

\noindent
For a pure $ABC$ state, we define the endpoint-subtracted quantity \cite{Iizuka:2025ioc}
\begin{equation}
\GM_n^{(3)}=S_n^{(3)}-\frac12\big[S_n(A)+S_n(B)+S_n(C)\big].
\label{eq:GMdef}
\end{equation}
Why one half? Each endpoint borders two regions, so it occurs twice in the sum of the single-region entropies. The factor of one half counts its contribution once. The following computation checks that this reasoning cancels every distance and cutoff term with our normalization~\cite{Iizuka:2025ioc}.
We choose ordered endpoints $x_1<x_2<x_3$ on the compactified spatial line with $A=(x_1,x_2)$, $B=(x_2,x_3)$, and $C$ the complementary arc through infinity.
The vacuum of all three regions is pure. For a pure state shared between two complementary regions, the two reduced density matrices have the same nonzero eigenvalues. Its Schmidt decomposition pairs a basis state on one side with a basis state on the other; the squared amplitudes give the same probabilities on both sides. Taking the two sides to be $C$ and $AB$ gives $S_n(C)=S_n(AB)$. We can therefore use the finite interval from $x_1$ to $x_3$ for the arc passing through infinity. The three lengths in the subtraction are $|x_{12}|$, $|x_{23}|$, and $|x_{13}|$.
In the same unit two-point convention, each interval of length $L_X$ has
\begin{equation}
\mathrm{Tr}\rho_X^n=\left(\frac{L_X}{\epsilon}\right)^{-\frac c6(n-1/n)},
\qquad
S_n(X)=\frac c6\left(1+\frac1n\right)\ln\frac{L_X}{\epsilon},
\qquad X\in\{A,B,C\}.
\label{eq:intervalsubtraction}
\end{equation}
These expressions refer to the universal vacuum terms with matched endpoint constants; an additional regulator-dependent constant must be treated consistently in both replica constructions.
Multiplying the three interval moments and raising the result to $n/2$ yields
\begin{equation}
\left[\prod_{X\in\{A,B,C\}}\mathrm{Tr}\rho_X^n\right]^{n/2}
=\left(\frac{|x_{12}x_{23}x_{31}|}{\epsilon^3}\right)^{-c(n^2-1)/12}.
\label{eq:momentcancellation}
\end{equation}
The power on the right is $-\Delta_n$, precisely the power multiplying $C_n$ in Eq.~\eqref{eq:Rmulti}. Dividing one replica expression by the other removes all three endpoint separations and the cutoff at once.
One can check the cancellation by doubling all three endpoint separations while holding the short-distance cutoff fixed. At $n=2$ the product of distances grows by a factor of eight, so the three-party entropy gains $(3c/8)\ln2$. Each single interval gains $(c/4)\ln2$. Half the sum of the three gains is again $(3c/8)\ln2$, and the subtracted quantity stays fixed. Eq.~\eqref{eq:momentcancellation} proves the statement for arbitrary separations at each integer $n\ge2$.
Thus, with these matched endpoint normalizations, we get \cite{Iizuka:2025ioc}
\begin{equation}
\GM_n^{(3)}=\frac{\ln C_n}{n(1-n)}.
\label{eq:GMcoefficient}
\end{equation}
\\

{
\noindent Two low indices show what survives the subtraction. Inserting $C_2=2^{-c/2}$ and Eq.~\eqref{eq:n3} gives
\[
\GM_2^{(3)}=\frac c4\ln 2,\qquad
\GM_3^{(3)}=\frac c4\ln 3-\frac16\ln Z_\C[T^2,\tau=e^{i\pi/3}].
\]
The first surface is a sphere, whose partition function cancels in our normalization. The next is a torus, whose partition function counts states with weights set by its shape. Thus two theories with the same $c$ agree on the endpoint-subtracted answer at $n=2$. At $n=3$ one must also compare their partition functions on the same torus. This difference is visible at integer replica index, before taking any limit.
}
\\

\noindent
For a general state, the invariant replacing $C_n$ is
\begin{equation}
\mathcal K_n=
\frac{\mathcal R_n^{(3)}}{\big[\mathrm{Tr}\rho_A^n\,\mathrm{Tr}\rho_B^n\,\mathrm{Tr}\rho_C^n\big]^{n/2}},
\qquad \GM_n^{(3)}=\frac{\ln\mathcal K_n}{n(1-n)}.
\label{eq:Kgeneral}
\end{equation}
The numerator joins all three parties; the denominator uses only their separate moments. Sec.~\ref{sec:7} compares two states whose denominators agree at every index even though their numerators differ at $n=3$.
\\

\noindent
To see what an observer can read from the numerator, consider a regulated pure state with density matrix $\rho_{ABC}$. Prepare $n^2$ identical copies and put $\rho_*=\rho_{ABC}^{\otimes n^2}$, for an integer $n\ge2$. Let $U_X(p)$ permute only the copies of party $X$ according to a sheet permutation $p$. Using the row shift $a$ and column shift $b$ defined above, the three-party grid pairs $A$ with the same copy, $B$ with copy $a$, and $C$ with copy $ab$. The relative moves at the endpoints are therefore $a$, $b$, and $(ab)^{-1}$. Contracting each ket with its paired bra is precisely the measurement
\[
U_n=U_A(1)\,U_B(a)\,U_C(ab),\qquad
\mathcal R_n^{(3)}=\mathrm{Tr}(\rho_* U_n).
\]
Here $1$ denotes the identity permutation. The expectation of $(U_n+U_n^\dagger)/2$ gives the real part of this number; the expectation of $(U_n-U_n^\dagger)/(2\mathrm i)$ gives its imaginary part. Both operators are Hermitian. The dagger denotes the adjoint, which reverses a permutation.
\par\smallskip\noindent
The imaginary-part measurement has a definite prediction. Relabel \emph{every} copy, for all three parties at once, by $(r,s)\mapsto(-r,-s)$ modulo $n$, and call this relabeling $J$. The prepared state is unchanged because all copies are identical. Every row or column step becomes a step in the opposite direction; hence $J U_n J^{-1}=U_n^\dagger$. Invariance of the trace under relabeling gives
\[
\mathrm{Tr}(\rho_* U_n)=\mathrm{Tr}(\rho_* U_n^\dagger)
=\overline{\mathrm{Tr}(\rho_* U_n)} .
\]
Thus, the integer grid amplitude is real even if the state amplitudes in the chosen basis are complex. The denominator in Eq.~\eqref{eq:Kgeneral} is positive since each reduced density matrix has nonnegative eigenvalues summing to one, so the ratio $\mathcal K_n$ is real as well. This argument establishes reality, not strict positivity.
\\

\noindent
Reality at an integer index is a statement about the original state and measurement. Along a nonzero continuation of the \emph{physical} ratio at fixed $n\ge2$ its logarithm instead gives $\Ima\GM_n^{(3)}=\operatorname{Arg}_{\mathrm{cont}}\mathcal K_n/[n(1-n)]$, with the starting logarithm specified. Changing just a reference metric keeps this ratio fixed by Eq.~\eqref{eq:phasecancel}. We can test a change of physical data by following the endpoints themselves.
\\

\noindent
Consider the three distinct vacuum endpoints $x_1<x_2<x_3$ used above. We keep the positive cutoff $\epsilon$ fixed and continue the two coordinates $z$ and $\bar z$ independently. Write $z_j=x_j$ and $\bar z_j=\exp(\mathrm i\theta)x_j$ for $j=1,2,3$, starting at $\theta=0$ and ending at $\theta=+\pi$ or $-\pi$. The bar labels the second coordinate; only at the starting Euclidean configuration is it required to be the complex conjugate of the first. Every squared separation becomes $(z_i-z_j)(\bar z_i-\bar z_j)=\exp(\mathrm i\theta)x_{ij}^2$ for $1\le i<j\le3$. None passes through zero. At either endpoint of the path the Lorentzian coordinates $z=x_{\mathrm L}+t_{\mathrm L}$ and $\bar z=x_{\mathrm L}-t_{\mathrm L}$ put all three events at $x_{\mathrm L}=0$, at the distinct times $t_{\mathrm L}=x_j$.
\par\smallskip\noindent
Take the square root $L_{ij}$ of each squared separation continuously from $L_{ij}(0)=|x_{ij}|$. The logarithm of $L_{ij}/\epsilon$ changes by $\mathrm i\theta/2$. Define $\delta_\theta Y=Y(\theta)-Y(0)$ and follow all logarithms from the same initial choices. The single-region entropy in Eq.~\eqref{eq:intervalsubtraction} has one length logarithm; the three-party entropy obtained from Eq.~\eqref{eq:Rmulti} has three, each with half that coefficient. The three-point constant $C_n$ is position independent. For nonzero amplitudes and every integer $n\ge2$ we therefore get
\[
\begin{aligned}
\delta_\theta S_n(X)&=\frac{\mathrm i c\theta}{12}\left(1+\frac1n\right)
\quad(X=A,B,C),\qquad
\delta_\theta S_n^{(3)}=\frac{\mathrm i c\theta}{8}\left(1+\frac1n\right),\\
\delta_\theta\GM_n^{(3)}&=\delta_\theta S_n^{(3)}
-\frac12\sum_{X\in\{A,B,C\}}\delta_\theta S_n(X)=0 .
\end{aligned}
\]
For example, at $n=2$ and $\theta=\pi$, each interval gains $\mathrm i\pi c/8$. The three-party quantity gains $3\mathrm i\pi c/16$, precisely half the sum of the three interval gains. In the usual $n\to1$ interval formula, the upper path gives $\mathrm i\pi c/6$; the lower path gives the opposite sign. This calculation continues a vacuum correlator. It does not treat a segment of time as a subsystem with an ordinary density matrix.
\\

\noindent
The same path also tests an entanglement quantity on a cover at \emph{every} positive genus. In the odd partial-transpose moment $M_m=\mathrm{Tr}[(\rho_{A_1A_2}^{T_2})^m]$ of Sec.~\ref{sec:4}, $m=2g+1$ and each of the three twists is a single $m$-cycle. Its dimension from Eq.~\eqref{eq:weights} is $\Delta_m^{\mathrm{cyc}}=c(m-1/m)/12$. Eq.~\eqref{eq:K} gives one power $-\Delta_m^{\mathrm{cyc}}$ of each separation in $M_m$. The ordinary interval moment $I_{ij}^{(m)}=\mathrm{Tr}\rho_{[x_i,x_j]}^m$ has power $-2\Delta_m^{\mathrm{cyc}}$ of the same separation. These moment symbols denote the continuations of their spatial values when $\theta\ne0$. With matched endpoint normalizations, define
\[
Q_m=\frac{M_m}{\left[\prod_{1\le i<j\le3} I_{ij}^{(m)}\right]^{1/2}},
\qquad
\delta_\theta\ln M_m=-\frac{3\mathrm i\theta}{2}\Delta_m^{\mathrm{cyc}},
\qquad
\delta_\theta\ln Q_m=0 .
\]
The square root follows the three interval logarithms continuously. Its logarithmic change is exactly the one in the numerator. If we express this ratio in entropy units, $\mathcal G_m=\ln_{\mathrm{cont}}Q_m/(1-m)$, it follows that $\delta_\theta\Ima\mathcal G_m=0$ at every $g\ge1$. For $m=5$, the cover has genus two: the normalized logarithm acquires no phase, whereas the geometric cone exponent has imaginary part $\mp\pi c/6$. The zero is a \emph{change} from a specified starting value. The surviving constant contains the seed partition function, whose starting phase we have not assumed to vanish.

\subsection{Replica Limit}
\noindent
For each integer $n\ge2$ the gluing gives a particular coefficient $C_n$. The entropy depends on a different quantity: the slope of $\ln C_n$ at one. Giving the sheet count $N=n^2$ and the corresponding genus adds no measurements between the integers. A simple example shows why even all integer coefficients together leave this slope undetermined~\cite{Gadde:2022cqi,Czech:2026jil}. For any real constant $A$, make the replacement
\begin{equation}
\ln C_n\longmapsto\ln C_n+A\sin[\pi(n-1)]
\end{equation}
At every integer, the added sine is zero, so the replacement preserves all integer coefficients, including $C_1=1$. At one, the added term has derivative $A\pi$, shifting the slope of $\ln C_n$ by that amount. To infer the entropy, we need a physical prescription for interpolating the coefficients or bounds strong enough to fix the derivative. No surface with a fractional number of handles is used here.
\\

\noindent
Near $n=1$, the real spherical branch of the geometric function is well-defined.
We set $\eta=(1-1/n)/2$.
Eq.~\eqref{eq:CL} becomes
\begin{align}
\PCL^{\sph}(n)=\frac{cn^2}{6}\big[&3F(2\eta)-3F(\eta)
+3(1-2\eta)\ln(1-2\eta)\nonumber\\
&+F(0)-F(3\eta)-2(1-3\eta)\ln(1-3\eta)\big].
\label{eq:Pspherical}
\end{align}
Although the derivative of $F$ is singular at the individual arguments that approach zero, the combination entering the entropy has a finite slope. The needed endpoint expansion is
\begin{equation}
F(x)=F(0)-x\ln x+x+{\mathcal O}(x^2).
\end{equation}
In $3F(2\eta)-3F(\eta)-F(3\eta)+F(0)$ the terms proportional to $\eta\ln\eta$ cancel: their coefficients are $-6$, $+3$ and $+3$. The remaining logarithms of $2$ and $3$ give $3\eta\ln(3/4)$. The two explicit logarithms in Eq.~\eqref{eq:Pspherical} give $-6\eta$ and $+6\eta$ at this order, so they also cancel. Since $\eta=(n-1)/2+{\mathcal O}((n-1)^2)$, we obtain
\begin{equation}
\PCL^{\sph}(n)=-\frac c2(n-1)\ln\frac2{\sqrt3}
+{\mathcal O}(c(n-1)^2).
\label{eq:Pderivative}
\end{equation}
This fixes the tangent of the geometric term at one. To ask whether it is the tangent of the full answer, we need a specified differentiable function $C_\C(n)$ that agrees with the normalized replica coefficients at the integers and has $C_\C(1)=1$. Choosing the logarithm that vanishes at one, we subtract the known geometric function and call the remainder
\begin{equation}
H_\C(n)\equiv\ln C_\C(n)-\PCL^{\sph}(n),\qquad H_\C(1)=0.
\end{equation}
At one, the numerator and denominator of Eq.~\eqref{eq:GMcoefficient} both vanish. The denominator $n(1-n)$ has derivative $-1$ there, so the limit is minus the derivative of $\ln C_\C(n)$. Adding the three interval terms gives
\begin{align}
S^{(3)}&=\frac c6\ln\frac{|x_{12}x_{23}x_{31}|}{\epsilon^3}
+\frac c2\ln\frac2{\sqrt3}-H_\C'(1),\label{eq:Slimit}\\
\GM^{(3)}&=\frac c2\ln\frac2{\sqrt3}-H_\C'(1).
\label{eq:GMlimit}
\end{align}
This separates two tests of the gravity prediction~\cite{Gadde:2022cqi,Harper:2024ker}. The geometric calculation gives the coefficient $(1/2)\ln(2/\sqrt3)$ multiplying $c$. For the full CFT entropy to have that same coefficient, the omitted slope must satisfy $H_\C'(1)/c\to0$ along the family of theories. The slope itself need not vanish: a finite correction contributes to the entropy but disappears after division by $c$.
\\

\noindent
On the hyperbolic branches at integer $n\geq4$,
\begin{equation}
\Ima\PCL^{(\pm)}(n)=\mp\frac{\pi c}{12}n(n-3).
\end{equation}
This formula cannot simply be evaluated at $n=1$ to infer an imaginary multi-entropy.
For example, $n=4$ gives $g=3$ and phase angle $\mp\pi c/3$ in the defect amplitude. To reach $n=1$ along real $n$, one first encounters the flat point $n=3$. Near one, the geometric derivative is calculated on the real spherical branch, Eq.~\eqref{eq:Pspherical}; the polynomial for the hyperbolic phase does not give that derivative.

\subsection{Higher Orders of Geometric Replica Expansion}
\noindent
The limiting entropy gives the value at one. Its next two terms tell us how the geometric answer changes if the replica index moves away from one. Since $n(1-n)$ has a simple zero there, we need the geometric exponent through cubic order to obtain the entropy through quadratic order.
We put $\xi=n-1$ and $\lambda_*=\ln(2/\sqrt3)$.
The convergent gamma-function expansion near the origin is
\begin{equation}
\ln\Gamma(1+x)=-\gamma_E x+
\sum_{m=2}^{\infty}\frac{(-1)^m\zeta(m)}{m}x^m,\qquad |x|<1,
\end{equation}
where $\gamma_E$ is Euler's constant and $\zeta(m)$ is the Riemann zeta function.
For completeness, the canonical product representation
\begin{equation}
\frac1{\Gamma(1+x)}=e^{\gamma_E x}
\prod_{r=1}^{\infty}\left(1+\frac{x}{r}\right)e^{-x/r}
\label{eq:gammaproduct}
\end{equation}
gives this series by taking its logarithm, expanding each $\ln(1+x/r)$, and summing $\sum_{r=1}^{\infty}r^{-m}=\zeta(m)$ for $m\geq2$.
The resulting double series is absolutely convergent for $|x|<1$, which justifies interchanging the sums.
Using
\bea
\Gamma(x)=\frac{\Gamma(1+x)}{x}
\eea
and subtracting the expansion at $-x$ gives
\begin{align}
\ln\gamma(x)&=-\ln x-2\gamma_E x-\frac23\zeta(3)x^3+{\mathcal O}(x^5),\\
F(x)&=F(0)-x\ln x+x-\gamma_E x^2-\frac16\zeta(3)x^4+{\mathcal O}(x^6).
\label{eq:Fseries}
\end{align}
The real expressions are first evaluated for $x>0$.
In the combination relevant to Eq.~\eqref{eq:Pspherical}, the logarithms of $\eta$ cancel, so the resulting local geometric function has a regular Taylor series.
In detail, we obtain
\begin{align}
3F(2\eta)-3F(\eta)+F(0)-F(3\eta)
&=3\eta\ln\bigg(\frac{3}{4}\bigg)+6\zeta(3)\eta^4+{\mathcal O}(\eta^6),\\
3(1-2\eta)\ln(1-2\eta)-2(1-3\eta)\ln(1-3\eta)
&=-3\eta^2-5\eta^3+{\mathcal O}(\eta^4).
\label{eq:seriescancellations}
\end{align}
In the first line, the coefficient of $\eta^2$ is proportional to $-3\cdot2^2+3+3^2=0$, and $ F$ has no cubic term.
The linear term is $-6\eta\ln2+3\eta\ln3=3\eta\ln(3/4)$. In the second line, the two linear terms cancel; expanding $(1-a\eta)\ln(1-a\eta)$ fixes the displayed quadratic and cubic coefficients.
\\

\noindent
Due to $\eta=\xi/[2(1+\xi)]$, multiplying these expansions by $n^2/6=(1+\xi)^2/6$ yields
\begin{equation}
\frac{\PCL^{\sph}(1+\xi)}{c}
=-\frac{\lambda_*}{2}\xi-\left(\frac{\lambda_*}{2}+\frac18\right)\xi^2
-\frac5{48}\xi^3+{\mathcal O}(\xi^4).
\label{eq:Pcubic}
\end{equation}
We define the geometric contribution to the endpoint-subtracted measure by
\begin{equation}
\GM_{n,\mathrm{geom}}^{(3)}=\frac{\PCL^{\sph}(n)}{n(1-n)}.
\end{equation}
Its next two terms are therefore
\begin{equation}
\boxed{\displaystyle
\GM_{1+\xi,\mathrm{geom}}^{(3)}
=c\left[\frac12\ln\frac2{\sqrt3}+\frac\xi8-\frac{\xi^2}{48}
+{\mathcal O}(\xi^3)\right].}
\label{eq:geomhigher}
\end{equation}
These coefficients follow from the normalized geometric function, not from an assumption about a holographic dual.
They do not determine the higher derivatives of the full CFT measure.
If $H_\C$ is twice differentiable at one, expansion of its prefactor gives
\begin{equation}
\left.\partial_n\GM_n^{(3)}\right|_{n=1}
=\frac c8+H_\C'(1)-\frac12H_\C''(1).
\label{eq:fullslope}
\end{equation}
The constant term of the entropy needs one derivative of the unknown $H_\C$; its slope needs two. A test of the geometric slope $c/8$ is therefore a stronger test of a theory than a match of the constant alone. The multipartite slope defined here has no assumed positivity property.

\subsection{Sufficient Controlled Bound}
\noindent
The geometric term grows in proportion to $c$. A theory-dependent contribution could be finite and survive in the entropy without changing this leading coefficient. The relevant test is therefore whether $H_\C'(1)/c$ tends to zero as $c$ grows. Knowing $H_\C(1)=0$ tells us only where the graph starts, not how steep it is there. We write $H_c$ for the remainder along a chosen family of theories; its value need not be the same in different theories with the same $c$.
One sufficient test uses a circle centered at $n=1$ in the complex $n$-plane. Suppose $H_c$ is holomorphic throughout a neighborhood of the closed disk $|n-1|\leq r$, the radius $r>0$ is independent of $c$, and
\begin{equation}
\sup_{|n-1|=r}|H_c(n)|\le M_c,\qquad M_c\ge0,\qquad
\lim_{c\to\infty}\frac{M_c}{c}=0.
\label{eq:uniformbound}
\end{equation}
The supremum is the largest magnitude of $H_c$ on the circle. The condition asks that this magnitude divided by $c$ tend to zero. A bound of order $c$ alone would still allow a correction as large as the geometric term. We also need the continued coefficient to stay nonzero throughout the disk so that its logarithm can be chosen continuously from $C_\C(1)=1$. Values at integer replica indices do not guarantee either condition.
\\

\noindent
Cauchy's integral formula reconstructs derivatives of a holomorphic function from its values on a surrounding closed contour.
With the circle traversed counterclockwise, it implies
\begin{equation}
H_c'(1)=\frac1{2\pi i}\oint_{|n-1|=r}\frac{H_c(n)}{(n-1)^2}\mathrm d n,
\qquad \frac{|H_c'(1)|}{c}\le\frac{M_c}{cr}\longrightarrow0.
\label{eq:Cauchybound}
\end{equation}
The circumference of the integration path is $2\pi r$, and the denominator has magnitude $r^2$ everywhere on it, giving the factor $1/r$. For a fixed radius, a bound on $H_c$ throughout the circle therefore controls the slope at its center. If $M_c$ stays of order one, the slope also stays bounded; if $M_c\to0$, it tends to zero even without dividing by $c$.
For example, the entire test function
\begin{equation}
H_c^{\mathrm{test}}(n)=c(n-1)e^{-c(n-1)}
\end{equation}
is holomorphic for all complex $n$, vanishes at one, and satisfies
\begin{equation}
\lim_{c\to\infty}\frac{H_c^{\mathrm{test}}(n)}c=0
\quad\text{for every fixed real }n>1,
\qquad
\frac{(H_c^{\mathrm{test}})'(1)}c=1.
\label{eq:nonuniformtest}
\end{equation}
For this example the ratio $H_c^{\mathrm{test}}(n)/[c(n-1)]$ is $\exp[-c(n-1)]$. At any fixed real $n>1$ it tends to zero as $c$ grows; at any fixed $c$ it tends to one as $n$ approaches one. The two limits give different answers.
On the circle, the point $n=1-r$ gives $\sup_{|n-1|=r}|H_c^{\mathrm{test}}(n)|=cr\exp(cr)$, violating Eq.~\eqref{eq:uniformbound}. The example is an ordinary analytic function, not a proposed CFT. It shows why estimates at separate replica indices do not establish a bound on the entropy derivative.
More generally, for each fixed positive integer $k$, the same Cauchy's formula gives
\begin{equation}
|H_c^{(k)}(1)|\le\frac{k!M_c}{r^k},\qquad
\frac{|H_c^{(k)}(1)|}{c}\longrightarrow0.
\label{eq:higherCauchy}
\end{equation}
Combining the $k=1,2$ cases with Eq.~\eqref{eq:fullslope}, we obtain the quantitative slope estimate
\begin{equation}
\left|\frac1c\left.\partial_n\GM_n^{(3)}\right|_{n=1}-\frac18\right|
\le\frac{M_c}{c}\left(\frac1r+\frac1{r^2}\right)\longrightarrow0.
\label{bound3}
\end{equation}
Under the same hypothesis, both the constant and the first slope of the entropy divided by $c$ agree with their geometric values. The estimate controls derivatives, which is stronger than agreement of values at individual replica numbers.
If the circle shrinks as $c$ grows, the same estimate is $k!M_c/r_c^k$ for a circle of radius $r_c$. One must then show $M_c/(cr_c^k)\to0$: agreement in an ever smaller neighborhood is weaker evidence about a slope.
\\

\noindent
One way to seek the missing bound is to approximate the path integral by a classical stationary configuration, or saddle, and then estimate the fluctuations around it. Small Gaussian fluctuations give a determinant. For Eq.~\eqref{eq:uniformbound}, a successful estimate would have to control the other saddles and the fluctuation factors throughout the stated neighborhood in the complex $n$ plane. It must also allow a logarithm to be chosen there: a zero of the amplitude would obstruct this step. Such an estimate has not been established here for the Fermat replica family.
A sparse spectrum means that relatively few states have low scaling dimension; how few are needed depends on the quantity being estimated. For example, vacuum dominance on the torus would give information about the grid at $n=3$. It says nothing on its own about the slope at one, and higher-genus partition functions require additional CFT data. Eqs.~\eqref{eq:Cauchybound},~\eqref{eq:higherCauchy}, and~\eqref{bound3} state what follows \emph{if} the bound holds. They do not establish it for the Fermat family.

\subsection{More Parties}
\noindent
In the three-party grid, two cyclic coordinates tell us which copy we occupy. With four parties, we need three such coordinates, as in a stack of square grids whose top and bottom are identified. In general, for $q\ge2$ adjacent regions on a circle and integer $n\ge2$, a sheet is labeled by $\mathbf r=(r_1,r_2,\ldots,r_{q-1})\in\mathbb Z_n^{q-1}$, with $r_j=0,\ldots,n-1$ for $j=1,\ldots,q-1$. Counting cycles of these labels gives the genus before any lengths have been assigned.
Let $\mathbf e_j$ have one in component $j$ and zero in every other component.
We choose the branch permutations
\begin{equation}
\sigma_j(\mathbf r)=\mathbf r+\mathbf e_j\quad(j=1,2,\ldots,q-1),
\qquad
\sigma_q(\mathbf r)=\mathbf r-\sum_{j=1}^{q-1}\mathbf e_j.
\label{eq:qmonodromy}
\end{equation}
All additions are modulo $n$.
The permutations commute, and their product is the identity.
The first $q-1$ translations reach any sheet from any other, proving transitivity and hence connectedness.
For each permutation, an orbit returns to its initial sheet after exactly $n$ steps: a nonzero step number smaller than $n$ cannot vanish in the component shifted by one.
There are consequently $N=n^{q-1}$ sheets and $N/n=n^{q-2}$ cycles at each of the $q$ branch values.
The Riemann--Hurwitz formula gives
\begin{equation}
2g(n,q)-2=-2n^{q-1}+\sum_{j=1}^{q}n^{q-2}(n-1)
=-2n^{q-1}+q n^{q-2}(n-1),
\end{equation}
or
\begin{equation}
g(n,q)=1-n^{q-1}+\frac q2n^{q-2}(n-1).
\label{eq:qgenus}
\end{equation}
For two parties, the formula gives a sphere for every $n$; for three, it gives the Fermat genus in Eq.~\eqref{eq:Fermatgenus}. Even at fixed $n=2$, a fourth party gives eight sheets and a torus, while a fifth gives sixteen sheets and a surface with five handles. Thus, the number of parties changes the topology without changing the length of any individual cycle.
In the hyperbolic cases, we have
\begin{equation}
d=\frac{g(n,q)-1}{n^{q-1}}=\frac q2\left(1-\frac1n\right)-1,
\qquad \sum_{j=1}^{q}\eta_j=1+d,
\qquad \eta_j=\frac12\left(1-\frac1n\right).
\end{equation}
For $q>3$, the branch values have cross ratios, and the projective connection has accessory parameters. Eqs.~\eqref{eq:manyfactor}--\eqref{eq:manyP} supply the normalized geometric factor with $\nu=q$ and $\beta_j=1/n$ for $j=1,2,\ldots,q$ whenever $g(n,q)>1$.
Counting sheets therefore answers one question for any number of parties: how many handles the replicas produce. Two further questions remain. What metric corresponds to the chosen endpoint positions? What number does the seed theory assign to that surface? With four or more endpoints, changing a cross ratio can change both answers while leaving the handle count fixed.


\section{Large-Central-Charge Example}
\label{sec:7}
\noindent
Can three observers change the entropy they measure without changing the central charge? Give them a label that all three can read. In one preparation its value is certain. In another, several values occur in a coherent superposition. Whenever one observer reads a value, the other two read the same one. We calculate the difference, compare it with the result for information shared separately by pairs, and then increase the number of available labels.
The label theory $\mathcal T_D$ is a direct sum of $D$ trivial two-dimensional topological quantum field theories (TQFTs). It has zero central charge and no propagating fields. On a connected surface, there is one choice to make: which label occurs everywhere. Every choice has weight one, giving
\begin{equation}
Z_{\mathcal T_D}[\Sigma_g]=\sum_{\alpha=1}^{D}1=D,
\qquad s_{\mathcal T_D}=D,
\qquad \mathcal Z_{\mathcal T_D}(\Sigma_g)=D^g.
\label{tqftpf}
\end{equation}
Only one label is chosen on a connected surface, regardless of its number of handles. Its partition function is therefore $D$ at every genus. The factor $D^g$ in the last expression has a different origin: we multiplied by $s_{\mathcal T_D}^{\,g-1}=D^{g-1}$ to use the normalization of Eq.~\eqref{eq:Zhatintro}. It does not assign an independent label to each handle. There are $D$ vacuum states on a circle even though the local central charge is zero. Writing the sphere factor as a partition function keeps track of this distinction; an expression $Q^c$ with finite nonzero $Q$ would miss it.
\\

\noindent
We now combine the labels with an ordinary seed theory. In a tensor product, the two systems are independent: their Hilbert spaces form a product, their partition functions multiply, and their central charges add. Taking the logarithm turns the product of their replica invariants into a sum. Additivity at the entropy limit requires that the chosen continuation also preserve the product.
We fix a seed $\C_0$ with $c_0>0$ and a differentiable replica prescription such that
\begin{equation}
a(c_0)=H_{\C_0}'(1).
\end{equation}
We assume this derivative is finite; its numerical value is not needed for the example.
We then choose the tensor products and define
\begin{equation}
\C_M\equiv\C_0^{\otimes M}\otimes\mathcal T_{D_M},\qquad c_M\equiv Mc_0.
\label{tensorproduct}
\end{equation}
At integer replica number, the bare path integrals and the two-point normalizations factorize.
Because the geometric exponent is linear in the central charge, the topological factor has zero geometric exponent.
Equivalently, this is a single reducible conformal theory:
\begin{equation}
\C_0^{\otimes M}\otimes\mathcal T_{D_M}
\cong\bigoplus_{\alpha=1}^{D_M}\C_0^{\otimes M}.
\end{equation}
The distinction between a product and a sum is physical here. The $M$ seed copies are present simultaneously, so their stress tensors add and give $Mc_0$. The $D_M$ blocks are alternatives, not additional fields present at the same point. The stress tensor is block diagonal; every block has the same central charge $Mc_0$ regardless of how many blocks are available.
{
To state which entanglement observable is being counted, take the equal-weight copying state on the three extended party spaces,
\[
|\psi_D\rangle=\frac1{\sqrt D}\sum_{\alpha=1}^{D}
|\alpha\rangle_A|\alpha\rangle_B|\alpha\rangle_C .
\]
The labels are orthogonal. If party $A$ measures the label, each value has probability $1/D$, and $B$ and $C$ will find that same value. Tracing out any two parties therefore gives $\rho_X=\mathbf 1_D/D$ for the remaining party $X\in\{A,B,C\}$. Its $D$ equal eigenvalues give $\mathrm{Tr}\,\rho_X^n=D(1/D)^n=D^{1-n}$. Each of the $n^2$ copies supplies a state and its conjugate, giving $(1/\sqrt D)^{2n^2}=D^{-n^2}$ in total. Along every edge of the grid, the labels must agree. Because any node can be reached from any other, choosing the label at one node fixes all the rest. Summing the $D$ possible choices gives $\mathcal R_n^{(3)}=D^{1-n^2}$. Substitution in Eq.~\eqref{eq:Kgeneral} gives
\[
\mathcal K_n=\frac{D^{1-n^2}}{D^{(3n/2)(1-n)}}
=D^{(n-1)(n-2)/2}=D^{g(n)} .
\]
At integer $n$, this independently reproduces the sphere-normalized surface answer in Eq.~\eqref{tqftpf}. For this example we continue the displayed powers as functions of $n$ and factorize the continuation for independent tensor factors. This is a prescription; agreeing at every integer would not make it unique. With this choice the three-party entropy of the labels is $(1+1/n)\ln D$, directly from $\mathcal R_n^{(3)}=D^{1-n^2}$. Each party alone has entropy $\ln D$. Subtracting half their sum leaves $(1/n-1/2)\ln D$. For two possible labels the difference is zero at $n=2$, is $-(\ln2)/6$ at $n=3$, and tends to $(\ln2)/2$ as $n\to1$ with our prescription. A negative value here means that half the sum of the three single-party entropies exceeds the three-party entropy at that index. All of these numbers are real. They come from counting labels in a sector with no Weyl phase. Combining it with the seed gives
}
\begin{align}
C_{\C_M}(n)&=C_{\C_0}(n)^M D_M^{g(n)},\\
H_{\C_M}(n)&=M H_{\C_0}(n)+g(n)\ln D_M,\\
H_{\C_M}'(1)&=Ma(c_0)-\frac12\ln D_M.
\label{eq:Hcounter}
\end{align}
At integer $n$, the first line follows from counting labels. Continuing the explicit powers and differentiating gives the next two lines, since $g'(1)=-1/2$. For comparison, prepare a definite label instead. Then each party knows it in advance, all the normalized label moments are one, and the contribution to the subtracted entropy is zero.
\\

\noindent
We can see which property of the state changes this answer without changing the list of labels. Let the parties share the pure state $\sum_{\alpha=1}^{D}\sqrt{p_\alpha}\,|\alpha\rangle_A|\alpha\rangle_B|\alpha\rangle_C$, where $p_\alpha>0$ and $\sum_{\alpha=1}^{D}p_\alpha=1$; unused labels can be omitted. Define $m(t)=\sum_{\alpha=1}^{D}p_\alpha^t$. A single party has moment $m(n)$. On the connected grid, all copies carry the same label, and each copy and its conjugate together supply one factor $p_\alpha$. Summing over the one free label gives
\[
\mathcal R_n^{(3)}=m(n^2),\qquad
\mathcal K_n=\frac{m(n^2)}{m(n)^{3n/2}} .
\]
These are moments of the specified state, not a change to the closed-surface partition function in Eq.~\eqref{tqftpf}. For this finite example, we continue the displayed sums to real $n$ near one. Since $m(1)=1$, the derivative of $\ln m(n^2)$ at one is $2m'(1)$, while that of $(3n/2)\ln m(n)$ is $3m'(1)/2$. With the minus sign from the entropy limit, their difference gives
\[
\GM^{(3)}[\boldsymbol p]=-\tfrac12 m'(1)
=-\frac12\sum_{\alpha=1}^{D}p_\alpha\ln p_\alpha .
\]
The right-hand side is half the Shannon entropy of the outcome that all three parties read. With two possible labels it gives $(\ln 2)/2$ for equal probabilities, but $(\ln 4-\tfrac34\ln 3)/2$ for probabilities $3/4$ and $1/4$. More generally it gives $(\ln D)/2$ for $D$ equally likely labels, and zero when one label is certain. Thus, keeping the same available labels and changing only their probabilities can change the entropy.
\\

\begin{figure}[htbp]
\centering
\includegraphics[width=\linewidth]{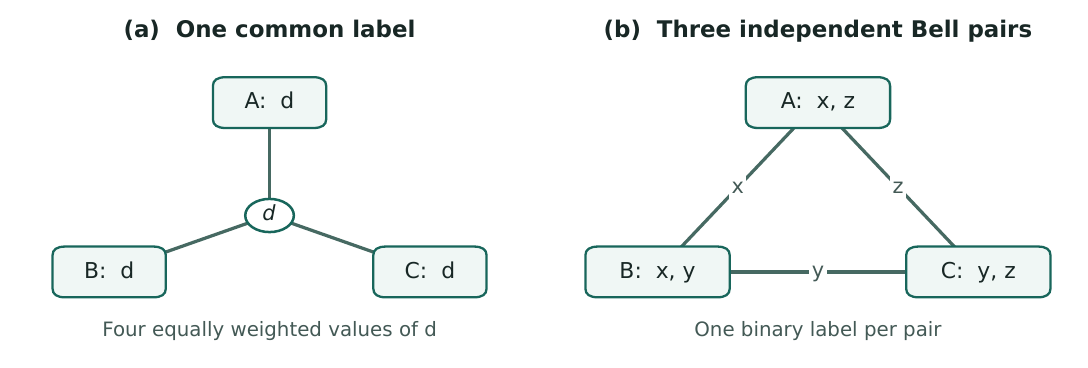}
\caption{Two pure states with the same statistics in every individual laboratory. At left, the three parties share one of four labels in a coherent superposition; the central symbol denotes the common index, not a fourth party. At right, each edge carries an independent Bell pair; a party holds the two qubits on its incident edges. The connections describe which basis labels agree in the amplitudes. The three-party replica contraction distinguishes these patterns at $n=3$.}
\label{fig:partystates}
\end{figure}

\noindent
Can one tell that a label is shared by all three parties using only measurements in the three separate laboratories? Consider two pure states, with four basis states per party in both cases (Fig.~\ref{fig:partystates}):
\[
|\psi_{\mathrm{common}}\rangle=\frac12\sum_{d=0}^{3}|d\rangle_A|d\rangle_B|d\rangle_C,\qquad
|\psi_{\mathrm{pairs}}\rangle=\frac1{\sqrt8}
\sum_{x=0}^{1}\sum_{y=0}^{1}\sum_{z=0}^{1}
|x,z\rangle_A|x,y\rangle_B|y,z\rangle_C .
\]
The first state is the $D=4$ copying state. The second is a product of three Bell pairs, one on each of $AB$, $BC$, and $CA$. A Bell pair here is $(|00\rangle+|11\rangle)/\sqrt2$. The three binary labels $x,y,z$ describe its three copies, not spatial coordinates. In the first state, the other two parties distinguish different values of $d$, so tracing them out gives $\mathbf 1_4/4$. In the second, each observer holds one qubit from each of two independent Bell pairs. Each local qubit has density matrix $\mathbf 1_2/2$, whose tensor square is again $\mathbf 1_4/4$. Thus \emph{every} local measurement, in any basis, has the same distribution for both states.
\\

\noindent
The grid distinguishes the two ways to share information. For one Bell pair, a bit must agree along a cycle of the relative sheet permutation at its two ends. To see the cycles, label the $n^2$ positions by $(r,s)$ modulo $n$, and use the row and column shifts $a,b$ of Sec.~\ref{sec:6}. Pair the indices of $A$, $B$, and $C$ with conjugate copies using, respectively, the identity, $a$, and $ab$. The relative moves from $A$ to $B$, $B$ to $C$, and $C$ back to $A$ are $a$, $b$, and $(ab)^{-1}$: one row step, one column step, or one diagonal step. Each returns after $n$ steps, giving $n$ disjoint loops of length $n$. There are two choices of bit on each loop and a factor $1/2$ from the normalized amplitude and its conjugate at each grid position. One Bell pair therefore contributes $2^n\,2^{-n^2}=2^{n(1-n)}$. The three independent pairs give the cube of this number.
\\

\noindent
For both states each local moment is $4^{1-n}$, so the denominator of Eq.~\eqref{eq:Kgeneral} is $2^{3n(1-n)}$. The common-label numerator is $4^{1-n^2}$, since the entire grid must carry one label. Dividing gives
\[
\mathcal K_{n,\mathrm{common}}=4^{(n-1)(n-2)/2},\qquad
\mathcal K_{n,\mathrm{pairs}}=1,\qquad n=2,3,\ldots .
\]
With two copies along each grid direction, the subtracted measures both vanish. At $n=3$ the common-label state gives $-\ln4/6$, while the state made of separate Bell pairs gives zero. Since the one-party entropies are identical, this difference comes entirely from how the three parties share their labels. It is measured at an integer index without any interpolation. The negative sign means that the subtracted single-party terms exceeded the raw three-party entropy. Continuing the displayed positive powers to real $n$ near one gives limiting measures $\ln2$ and zero, respectively, for this specified continuation of the two finite states.
\\

\noindent
To test the scale of this effect, keep the seed fixed and choose $D_M=2^{M^2}$. The shared label can then be read as a string of $M^2$ independent bits. Each bit contributes $(\ln2)/2$ to the limiting subtracted entropy and zero to the central charge. The seed contributes one copy of its entropy for each of the $M$ factors. Thus the choice of how many bits to share, rather than a change in the seed, produces the quadratic term:
\begin{align}
H_{\C_M}'(1)&=Ma(c_0)-\frac{M^2}{2}\ln2
=\frac {a(c_0)}{c_0}c_M-\frac{\ln2}{2c_0^2}c_M^2,\label{eq:largec}\\
\frac{H_{\C_M}'(1)}{c_M}
&=\frac{a(c_0)}{c_0}-\frac{\ln2}{2}\frac{M}{c_0}.
\label{eq:jointlimit}
\end{align}
There are two ways to make the total central charge large, and they yield different comparisons. At fixed seed $c_0>0$, adding copies sends $c_M=Mc_0$ to infinity. Because the number of binary labels was chosen to be $M^2$, their correction grows as $c_M^2$, faster than the geometric term. No separate large-$c_0$ limit is needed.
At fixed $M$, increasing $c_0$ instead keeps the label contribution fixed while the geometric term grows. If both parameters increase, Eq.~\eqref{eq:jointlimit} shows that the size of the label contribution divided by $c_M$ is $(\ln2/2)(M/c_0)$. A nonzero finite limit of $M/c_0$ gives a label term of the same order as geometry; a diverging ratio gives a larger one. The seed slope $a(c_0)$ remains an independent input in these last two limits.
\\

\noindent
The example answers a question about assumptions. A growing number of vacuum sectors can change entanglement while leaving the stress tensor unchanged. Large central charge alone therefore does not force $H'_\C(1)/c\to0$. The argument makes no claim about an irreducible CFT with a single vacuum and a sparse spectrum.

\section{Discussion and Conclusion}
\label{sec:8}
\noindent
For three branch values, the geometric calculation begins with a simple question: where does a path end if it goes once around a branch value? The three answers, expressed as sheet permutations, specify the cover. The cone angles determine the constant-curvature metric on the base, and lifting it gives the smooth reference metric on this cover. Matching lengths near the tips fixes the local scales. Varying a cone angle, fixing the action in the flat limit, and calibrating with two-point functions give the finite exponent in Eq.~\eqref{eq:mainintro}. The genus fixes the sign of the curvature; a unit of length fixes its nonzero magnitude. These steps do not determine the partition function of the quantum fields on the resulting surface.
\\

\noindent
An application is to compare two theories at the same central charge on the same cover. Their geometric exponents agree, including the area contribution $c(g-1)/6$ and the sphere conversion that together produce the $2d$ term in Eq.~\eqref{eq:mainintro}. If the second coefficient is nonzero, dividing the two three-point coefficients cancels that entire factor. What remains is the ratio of the sphere-normalized partition functions $\Z_\C$, each with its own factor $s_\C^{g-1}$. At grid index three, for example, any difference between the two subtracted entropies comes entirely from the two partition functions on the same torus.
\\

\noindent
The real part of the continued classical cone exponent is $\Pcov$, the \emph{logarithm} of the geometric factor. Thus the magnitude of the exponential agrees with the result from the real covering metric.
The complete correlator still includes $\Z_\C(\Sigma_g)$. Even if the branching orders agree, two covers can give different coefficients. Comparing the two explicit geometric functions does not tell us which classical geometries actually contribute to a gravitational path integral, or which one dominates.
Genus two makes the distinction particularly transparent. Changing from five sheets to eight changes $\Pcov/c$ from approximately $-1.31240$ to $-1.77192$. The continued exponents on either cover have the same imaginary parts $\mp\pi c/6$. Comparing each with its own real covering answer gives zero difference in the real part.
\\

\noindent
What phase could an entanglement measurement detect? For a change of reference metric alone, the state and physical distances are fixed. The geometric factor and the seed partition function acquire inverse phases. The Euclidean measurement is unchanged, regardless of the seed's initial phase. Reversing the sheet labels also proves the reality of the integer three-party grid; the odd partial-transpose moment is real on its cyclic cover at every positive genus.
\\

\noindent
Moving the endpoints is a different test. On the timelike path of Sec.~\ref{sec:6}, a single interval picks up an imaginary contribution. The raw three-party entropy also changes, but the same change appears in half the sum of the three interval entropies. The normalized odd partial-transpose moment has the corresponding cancellation for every $m=2g+1$. These are statements about specified vacuum correlators, a common cutoff, and a specified path. They neither identify the cone phase with a timelike entropy nor fix the phase for an arbitrary state or continuation.
\\

\noindent
For the three-region multi-entropy~\cite{Gadde:2022cqi}, index two gives a sphere, index three a torus, and index four a surface with three handles. All integer indices $n\ge4$ use the negative-curvature formula. The limit near one instead uses the real spherical expression for the geometric exponent. The negative of its slope at one gives the proposed gravity constant $(c/2)\ln(2/\sqrt3)$ in the subtracted entropy. The full entropy also contains $-H'_\C(1)$.
Integer replica values alone do not fix that additional slope~\cite{Gadde:2022cqi,Czech:2026jil}. Eq.~\eqref{eq:uniformbound} is a sufficient condition for it to be smaller than the gravity term at large $c$. Whether this bound holds for an interacting CFT on the replica family remains open.
Two finite states sharpen this distinction. Giving every party the same local density matrix does not fix the integer three-party invariant: one common four-valued label and three independent Bell pairs already give different values at $n=3$. In the shared-label theory, changing the probabilities changes the limiting entropy without changing the central charge. With equal weights, the correction can grow more slowly than $c$, proportionally to $c$, or as $c^2$, depending on the number of sectors. What remains open is the behavior of $H_\C'(1)$ in an irreducible theory with one vacuum, few low-lying states and a controlled replica continuation.
\\

\noindent
Several questions now admit concrete tests.
First, one could compute $Z_\C(\Sigma_n)$ for a specified theory. Sewing cuts a surface into pieces, evaluates the amplitude of each piece, and sums over the states matched along each cut. Conformal blocks organize this sum by the family of intermediate states. For free fields, a determinant instead combines the contributions of the fluctuation modes once the operator and regularization are fixed. Each route supplies information about the theory that the covering map by itself cannot give.
The large automorphism groups of Fermat curves---the invertible holomorphic self-maps---may simplify these calculations without removing the CFT dependence.
\\

\noindent
Second, to control the replica derivative one can seek a bound on a fixed neighborhood of one, or a stronger bound if the neighborhood shrinks. In a calculation based on a dominant saddle, that route would require estimating its competitors and checking for zeros before taking the logarithm.
A large-$c$ estimate at each separate integer index does not supply these conditions.
Establishing them would provide a CFT derivation of the full holographic constant without assuming its value.
\\

\noindent
Third, Eq.~\eqref{eq:manyP} extends the normalization to additional uniformly ramified branch values through the on-shell cone action.
The remaining inputs are the $\nu-3$ accessory parameters and the seed partition function.
For four points, one can fix three positions and vary the fourth. At each new position one adjusts a single coefficient until continuation around \emph{each} independent loop preserves the disk metric. The normalized action then tests how the geometric factor varies with the cross ratio. If the branching above one point is nonuniform, two different cycle lengths there would require two different cone angles to give a full $2\pi$ turn upstairs. A single metric on the base cannot do both, so that case needs a different construction.
\\

\noindent
Finally, Eq.~\eqref{eq:geomhigher} predicts both the limiting geometric entropy and how it changes for nearby replica index. The full slope in Eq.~\eqref{eq:fullslope} also contains derivatives of $H_\C$. Computing the partition function on a specified Fermat surface would test the answer at one integer index. Controlling a continuation near one would test the entropy; controlling its next derivative would test the slope. A match at an integer and a match of derivatives at one are distinct predictions a CFT calculation can address.

\subsection*{Acknowledgments}
\noindent 
The author thanks Xin-Xiang Ju and Ma-Ke Yuan for their helpful discussion and thanks Nan-Peng Ma for his encouragement.


\providecommand{\href}[2]{#2}

\address{
Hetao Institute of  Mathematics and Interdisciplinary Sciences, Shenzhen, Guangdong 518017, China.\\
\email{yefgst@gmail.com}\\
}

\end{document}